\documentclass[11pt,twocolumn,tight,times]{aastex62}
\pdfoutput=1
\usepackage{natbib}
\usepackage{multirow} 
\usepackage{graphicx}
\usepackage{float}
\usepackage{amsmath}
\usepackage{mathrsfs}%
\usepackage{appendix}
\usepackage{CJK}

\def\be{\begin{equation}}
\def\ee{\end{equation}}
\def\bea{\begin{eqnarray}}
\def\eea{\end{eqnarray}}

\shorttitle{Detecting Nearby GW sources via PTA}
\shortauthors{Guo, Lu, \& Yu}

\begin{document}
\begin{CJK*}{UTF8}{gbsn}
\title{On Detecting Nearby Nano-Hertz Gravitational Wave Sources via Pulsar Timing Arrays}

%\correspondingauthor{Youjun Lu}
%\email{luyj@nao.cas.cn}

\author[0000-0001-5174-0760]{Xiao Guo (郭潇)}

\affil{
CAS Key laboratory for computational Astrophysics, 
National Astronomical Observatories, Chinese Academy of Sciences, 20A Datun Road, Beijing 100101, China; $^\dagger$\,luyj@nao.cas.cn}
\affil{
School of Astronomy and Space Science, University of Chinese Academy of Sciences, 19A Yuquan Road, Beijing 100049, China}

\author[0000-0002-1310-4664]{Youjun Lu (陆由俊)$^{\dagger}$}
\affil{
CAS Key laboratory for computational Astrophysics, 
National Astronomical Observatories, Chinese Academy of Sciences, 20A Datun Road, Beijing 100101, China; $^\dagger$\,luyj@nao.cas.cn}
\affil{
School of Astronomy and Space Science, University of Chinese Academy of Sciences, 19A Yuquan Road, Beijing 100049, China}

\author[0000-0002-1745-8064]{Qingjuan Yu (于清娟)$^{\ddagger}$}
\affil{
Kavli Institute for Astronomy and Astrophysics and School of Physics, Peking University, Beijing 100871, China; $^\ddagger$\,yuqj@pku.edu.cn
}
%
%}

\begin{abstract}
Massive binary black holes (MBBHs) in nearby galactic centers, if any, may be nano-Hertz gravitational wave (GW) sources for pulsar timing arrays (PTAs) to detect.  Normally the objective GWs for PTA experiments are approximated as plane waves because its sources are presumably located faraway. For nearby GW sources, however, this approximation may be inaccurate due to the curved GW wave front and the GW strength changes along the paths of PTA pulsar pulses. In this paper, we analyze the near-field effect in the PTA detection of nearby sources and find it is important if the source distance is less than a few tens Mpc, and ignoring this effect may lead to a significant signal-to-noise underestimation especially when the source distance is comparable to the pulsar distances. As examples, we assume a nano-Hertz MBBH source located at either the Galactic Center (GC) or the Large Magellanic Cloud (LMC) according to the observational constraints/hints on the MBBH parameter space,
and estimate its detectability by current/future PTAs. We find that the GC MBBH may be detectable by the Square Kilometer Array (SKA) PTA. It is challenging for detecting the LMC MBBH; however, if a number ($N\gtrsim10$) of stable millisecond pulsars can be found in the LMC center, the MBBH may be detectable via a PTA formed by these pulsars. We further illustrate the near-field effects on the PTA detection of an isotropic GW background contributed mainly by nearby GW sources, and the resulting angular correlation is similar to the Hellings-Downs curve. 
\end{abstract}

\keywords{black hole physics (159),  Galaxy Center (565), gravitational waves (678), Magellanic Clouds (990),  pulsars (1306), supermassive black holes (1663)}

\section{Introduction}
\label{sec:intro}

Pulsar Timing Arrays (PTAs) are aiming at detecting low frequency gravitational waves (GWs) emitting from massive binary black holes (MBBHs; \citealt[][]{1980Natur.287..307B, 2002MNRAS.331..935Y}) and cosmic strings, etc. \citep[e.g.,][]{1978SvA....22...36S, 1979ApJ...234.1100D, mingarelli2015gravitational, 2014gwdd.book.....V, 2011gwpa.book.....C, book:1417639, 2015SCPMA..58.5748B, 2019arXiv190308183T, 2009MNRAS.394.2255S,2010CQGra..27h4016S, 2013CQGra..30x4009S}. Current PTAs include the Parkes PTA (PPTA; \citealt{manchester2013the})\footnote{\url{http://www.atnf.csiro.au/research/pulsar/ppta/}}, the European PTA (EPTA; \citealt{2013CQGra..30v4009K})\footnote{\url{http://www.epta.eu.org/}}, the North American Nanohertz Observatory for Gravitational Waves (NANOGrav; \citealt{2013CQGra..30v4008M,2019BAAS...51g.195R})\footnote{\url{http://nanograv.org/}},  the Indian Pulsar Timing Array (InPTA; \citealt{2018JApA...39...51J}), and the Chinese pulsar timing array (CPTA). The first four combined together to form the International PTA (IPTA; \citealt{2013CQGra..30v4010M,2019MNRAS.490.4666P})\footnote{\url{http://www.ipta4gw.org/}}. NANOGrav, PPTA, EPTA, and IPTA have all shown the existence of a signal from common-spectrum process in the data, which might be due to the GW background (GWB) but lack significant evidence for quadrupolar spatial correlation \citep{2020ApJ...905L..34A, 2021ApJ...917L..19G, 2021MNRAS.508.4970C, 2022MNRAS.510.4873A}. 
This signal is possibly (partly) due to the ephemeris systematics and/or a single pulsar in the PTA data sets \citep{2021ApJ...923L..22A}. It was proposed to be even due to a non-Einsteinian polarization mode (scalar-transverse mode) signal \citep[][]{2021SCPMA..6420412C}, but one should be cautious with the detailed data analysis and the probability for the existence of the scalar-transverse mode could be insignificant \citep{2021ApJ...923L..22A}. Nevertheless, it may suggest that the nano-Hertz GWB is close to be detected in the near future.

MBBHs with mass ratio $\gtrsim 0.01$ are predicted to exist in about a fraction of a few to ten percent of nearby galaxies \citep[][]{chen2020dynamical}, some of which are also expected to be detected by PTAs in the future. These individual MBBHs (with distances at least many Mpcs away; \citealt{2010CQGra..27h4016S, 2011MNRAS.414...50D, 2016MNRAS.459.1737S, 2019MNRAS.490.4666P, 2022MNRAS.510.5929C, 2016ApJ...817...70T,2021ApJ...914..121A}) are usually much more distant than those of the stable millisecond pulsars (MSPs) in the Milky Way adopted in the PTAs (typically hundreds to ten thousands of pc away from the Earth; \citealt{10.1093/mnras/stw347, 2005AJ....129.1993M}). In this case, GW emitted from an individual source can be regarded as the plane wave in the data analysis as done in many previous studies \citep[e.g.,][]{2017LRR....20....2R}. 

It has been proposed that MBBHs may even exist in our Galactic center (GC), or some nearby galaxies, such as Large Magellanic Cloud (LMC), etc. \citep[e.g.,][]{2017ApJ...850L...5T, 2019ApJ...871L...1T, 2003ApJ...599.1129Y, 2006ApJ...641..319P,2007ApJ...666..919Y, RevModPhys.82.3121, 2018MNRAS.tmp.2529G,2021ApJ...914..121A,2017NatAs...1..886M}, which can also be potential sources for future PTAs. However, these MBBHs are quite close to the Earth, with distances less than a few tens kpc. Therefore, the conventional plane-wave assumption is probably inaccurate or even invalid when considering the detectability of these nearby MBBHs, if exist, via PTAs \citep[e.g.,][]{2011MNRAS.414...50D, 2012ApJ...752...67K, 2021MNRAS.505.4531M}. In this paper, we construct a general framework for studying the detectability of nano-Hertz GWs emitted from nearby MBBHs, if any, via PTAs, by considering that the propagation directions and amplitude of the GW from nearby sources are different at different locations along the path of pulses from a pulsar to the Earth \citep[for comparison, see][for distant GW sources]{2009PhRvD..79h4030A,book:1417639,mingarelli2015gravitational,2021arXiv210513270T,2014gwdd.book.....V}. 

\citet{2012ApJ...752...67K} discussed the problem to detect the GW from a hypothetical MBBH in the GC. They mainly considered the case where all PTA pulsars were assumed to be located in the neighborhood of the GC. However, almost all known MSPs adopted in the current PTAs are not that close to the GC  \citep{2013CQGra..30v4010M, 10.1093/mnras/stw347, 2019MNRAS.490.4666P}, and no MSP is found in directions close to the GC, yet \citep[e.g.,][]{2005AJ....129.1993M}. Therefore, it is interesting to consider more realistic cases, in which MSPs adopted are the same as those adopted in current PTAs or similar to those expected from future surveys by Five-hundred meter Aperture Spherical Telescope (FAST; \citealt{2011IJMPD..20..989N,2009A&A...505..919S}) and/or Square Kilometer Array (SKA; \citealt{Lazio2013SKA, 2017PhRvL.118o1104W}). In such a study, the GWs emitted from the hypothetical MBBHs cannot be approximated as plane waves because they are so close to PTA(s) and thus the ``near-field'' effect must be considered. Here the ``near-field'' effect mean the effects of GWs from nearby sources by including both the curvature of the GW wavefront and the change of the GW amplitude and phase along the paths of PTA pulsar pulses. It is worthy to note that the definition of the ``near-field'' effect considered in \citet{2012ApJ...752...67K} is different from ours, which refer to the post-Newtonian effect (or Einstein delay) and tidal effects due to the MBBH on the motion of nearby pulsars (or Roemer delay) that was not considered in our paper for simplicity. Nevertheless, the PTA geometrical configuration considered in \citet{2012ApJ...752...67K} can be regarded as a special case of those in the present paper (see Appendix~\ref{sec:Kocsis}).

\citet{2021MNRAS.505.4531M} recently developed, for the first time, a Fresnel formalism to consider the non-planar wave front for nearby GW sources, which is a treatment closer to the reality compared to the plane-wave approximation. For a nearby GW source, if any, the Fresnel approximation is even not sufficient. The reason is that the Fresnel formalism is still only valid under the far-field approximation, even though it improves the plane-wave approximation. In the present paper we consider the accurate geometrical configuration without making those approximations and calculate the near-field effect for assumed nearby GW sources numerically, which is distinguished from that presented in \citet{2021MNRAS.505.4531M}. 

This paper is organized as follows. We provide a general framework for considering the detectability of both nearby and distant nano-Hertz GW sources via PTAs in Section~\ref{sec:framework}. Then we consider the cross correlation between the signals from two MSPs in the near-field regime in Section~\ref{sec:Two} both for individual sources and a GWB contributed mostly by nearby sources. Then we investigate detection strategies for PTAs (the matched-filtering and cross-correlation method) in Section~\ref{sec:PTA} and calculate the influence of the near-field effect for PTA experiments in Section~\ref{sec:Effect}. In Section~\ref{sec:Application}, we apply the framework to a hypothetical MBBH in the GC or nearby galaxies to calculate the signal-to-noise ratio (SNR) of the GWs emitted from these MBBHs. Conclusions and discussions are given in Section~\ref{sec:Conclusion}. 

\section{Perturbations on the propagation of pulses from pulsars by the GWs from an MBBH}
\label{sec:framework}

In this section, we introduce a general framework for calculating the redshift of frequency of pulses radiated from distant MSPs due to metric perturbations by GWs from distant sources. It can be reduced to the far-field approximation that is generally adopted in the PTA analysis.  

\subsection{General Framework}
\label{sec:General}

\begin{figure*}
\centering
\includegraphics[width=\textwidth]{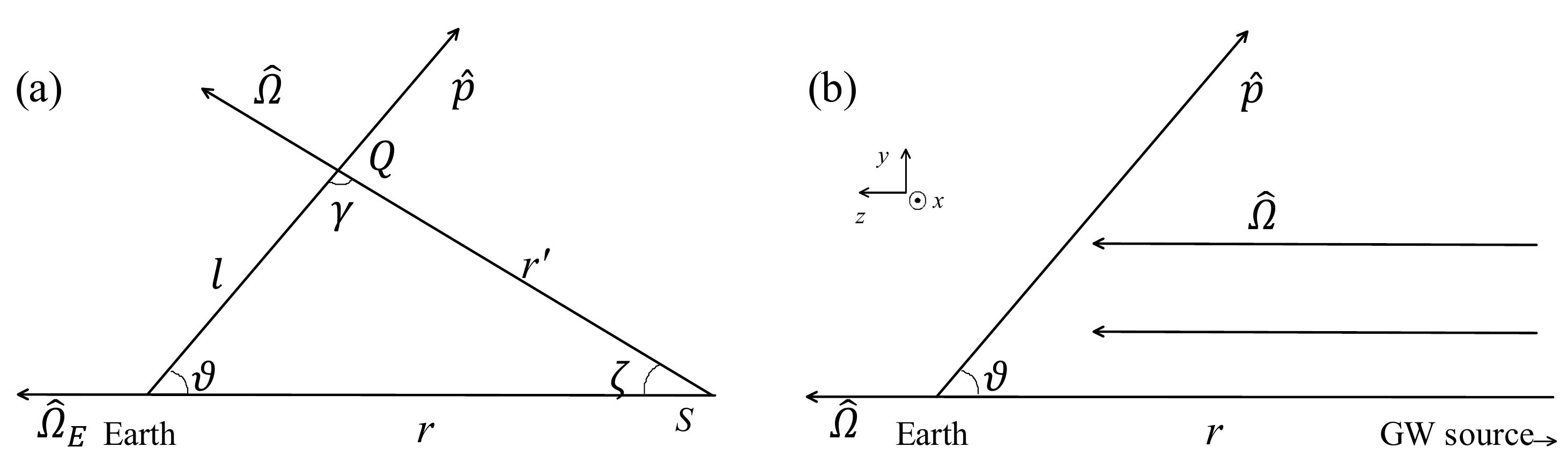}
\caption{Schematic diagrams for PTA experiments with a single MSP (geometrical configuration). Left diagram (a) is for those cases that the distance from the target MSPs to Earth ($L$) is comparable to or not too much smaller than the distance of the GW source (point $S$) to Earth ($r$). Right diagram (b) for those cases with $L\ll r$, in which the GW can be approximated as plane wave. In diagram (a), $Q$ denotes any point on the path of pulses from the MSP to Earth, $l$ and $r'$ denote the distance between $Q$ and Earth, and between $Q$ and GW source, respectively. $\gamma$, $\zeta$, and $\vartheta$ represent the angles between GW propagation direction $\hat{\Omega}$ and that from Earth to MSP $\hat{p}$, between $\hat{\Omega}$ and that from GW source to Earth $\hat{\Omega}_{\rm E}$, and between $-\hat{\Omega}_{\rm E}$ and $\hat{p}$, respectively. In diagram (b), the GW source is far from the Earth,  the GW propagation directions $\hat{\Omega}$ are assumed to be the same at different points $Q$. For the far-field regime, we adopt a Cartesian coordinate system shown at the top-left of diagram (b), where $x$-axis is perpendicular to the paper, and its positive direction $\hat{e}_1$ points to us. 
} 
% 
%\label{fig:schematic}
\label{fig:f1}
\end{figure*}

Figure~\ref{fig:f1} shows the schematic diagrams for PTA experiments with a single MSP, for both the general case [left diagram (a)] and the far field approximation [right diagram (b)]. In the general case, the distances of GW sources $r$ could be comparable to, smaller than, or larger than distances of PTA MSPs $L$. It includes the near field case where $r\lesssim L$. In the far-field approximation, the distances of GW targets are much larger than those of the PTA MSPs and the GW radiation can be securely approximated as the plane wave [see diagram (b)].

The redshift of frequency $\nu$ of pulses from the MSP, received by an observer at time $t_{\rm E}$, due to the perturbation of GWs can be expressed as \citep[e.g.,][]{2009PhRvD..79h4030A}
\be
z(t_{\rm E})=\sum_a\int^{L}_{0}\mathscr{F}^a(l) \frac{d h_a|_Q}{dl}dl,
\label{eq:znear}
\ee
where $z(t_{\rm E})=(\nu(t_{\rm E})-\nu_0)/\nu_0$ represents the frequency shift and the integral is from the Earth ($l=0$) to MSP ($l=L$),  $\nu_0$ and $\nu(t_{\rm E})$ represent the received frequency of pulses without and with including GW induced redshift, respectively, $a=+,\times$, and the antenna pattern functions in the source frame
\bea
\mathscr{F}^+&=&F^+\cos2\psi-F^\times\sin2\psi, \\
\mathscr{F}^\times&=&F^+ \sin2\psi+F^\times \cos2\psi.
\label{eq:patternf}
\eea
In the above equations, $F^+$ and $F^\times$ are the antenna pattern functions for PTA in the detector frame, $\psi$ is the polarization angle \citep[defined in][see Fig.~1 therein]{1994PhRvD..49.6274A}, i.e., the rotation angle ($-\pi<\psi\leq\pi$) of the coordinate axis from a basis vector $\hat{e}_1$ to a principal reference direction in the source frame. Here $\hat{e}_1$ is defined to be the unit vector perpendicular to the pulsar-Earth-GW source plane, and the reference direction is defined as  $\hat{n}\times\hat{\Omega}$ in the plane perpendicular to the GW propagation direction $\hat{\Omega}$, with $\hat{n}$ representing the normal vector of the MBBH orbital plane. We also define a unit vector $\hat{e}_2$, which is in the pulsar-Earth-GW source plane and perpendicular to $\hat{\Omega}$, thus $(\hat{e}_1, \hat{e}_2, \hat{\Omega})$ can be taken as the basis vectors of a rectangular coordinate system. The inclination angle $\iota$ is defined as the angle between the GW propagation direction $\hat{\Omega}$ and $\hat{n}$ \citep[e.g., see][]{2015CQGra..32e5004M, 2014MNRAS.444.3709Z, 2015MNRAS.449.1650Z, 2016MNRAS.461.1317Z, 2012ApJ...756..175E, 2012PhRvD..85d4034B} and $\iota=\iota_{\rm E}$ when $\hat{\Omega}=\hat{\Omega}_{\rm E}$, $L$ is the distance from MSP to  Earth, $l$ is the distance between the Earth and point $Q$ on the path from MSP to the Earth, and $h_a|_Q$ denotes the GW signal at point $Q$ encoded in the pulsar pulses received by an observer at a given time $t_{\rm E}$. For definitions of these relevant geometric quantities, see Figure~\ref{fig:f1}.

For continuous GW, we have
\begin{eqnarray}
h_+ & = & A_+ h_0\cos\Phi(t), 
\label{eq:hplus}\\
h_\times & = & A_\times h_0\sin\Phi(t).
\label{eq:hcross}
\end{eqnarray}
where $h_0=\frac {4{G}^{5/3}\mathcal{M}^{5/3}(\pi f)^{2/3}}{{c}^4r}$, $A_+(\iota)=\frac{1+\cos^2\iota}{2}$, $A_\times(\iota)=\cos\iota$, $\mathcal{M}$ is the chirp mass of this system and $\Phi(t)$ is the phase of GW. For convenience, we put the dependence of the GW signal on $\iota$ out of $h_a|_Q$ (along with $\mathscr{F}^a$) rather than directly in $h_a|_Q$ in our following analysis. The antenna pattern functions are given by
\be
F^+=\frac{1}{2}\frac{\hat{p}^i\hat{p}^j}{1+\hat{p}\cdot\hat{\Omega}}e^+_{ij}=\frac{1}{2}\frac{(\hat{p}\cdot\hat{e}_1)^2-(\hat{p}\cdot\hat{e}_2)^2}{1+\hat{p}\cdot\hat{\Omega}},
\label{eq:Fplus}
\ee
and
\be
F^\times=\frac{1}{2}\frac{\hat{p}^i\hat{p}^j}{1+\hat{p}\cdot\hat{\Omega}}e^\times_{ij}=\frac{(\hat{p}\cdot\hat{e}_1)(\hat{p}\cdot\hat{e}_2)}{1+\hat{p}\cdot\hat{\Omega}},
\label{eq:Fcross}
\ee
respectively, where $\hat{p}$ is a unit vector pointing from the Earth to MSP, and $\hat{p}^i$ and $\hat{p}^j$ represent the  components of $\hat{p}$ with $i=1, 2$, $j=1, 2$. The components of the basis tensor $\mathbf{e}^a$ represent by $e^{a}_{ij}$, where 
\begin{eqnarray}
\mathbf{e}^+ & =& \hat{e}_1\hat{e}_1-\hat{e}_2\hat{e}_2, \\
\mathbf{e}^\times & = & \hat{e}_1\hat{e}_2+\hat{e}_2\hat{e}_1,
\end{eqnarray}
where
$$
\hat{e}_1=\frac{\hat{p}\times\hat{\Omega}}{|\hat{p}\times\hat{\Omega}|},
$$
$$
\hat{e}_2=\hat{e}_1\times\hat{\Omega}.
$$
In the near-field regime, $\hat{\Omega}$ are different at different $Q$ between the Earth and MSP, the inclination and polarization angles ($\iota$, $\psi$) vary with $l$ significantly. Thus GW cannot be regarded as the plane wave in the near-field regime with $L$ comparable to or smaller than $r$.
Hence $\mathscr{F}^a$ is also a function of $l$, thus it cannot be separated from the integral, which is different from that adopting the far-field approximation  \citep[e.g., see][]{2009PhRvD..79h4030A}.

We denote the phase of the GW received by an observer at time $t_{\rm E}$ as $\Phi_{\rm E}(t_{\rm E})$. The phase of the GW at point $Q$ ($\Phi|_Q$) encoded in pulsar pulses received by the observer at $t_{\rm E}$ is thus related to $\Phi_{\rm E}$ due to the time delay $(l+r'-r)/c$ as
\be
\Phi|_Q =\Phi_{\rm E}\left(t_{\rm E}-\frac{l+r'-r}{c}\right),
\ee
where $r$ ($r'$) is the distance between the GW source and Earth ($Q$ point). Since amplitude $h_0(r)\propto 1/r$, we have
\be
h_0|_Q = h_0\left.\left(t_{\rm E}-\frac{l+r'-r}{c}\right)\right|_{\rm E} \frac{r}{r'}
\ee
Therefore, $\left.h_{a}\right|_Q$ can be expressed in the inverse Fourier transform as
\be
\left.h_{a}\right|_Q=\frac{rA_a(\iota)}{r'A_a(\iota_{\rm E})}\int_{-\infty}^{\infty} {\rm d}f e^{i2\pi f \left(t_{\rm E}-\frac{l+r'-r}{c} \right)}\left.\tilde{h}_a(f)\right|_{\rm E},
\label{eq:hFourier}
\ee
where $\left. \tilde{h}_a(f)\right|_{\rm E}$ is the strain spectrum of the GW signal $\left. h_a(t_{\rm E})\right|_{\rm E}$ at the Earth.

Combining Equations~\eqref{eq:hFourier} and \eqref{eq:znear}, the Fourier transform of the redshift is given by
\bea
\tilde{z}(f) & = & \sum_a\frac{\left. \tilde{h}_a(f)\right|_{\rm E}}{A_a(\iota_{\rm E})}\int^{L}_0 dl \mathscr{F}^a(l) {\frac{d}{dl}} \left(\frac{rA_a(\iota)}{r'}e^{-i2\pi f(\frac{l+r'-r}{c})}\right) \nonumber \\
& \equiv & \sum_a\mathscr{P}^a(f)\tilde{h}_{a,\rm o}(f).
\label{eq:z=ph}
\eea
In the above Equation, $\left. \tilde{h}_a(f)\right|_{\rm E}$ and $A_a(\iota_{\rm E})$ are independent of $l$, and the integral
\be
\mathscr{P}^a(f) \equiv \int^{L}_0 dl \mathscr{F}^a(l){\frac{d}{dl}}\left(\frac{rA_a(\iota)}{r'}e^{-i2\pi f(\frac{l+r'-r}{c})}\right),
\label{eq:pdefine}
\ee
is integrated over $l$ from $0$ to $L$, $\frac{rA_a(\iota)}{r'}e^{-i2\pi f(\frac{l+r'-r}{c})}$ is a complex function of $l$, and its differential expression is too tedious to be explicitly shown here; $\mathscr{P}^a(f)$ represents the response of a PTA pulsar to the GW signal, $\tilde{h}_{a,\rm o}(f)\equiv \left. \tilde{h}_a(f)\right|_{\rm E}/A_a(\iota_{\rm E})$ equals the GW strain in the case with an optimal orientation ($\iota_{\rm E}=0$) and it is invariant for any $\iota_{\rm E}$ ($\neq0$). (For the expression of $r'$ in the above equation, see Appendix~\ref{sec:concrete}.) 

According to Equation~\eqref{eq:z=ph}, redshift $\tilde{z}(f)$ can be obtained given known GW spectrum, distances (which can be measured accurately with timing parallax as proposed in \citealt{2011MNRAS.414.3251L}, see also \citealt{2021PhRvD.104f3015D}), and directions to PTA pulsars, which suggests that the standard matched-filtering method \citep{2015CQGra..32a5014M,book:1417639, 2011gwpa.book.....C} can be adopted to extract GW signals and properties of GW systems. The optimum filter can be defined as $\tilde{z}(f)/S_{\rm n}(f)$, with $S_{\rm n}(f)$ describing the power spectrum density (PSD) of the noise for a given PTA. The SNR $\varrho$ is then given by
\be
\varrho^2=\int^{\infty}_{0}df \frac{4|\tilde{z}(f)|^2}{S_{\rm n}(f)}.
\ee

In the literature, for the detection of the GWB, it is straightforward to prove that \citep{1989thyg.book.....H, book:1417639, 2018arXiv180301944R}
\be
\overline{(\mathscr{F}^{+})^2}= \overline{(\mathscr{F}^{\times})^2},
\ee
\be
\overline{\mathscr{F}^{+}\mathscr{F}^{\times}}=
\overline{\mathscr{F}^{\times}\mathscr{F}^{+}}=0.
\ee
The long overbar symbol $\overline{\cdots }$ in the above Equations represent the sky and polarization average defined by
\be
\overline{X} \equiv \frac{1}{4 \pi^{2}} \int_{0}^{\pi} d \psi_{\rm E} \int d^2 \hat{\Omega}_{\rm s} X, 
\ee
where $\hat{\Omega}_{\rm s}$ represents the position of GW source. 

For the detection of individual MBBHs, the position of the GW source is fixed, and the average should be taken over the sky for the directions of PTA MSPs ($\hat{\Omega}_{\rm p}$).  
Note here that the cases for individual MBBHs and GWB are somewhat symmetric, which are an average over many pulsars (for a single source) and an average over many GW sources (background), respectively.
According to the definition of $\mathscr{P}^{a}(f)$ in Equation~(\ref{eq:pdefine}), we have
\be
\overline{\mathscr{P}^{+*}\mathscr{P}^{\times}}=
\overline{\mathscr{P}^{\times*}\mathscr{P}^{+}}=0,
\ee
if the PTA MSPs are uniformly distributed. In the estimation of SNR, we define a mean quantity as
\be
\chi^2\equiv \frac{\overline{ |\mathscr{P}^{+}|^2+|\mathscr{P}^{\times}|^2}}{2},
\label{eq:chidefine}
\ee
which represents the geometrical effect of the spatial distribution of PTA pulsars relative to the GW propagation direction, and $\chi$  usually depends on $\iota_{\rm E}$ and may also depend on $L$.\footnote{In some literature, $\chi$ is denoted as the signal response function $\mathcal{R}(f)$.} We use $\bar{\chi}$ to represent the average of $\chi$ over $\iota_{\rm E}$ in section~\ref{subsec:chi} (Eqs.\ref{eq:chibar1} and \ref{eq:chibar2}).

The root-mean-square (RMS) value of the GW strain in the frequency domain is defined as \citep{2019arXiv190302049G} 
\be
\left|\tilde{h}_{\rm o}(f)\right|^2 \equiv \left|\tilde{h}_{+,\rm o}(f)\right|^2 + \left|\tilde{h}_{\times,\rm o}(f)\right|^2,
\label{eq:hrms}
\ee
which is independent of $\iota_{\rm E}$, and then we have
\be
\left|\tilde{z}(f)\right|^2\approx \chi^2(f)\left|\tilde{h}_{\rm o}(f)\right|^2.
\ee
The averaged SNR can thus be estimated as
\be
\varrho^2=\int^{\infty}_{0} df  \frac{4\chi^2(f)|\tilde{h}_{\rm o}(f)|^2}{S_{\rm n}(f)}.
\ee

\subsection{Reduction to the Far-Field Approximation}
\label{sec:FarField}

Targets of nano-Hertz GWs are mostly inspiral MBBHs in galactic centers far away from the Earth and the distances from the Earth to those GW sources ($\gg 1$\,Mpc) are much larger than the distances from the Earth to those MSPs (on the order of kpc) that are monitored by PTAs. Therefore, one may approximate the GWs emitted from those distant MBBHs as plane waves when considering its perturbation on the propagation of pulsar pulses to Earth.

If $r\gg L$, $\zeta\approx0$, then $\hat{\Omega}$ and $\iota$ can be approximated as non-variable constants $\hat{\Omega}_{\rm E}$, $\iota_{\rm E}$, $r'=r-l\cos\vartheta$, $r'/r\approx1$, $A_a(\iota)/A_a(\iota_{\rm E})\approx1$. Therefore, Equation \eqref{eq:z=ph} can be reduced to
\bea
\tilde{z}(f) &  = & \sum_a\tilde{h}_a(f)\mathscr{F}^a \cdot \left(e^{-i2\pi fL(1-\cos\vartheta)/c}-1\right) \nonumber \\
& \equiv & \sum_a\mathscr{P}^a_{\rm f}(f)\tilde{h}_{a,\rm o}(f),
\eea
and 
\be
\mathscr{P}^{a}_{\rm f}(f)=\left(e^{-i2\pi fL(1-\cos\vartheta)/c}-1\right) \mathscr{F}^{a}A_a.
\label{eq:Pfar}
\ee
This is the expression resulting from the far-field approximation adopted in many previous works  \citep[e.g.,][]{1978SvA....22...36S,1979ApJ...234.1100D}.
 
In the near-field regime, however, both the amplitudes and phases of GWs at different points $Q$ may vary significantly, different from that in the case adopting the far-field approximation. The relative difference of GW amplitudes at $r$ and $r'$ is $[h_0(r) -h_0(r')] /h_0(r)= 1-r/r' \lesssim 10^{-3}$ when $r> 10^3 L\sim O({\rm Mpc})$. Therefore, the amplitude difference is negligible if $r \gg 1$\,Mpc. If the pulsar-to-Earth line is perpendicular to the Earth-to-GW source line, the difference between GW propagation direction at pulsar and that at Earth is the largest. The maximum distance difference between $r'$ and $r$ is then $ |\sqrt{r^2+L^2} -r |\sim L^2/2r$. If $L^2/2r \ll \lambda_{\rm GW}/2$ (e.g., $\sim O(1 {\rm pc})$ for $f=10^{-8}$\,Hz), the phase difference can be ignored since it leads to a distortion of wave front much less than a half wavelength assuming pulsar distance $\sim $\,kpc. Therefore, the phase difference can be nearly ignored as well if $r\gg 10$\,Mpc. We conclude that the far-field approximation can be safely adopted if the distances of GW sources are much larger than $10$\,Mpc and the distances of PTA pulsars $\sim 1$\,kpc, while the near-field effect must be considered if otherwise.

\section{Cross-Correlation of GW Signals}
\label{sec:Two}

In the previous section, we have considered the case of single MSP in the near-field regime. Below we consider the cross-correlations of GW signals in the time of arrival (TOA) data series of two MSPs for  individual monochromatic GW  sources (Section~\ref{subsec:monofGW}), non-monochromatic individual GW sources (Section~\ref{subsec:nonmonofGW}) in the near-field regime, and the near-field effect on the Hellings-Downs curve for a GWB (Section~\ref{sec:HD}).  

\subsection{Individual monochromatic GW sources}
\label{subsec:monofGW}

The cross-correlation method can be also adopted to detect individual sources by two MSPs (or more) as an analogy to the method for the stochastic GWB  \citep{2009PhRvD..79h4030A,1983ApJ...265L..39H,book:1417639,2015MNRAS.451.2417R,2021arXiv210513270T}. If
\be
\left<\tilde{h}^*_{a,\textrm{o}}(f)\tilde{h}_{a',\textrm{o}}(f')\right>=\frac{1}{2}\delta_{aa'}\delta(f-f')S_h(f)
\label{eq:Sh}
\ee
can be applied to an individual source ( e.g., individual monochromatic GW sources), where $\tilde{h}_a(f)$ and $\tilde{h}_{a'}(f')$ are the GW frequency spectra at the Earth encoded in the TOA data series of these two MSPs, respectively, $S_h(f)$ is the GW PSD, and $\left<\cdots \right>$ represents an ensemble average over many noise realizations (in reality, it can be replaced by a time average for a stationary stochastic noise). From Equation~\eqref{eq:z=ph}, we have
\bea
\left< \tilde{z}^*_1(f)\tilde{z}_2(f')\right> & = & \left< \sum_a\sum_{a'}\mathscr{P}_1^{*a}(f)\mathscr{P}_2^{a'}(f')
\tilde{h}^*_{a,\textrm{o}}(f)\tilde{h}_{a',\textrm{o}}(f')\right>  \nonumber \\
& = & \frac{1}{2\beta_{12}}\delta(f-f')S_h(f)\Gamma_{12}(f),
\label{eq:z12=ph}
\eea
for two MSPs, where the overlap reduction function (ORF)\footnote{Here the ORF in equation~\eqref{eq:ORF} is defined for individual sources, not for GWBs to be discussed in Section~\ref{sec:HD}.} in the near-field regime is defined as
\be
\Gamma_{12}(f) \equiv\beta_{12}\sum_a\mathscr{P}^{*a}_1(f)\mathscr{P}^a_2(f),
\label{eq:ORF}
\ee
and a normalization constant $\beta_{12}$ is chosen for making $\Gamma_{12}(f)=1$ for coincident co-aligned detectors.

Similar to \citet{2009PhRvD..79h4030A}, the cross-correlation statistic can be defined as
\be
S=\int^{\infty}_{-\infty}df\int^{\infty}_{-\infty}df'\delta_T(f-f')\tilde{s}^{*}_1(f)\tilde{s}_2(f')\tilde{Q}(f'),
\ee
where $\tilde{s}(f)$ is the Fourier transform of $s(t)=z(t)+n(t)$ with $n(t)$ representing the stochastic noise, and $\tilde{Q}(f')$ is a filter, $\delta_T(f)=\sin(\pi fT)/(\pi f)$, and $T$ is the observation time span. If the noise is stationary and Gaussian, then the mean of $S$ is
\bea
\mu=\langle S\rangle& =& \int^{\infty}_{-\infty}df\int^{\infty}_{-\infty}df'\delta_T(f-f')\langle\tilde{z}^{*}_1(f)\tilde{z}_2(f')\rangle\tilde{Q}(f') \nonumber \\
& = &\frac{T}{2\beta_{12}}\int^{\infty}_{-\infty}df S_h(f)\Gamma_{12}(|f|)\tilde{Q}(f).
\eea

Assuming that the noise $n(t)$ is much greater than the signal $z(t)$, the variance is
\be
\sigma^2=\langle S^2\rangle-\langle S\rangle^2\approx\langle S^2\rangle=\frac{T}{4}\int^{\infty}_{-\infty}dfS_{\rm n1}(f)S_{\rm n2}(f)|\tilde{Q}(f)|^2,
\ee
where 
\be
\left< \tilde{n}^{*}_i(f)\tilde{n}_i(f')\right>=\frac{1}{2}\delta(f-f')S_{{\rm n}i}(f) \nonumber 
\label{eq:Sn}
\ee
for $i=1,2$.\footnote{According to the symmetry of Eqs.~\eqref{eq:Sh} and \eqref{eq:Sn}, $S_h(f)$ and $S_{{\rm n}i}(f)$ must be real functions, $S_h(-f)=S_h^*(f)=S_h(f)$, and $S_{{\rm n}i}(-f)=S_{{\rm n}i}^*(f)=S_{{\rm n}i}(f)$. However, $\Gamma_{12}(f)$ is a complex function, and $\Gamma_{12}(-f)=\Gamma^*_{12}(f)\neq\Gamma_{12}(f)$ in general.}
Defining an inner product as $$(A,B)\equiv \int^{\infty}_{-\infty}dfA^*(f)B(f)S_{{\rm n}1}(f)S_{{\rm n}2}(f),$$ then the mean and its variance can be rewritten as
\be
\mu=\frac{T}{2\beta_{12}}\left(\tilde{Q}^*(f),\frac{S_h(f)\Gamma_{12}(|f|)}{S_{{\rm n}1}(f)S_{{\rm n}2}(f)}\right),
\ee
and
\be
\sigma^2\approx\frac{T}{4}\left(\tilde{Q}^*(f),\tilde{Q}^*(f)\right),
\ee
and the SNR is defined as
\be
\varrho^2=|\mu|^2/\sigma^2.
\ee
According to the Schwartz inequality $|(A,B)|^2\le(A,A)(B,B)$, the optimum filter is
\be
\tilde{Q}^*(f)=\frac{S_h(f)\Gamma_{12}(|f|)}{S_{{\rm n}1}(f)S_{{\rm n}2}(f)},
\ee
and the maximum SNR is
\be
\varrho^2=\frac{T}{\beta_{12}^2}\left(\frac{S_h(f)\Gamma_{12}(|f|)}{S_{\rm n1}(f)S_{\rm  n2}(f)},\frac{S_h(f)\Gamma_{12}(|f|)}{S_{\rm n1}(f)S_{\rm n2}(f)}\right)
\ee
i.e.
\be
\varrho^2=\frac{2T}{\beta_{12}^2}\int^{\infty}_{0}df \frac{S_h^2(f)\Gamma_{12}^2(|f|)}{S_{\rm n1}(f)S_{\rm n2}(f)}.
\label{eq:SNR_Gamma12}
\ee

\subsection{Individual non-monochromatic GW sources}
\label{subsec:nonmonofGW}

For individual non-monochromatic GW sources [not satisfying Equation~\eqref{eq:Sh}], the mean of the cross-correlation statistics may be then generally defined as
\bea
\mu & \equiv & \int^{\infty}_{-\infty}df \langle\tilde{z}^{*}_1(f)\tilde{z}_2(f)\rangle\tilde{Q}(f) \nonumber \\
&=& \left(\tilde{Q}^*(f),\frac{\langle\tilde{z}^{*}_1(f)\tilde{z}_2(f)\rangle}{S_{\rm n1}(f)S_{\rm n2}(f)}\right).
\eea
Similarly, the optimum filter and the maximum SNR are given by
\be
\tilde{Q}^*(f)=\frac{\langle\tilde{z}^{*}_1(f)\tilde{z}_2(f)\rangle}{S_{\rm n1}(f)S_{\rm n2}(f)},
\ee
and 
\be
\varrho^2=\frac{8}{T}\int^{\infty}_{0}df \frac{|\tilde{z}_1(f)|^2|\tilde{z}_2(f)|^2}{S_{\rm n1}(f)S_{\rm n2}(f)},
\ee
respectively, consistent with those given in \citet{2015CQGra..32e5004M}. According to Equations~\eqref{eq:chidefine}, \eqref{eq:hrms}, and \eqref{eq:z12=ph}, the mean SNR can be roughly estimated as 
\be
\varrho^2=\frac{8}{T}\int^{\infty}_{0}df \frac{\chi_1^2\chi_2^2|\tilde{h}_{\rm o}(f)|^4}{S_{\rm n1}(f)S_{\rm n2}(f)}.
\label{eq:SNRchi12}
\ee
If the different polarization states of GW are independent from each other $\langle\tilde{h}^*_a(f)\tilde{h}_{a'}(f)\rangle=\delta_{aa'}\langle\tilde{h}^*_a(f)\tilde{h}_a(f)\rangle$, Equation~\eqref{eq:SNRchi12} can be replaced by 
\be
\varrho^2=\frac{8}{T\beta_{12}^2}\int^{\infty}_{0}df \frac{|\Gamma_{12}(f)|^2|\tilde{h}_{\rm o}(f)|^4}{S_{\rm n1}(f)S_{\rm n2}(f)}.
\label{eq:SNRgamma}
\ee

\subsection{Near-field effects on the Hellings-Downs curve}
\label{sec:HD}

In the traditional PTA data analysis for the GWB detection, it is assumed that the GWB is due to faraway GW sources and the angular correlation between the responses of different pulsars to the GWB should provide critical evidence for the existence of such a GWB (if any) \citep[][the so-called Hellings-Downs curve]{1983ApJ...265L..39H}. If the GWB is mainly contributed by many nearby GW sources, one may think there might be some near-field effects on the detection of such a GWB, and in this case the resulting angular correlation may be different from the Hellings-Downs curve. Below we estimate the near-field effect on the angular correlation between the responses of different PTA pulsars to the GWB, in addition to the main goal of the present paper that is to consider the near-field effect for the PTA detection of individual sources.

The total redshift due to the GWB $z^{\rm b}(t)$ can be expressed as the superposition of redshift $z(t,\hat{\Omega}_{\rm E})$ of many individual sources from all directions $\hat{\Omega}_{\rm E}$, i.e.,
\be
z^{\rm b}(t)=\int d^2\hat{\Omega}_{\rm E}z(t,\hat{\Omega}_{\rm E}).
\label{eq:z_background}
\ee
For a stationary, Gaussian, isotropic, unpolarized GWB, we have \citep{2009PhRvD..79h4030A, 1983ApJ...265L..39H, book:1417639, 2015MNRAS.451.2417R, 2021arXiv210513270T}
\be
\left<\tilde{h}^*_a(f,\hat{\Omega}_{\rm E})\tilde{h}_{a'}(f',\hat{\Omega}_{\rm E}')\right>=\frac{1}{8\pi}\boldsymbol{ \delta}(\hat{\Omega}_{\rm E}-\hat{\Omega}_{\rm E}')\delta_{aa'}\delta(f-f')S_h(f).
%
%\label{eq:ShGWB}
%
\ee
Combining Equations~\eqref{eq:z=ph} and \eqref{eq:z_background}, we obtain
\bea
\left< \tilde{z}^{\rm b*}_1(f)\tilde{z}^{\rm b}_2(f')\right>=  \frac{1}{2\beta_{12}}\delta(f-f')S_h(f)\Gamma_{12}^{\rm b}(f),
%
%\label{eq:z12=phGWB}
%
\eea
for two MSPs, i.e., 1 and 2. Here the ORF for the GWB in the near-field regime is defined as
\be
\Gamma^{\rm b}_{12}(f) \equiv\beta^{\rm b}_{12}\int d^2\hat{\Omega}_{\rm E}\sum_a\frac{\mathscr{P}_1^{*a}(f,\hat{\Omega}_{\rm E})\mathscr{P}_2^{a}(f,\hat{\Omega}_{\rm E})}{A^{2}_{a}(\iota_{\rm E})},
\label{eq:ORFGWB}
\ee
and a normalization constant $\beta^{\rm b}_{12}$ is chosen to make $\Gamma^{\rm b}_{12}(f)=1$ for coincident co-aligned detectors (two identical MSPs located at the same position). Some detailed formulas for the calculations of the ORF are listed in Appendix~\ref{sec:ORF_GWB}. Different from the calculation of the ORF with the far-field approximation, the ORF in the near-field regime depends on the inclination angle $\iota_{\rm E}$, the pulsar distances $L$, and the sources' distances $r$. Since $\iota_{\rm E}$ for GW sources may be randomly distributed, we can obtain the average ORF, i.e., 
\be
\bar{\Gamma}^{\rm b}_{12}(f)=\frac{\beta^{\rm b}_{12}}{4\pi}\int d^2\hat{n}\int d^2\hat{\Omega}_{\rm E}\sum_a\frac{\mathscr{P}_1^{*a}(f,\hat{\Omega}_{\rm E})\mathscr{P}_2^{a}(f,\hat{\Omega}_{\rm E})}{A^{2}_{a}(\iota_{\rm E})},
\label{eq:aveORF}
\ee
by averaging over $\hat{n}$. Denoting the angle between two PTA MSPs as $\theta_{12}$, $\bar{\Gamma}^{\rm b}_{12}$ is a function of $\theta_{12}$. To estimate $\bar{\Gamma}^{\rm b}_{12}$ with consideration of the near-field effect, it needs to know the number distribution of the ``nearby GW sources'' as a function of $r$ and the distances of PTA pulsars $L$. For simplicity, we assume that all the ``nearby GW sources'' are located at the same distance to the observer (fixed $r$) and all the PTA pulsars have the same $L$. In this way, $\bar{\Gamma}^{\rm b}_{12}$ can be calculated for each given set of $r$ and $L$. We take $\bar{\Gamma}^{\rm b}_{12}$ as the angular correlation function corresponding to the Hellings-Downs curve with considering the near-field effects.

\begin{figure}
\centering
\includegraphics[width=0.45\textwidth]{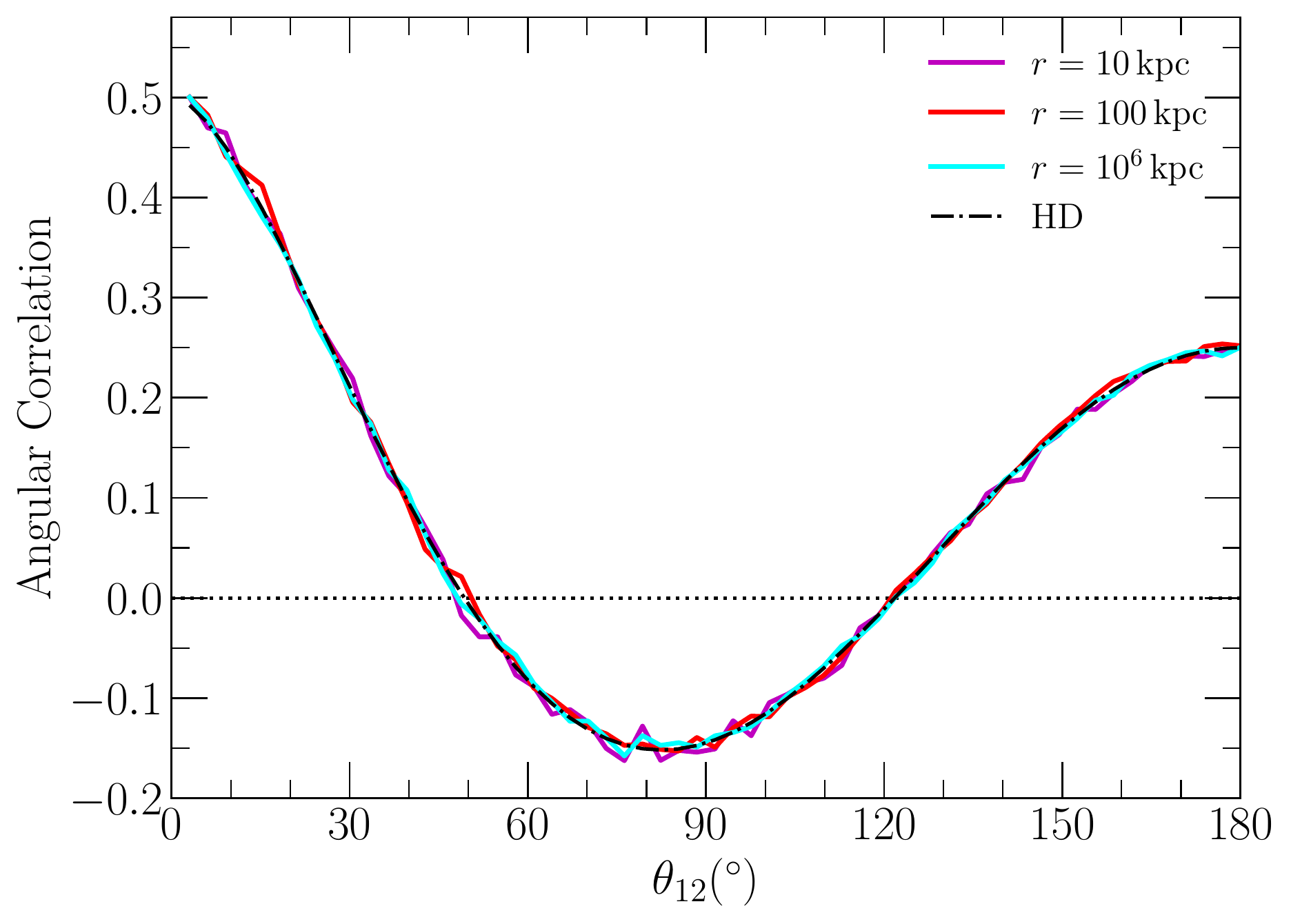}%ORF_theta12.ipynb
\caption{The angular correlation obtained by a PTA with all pulsars at distances of $L=2$\,kpc for a GWB due to nearby GW sources at a fixed distance of $10$\,kpc (magenta solid line), $100$\,kpc (red solid line), and $10^6$\,kpc (cyan solid line), respectively. The black dot-dashed line shows the Hellings-Downs curve obtained by adopting the far-field approximation \citep[Eq.~\ref{eq:HD};][]{1983ApJ...265L..39H}. Note here all these curves are normalized to $1/2$ at $\theta_{12}\rightarrow 0\arcdeg$ as done for the Hellings-Downs (HD) curve (see Eq.~\ref{eq:HD}).
}
% 
%\label{fig:ORF_theta_GWB}
\label{fig:f2}
\end{figure}

Figure~\ref{fig:f2} illustrates the resulted $\bar{\Gamma}^{\rm b}_{12}$ as a function of $\theta_{12}$ for $L=2$\,kpc, and $r=10$\,kpc (magenta solid line), $100$\,kpc (red solid line), and $10^6$\,kpc (cyan solid line), respectively. For comparison, we also show the standard Hellings-Downs curve obtained for a GWB contributed by faraway sources (black dot-dashed line) \citep{1983ApJ...265L..39H}, with which the far-field approximation is suitable to be adopted, i.e., 
\be
\bar{\Gamma}^{\rm b}_{12}|_{r=\infty} =\frac{1}{2}-\frac{x}{8}+\frac{3}{4} x\ln\frac{x}{2}, 
\label{eq:HD}
\ee
where $x=1-\cos \theta_{12}$ and $\bar{\Gamma}^{\rm b}_{12}$ is renormalized to $1/2$ at $\theta_{12} \rightarrow0\arcdeg$. We also renormalize those $\bar{\Gamma}^{\rm b}_{12}$ obtained by limiting $r$ to nearby sources to $1/2$ at $\theta_{12}\rightarrow 0\arcdeg$.\footnote{When $\theta_{12}=0$, its value is 1 but we do not show it in the figure and Eq.~\eqref{eq:HD}. The unnormalized $\bar{\Gamma}^{\rm b}_{12}$ values at $\theta_{12} \neq 0$ are  basically consistent with each other.}  As seen from this figure, $\bar{\Gamma}^{\rm b}_{12}$ in the near field regime is similar to the standard Hellings-Downs curve obtained with the far-field approximation.
{
This similarity can be understood by the following argument. Along the propagation path of the pulses from a pulsar to the observer, the metric perturbation due to the GWB should be uncorrelated with that at a different spacetime point along a different propagation direction of the pulses from another pulsar to the observer, if the distance of that point to the observer is much large than the GW wavelength. The contribution to the ORF comes mainly from the angular correlation of the effective metric perturbations at the same observer's spacetime point (where the effective metric perturbation means
the product of the metric perturbation and the pulsar antenna pattern function), similar to the case of the far-field approximation. That similarity also suggests that the near-field effect on the GWB SNR estimation is negligible. 
We further note that the ORF value $\bar{\Gamma}^{\rm b}_{12}(\theta)$ at $\theta_{12}=0^\circ$ (not shown in Fig.~\ref{fig:f2}) in the near-field regime is different from that in the far-field regime. For example, the un-normalized value $\bar{\Gamma}^{\rm b}_{12}(\theta=0^\circ)/\beta^{\rm b}_{12}$ is $1.64$ (or $1.00$) when the sources contributed to the GWB are all at $r=3 (\textrm{or}\,8)$\,kpc, while the un-normalized value $\bar{\Gamma}^{\rm b}_{12}(\theta=0^\circ)/\beta^{\rm b}_{12}$ is $2/3$ under the far-field approximation. This may indicate the difference between the near-field regime and the far-field regime. 

Note also in the above analysis, the GWB from nearby sources is assumed to be isotropically distributed. One should be cautious about this assumption as a GWB produced by nearby sources may be anisotropic and thus the analysis should be significantly different, which deserves a further study. 
}

\section{Pulsar Timing Arrays}
\label{sec:PTA}

For a PTA with $N_{\rm p}$ ($\geq3$) MSPs, two different methods may be adopted to deal with data, which give different SNR estimates. Below we introduce the formulas for SNR estimates via the matched-filtering and the cross-correlation methods in section~\ref{subsec:MFM} and section~\ref{subsec:CCM}, respectively.

\subsection{The Matched-Filtering Method}
\label{subsec:MFM}

Coherent network analysis has been developed for detecting individual sources via PTA  \citep[e.g.,][]{2012ApJ...756..175E,2014ApJ...794..141A,wang2014a, wang2015coherent, 2015MNRAS.451.2417R}, which is similar to that for the network of ground-based GW detectors \citep{1996CQGra..13.1279J}. For the TOA data from each MSP, the standard matched-filtering method can be used to estimate SNR. With this method, the total SNR obtained from the PTA observations can be given by (see Section~\ref{sec:General})
\be
\varrho^2=\sum^{N_{\rm p}}_{i=1}\int^{\infty}_{0}df \frac{4\chi^2(f)|\tilde{h}_{\rm o}(f)|^2}{S_{{\rm n}i}(f)},
\ee
where the summation is over all $N_{\rm p}$ MSPs. For convenience, the total SNR used for theoretical analysis may be approximated as
\be
\varrho^2 \simeq N_{\rm p}\int^{\infty}_{0}df \frac{4\chi^2(f)|\tilde{h}_{\rm o}(f)|^2}{S_{\rm n}(f)},
\label{eq:SNRmatched}
\ee
by assuming that all MSPs contribute to the SNR equally \citep[see also][]{2015CQGra..32e5004M}. 

The SNR estimate given by the above Equation can be treated as an effective SNR, though the real SNR for a PTA observations can be obtained only by considering the detailed properties of each MSP adopted in the PTA. For example, the effective SNR was adopted in \citet{2015CQGra..32e5004M}, \citet{2015PhRvD..92f3010H}, and \citet{2013PhRvD..88l4032T}, to estimate sensitivity curves for PTAs. In reality, different MSPs adopted in the PTA may have quite different properties and thus contribute to SNR differently. The contributions to SNR may be led by several close-to-source MSPs with small timing RMS noise. Therefore, more careful estimation of SNR should consider the properties of each MSP adopted in the PTA observations.

\subsection{The Cross-Correlation Method}
\label{subsec:CCM}

Cross-correlation method has also been introduced for detecting individual sources via PTA, similarly to that for the detection of a stochastic GWB \citep{2015CQGra..32e5004M,book:1417639}, particularly when the redshift $\tilde{z}(f)$ is difficult to obtain. According to Section~\ref{sec:Two}, the SNR estimated from the cross-correlation method for any two PTA MSPs ($i$, $j$, and $i\neq j$) is
\bea
\varrho_{ij}^2 & = &\frac{2T}{\beta_{ij}^2}\int^{\infty}_{0}df \frac{S_h^2(f)|\Gamma_{ij}(f)|^2}{S_{{\rm n}i}(f)S_{{\rm n}j}(f)} \nonumber \\
\label{eq:rhoij}
\eea
or
\bea
\varrho_{ij}^2 & = & \frac{8}{T}\int^{\infty}_{0}df \frac{\chi_i^2\chi_j^2|\tilde{h}_{\rm o}(f)|^4}{S_{{\rm n}i}(f)S_{{\rm n}j}(f)}.
\eea
In equation~\eqref{eq:rhoij}, the ORF of the $i$- and $j$-MSPs is defined as
\be
\Gamma_{ij}(f) \equiv\beta_{ij}\sum_a\mathscr{P}^{*a}_i(f)\mathscr{P}^a_j(f),
\ee
where $\beta_{ij}$ is adopted to make $\Gamma_{ij}(f)=1$ for coincident co-aligned detectors (see also in Eq.~\ref{eq:ORF} for $\Gamma_{12}$). The total SNR is the summation of $\varrho_{ij}$ over all MSP pairs, i.e.,  \citep{2015CQGra..32e5004M,book:1417639}
\be
\varrho^2=\sum^{N_{\rm p}}_{i=2}\sum^{i-1}_{j=1}\varrho_{ij}^2.
\ee
If also assuming that different MSPs contribute equally, the total SNR can be roughly estimated as
\be 
\varrho^2=\frac{4 N_{\rm p}(N_{\rm p}-1)}{T}\int^{\infty}_{0}df \frac{\chi^4|\tilde{h}_{\rm o}(f)|^4}{S^2_{\rm n}(f)}.
\label{eq:SNRcross}
\ee

\section{The Near-Field Effect}
\label{sec:Effect}

In this Section, we compare the differences of some characteristic quantities in the cases adopting the far-field approximation from that in the general framework by including the near-field effect. We then estimate the values of $\chi$ for some example systems in these two cases, of which the difference mainly represents the importance of near-field effect on average.

\subsection{The function $\mathscr{P}^{a}(f)$}

We first analyze the function $\mathscr{P}^{a}(f)$ in both the far-field and near-field regimes in some special cases to illustrate their differences.

\subsubsection{The Far-Field Approximation}
\label{subsubsec:ff}

If denoting $\Delta \Phi$ as $\Delta\Phi=-2\pi fL(1-\cos\vartheta)/c$, then $e^{i\Delta\Phi}-1= 2\sin(\Delta\Phi/2)e^{i(\Delta\Phi/2+\pi/2)}$. According to Equation~\eqref{eq:Pfar},  we have the far-field approximation of $\mathscr{P}^a(f)$ as
\be
\mathscr{P}^{a}_{\rm f}(f)=2\sin\left(\frac{\Delta\Phi}{2}\right)
e^{i(\Delta\Phi/2+\pi/2)}\mathscr{F}^{a}.
\ee
If $L\sim1\rm kpc$, $f\sim10^{-8}\rm Hz$, then $fL/c\sim10^3 \gg 1$, and the exponential factor oscillates with frequency $f$ rapidly. Hence $|\mathscr{P}^{a}_{\rm f}(f)|$ also oscillates with $f$ in the range $[0, 2\mathscr{F}^{a}]$ rapidly, and the average of its absolute value is $\overline{|\mathscr{P}^{a}_{\rm f}(f)|}=\sqrt{2} \mathscr{F}^{a}$. This oscillation of $\mathscr{P}^a(f)$ modulates the waveform detected by the PTA and thus it should be carefully considered in the SNR estimates. 

An example is provided below to show how to calculate $\mathscr{F}^{a}$. We define a coordinate system, in which the GW source is located at the negative direction of $z$-axis and PTA pulsars are located in the $yOz$ plane (see Fig.~\ref{fig:f1}b). In this coordinate system, $\hat{\Omega}=(0,0,1)$, $\hat{p}= (0,\sin\vartheta, -\cos \vartheta)$ if adopting the far-field approximation. According to Equations~\eqref{eq:Fplus} and \eqref{eq:Fcross}, one should have
\begin{eqnarray}
F^{+}& =& -\frac{\sin^2\vartheta}{2(1-\cos\vartheta)}=-\frac{1+\cos\vartheta}{2}, \\
F^{\times}& = & 0,
\end{eqnarray}
and also
\begin{eqnarray}
|\mathscr{P}^{+}_{\rm f}(f)| & \propto & (1+\cos\vartheta)\left|\sin\left(\frac{\Delta\Phi}{2}\right)\right|, \\
\label{eq:pfa}
|\mathscr{P}^{+}_{\rm f}(f)| & = & 0.
\end{eqnarray}
This result is consistent with that derived in \citet{2011MNRAS.414.3251L}. Furthermore, if the inclination angle $\iota=0$ and polarization angle $\psi=0$, then
\be 
|\mathscr{P}^{+}_{\rm f}(f)|=(1+\cos\vartheta)\left|\sin \left(\frac{\Delta\Phi}{2}\right)\right|.
\label{eq:huaP_far_plus}
\ee

\subsubsection{The Near-Field Regime}
\label{subsubsec:nf}

\begin{figure}
\centering
\includegraphics[width=0.5\textwidth]{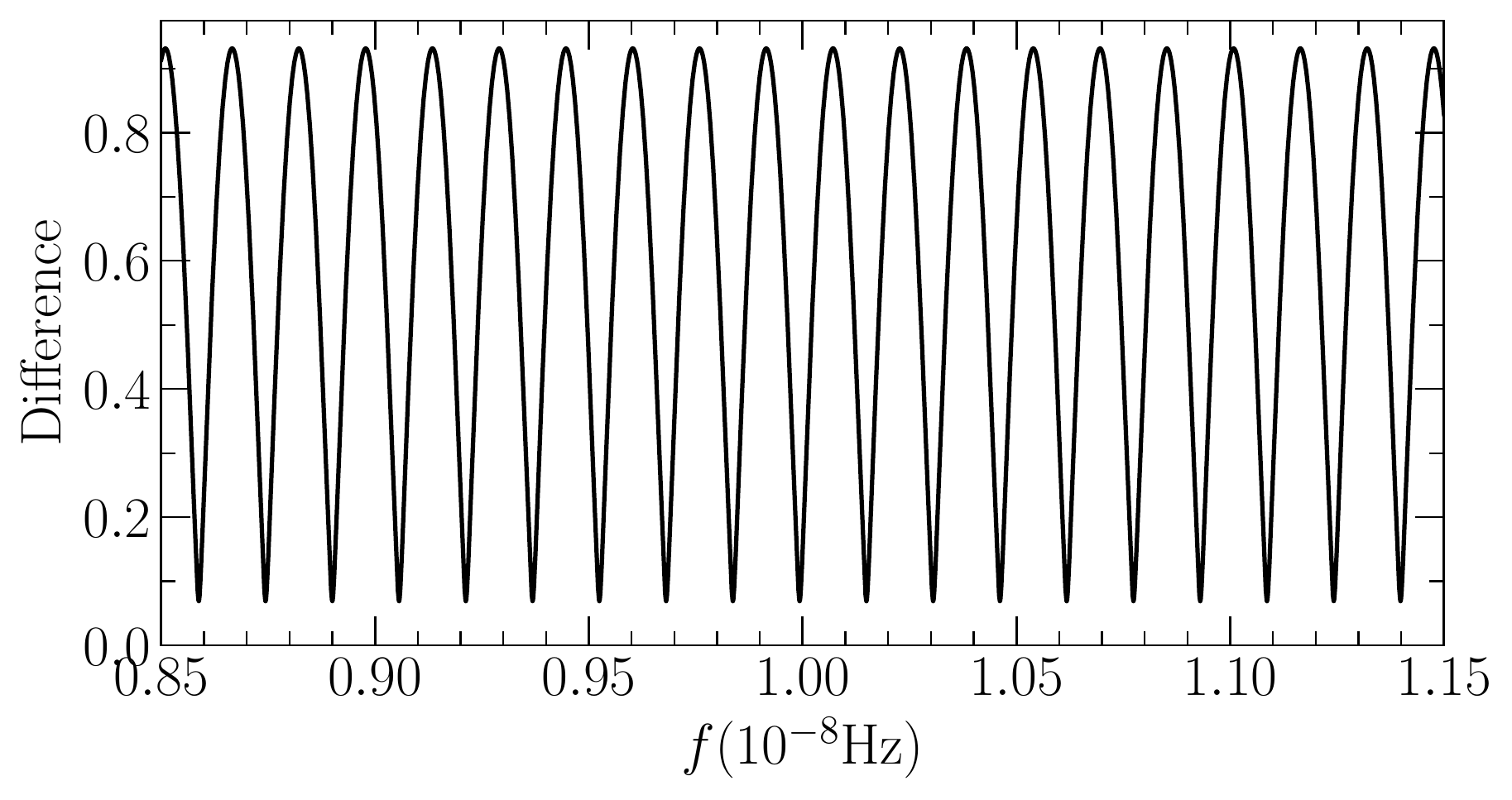}%PTAresidual FFT-8.ipynb
\caption{Relative difference of $\mathscr{P}^{+}(f)$ resulting from the near-field regime and that from the far-field approximation at different frequencies, defined as $|\mathscr{P}^{+}(f)-\mathscr{P}^{+}_{\rm f}(f)|/{\rm max}(|\mathscr{P}^{+}_{\rm f}(f)|)$. The GW source is assumed to be located at $r=8\rm kpc$ and the pulsar distance is assumed to be $L=1\rm kpc$. This figure illustrates that the near-field effect is important at almost all frequencies. See details in section~\ref{subsubsec:nf}.
}
% 
%\label{fig:diffP_f}
\label{fig:f3}
\end{figure}

\begin{figure*}
\centering
\includegraphics[width=\textwidth]{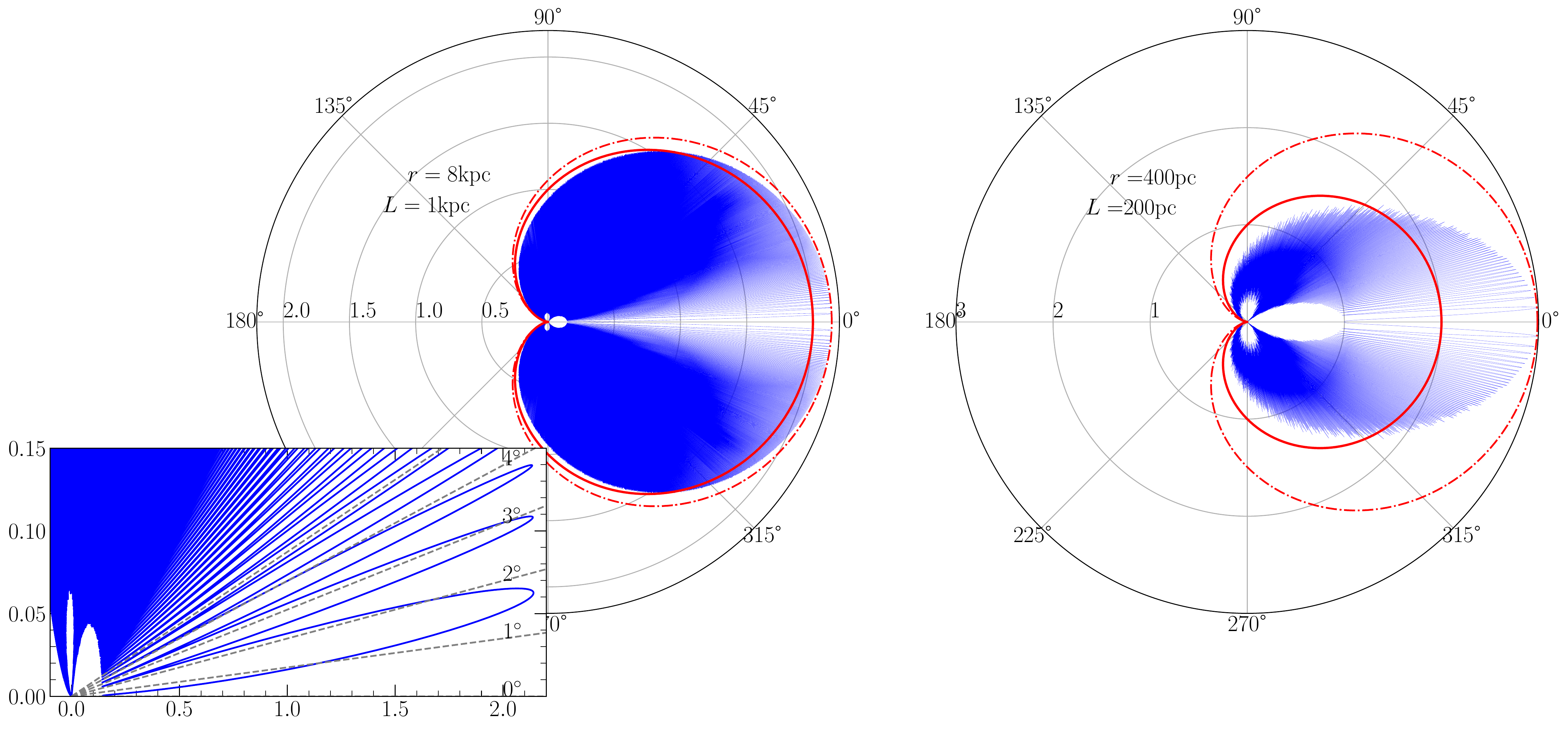}%PTAresidual FFT.ipynb
\caption{ $|\mathscr{P}^{+}(f)|$ as a function of $\vartheta$
(Eqs.~\ref{eq:pdefine} and \ref{eq:vartheta}).
Left panel: the GW source is located at $r=8$\,kpc (e.g., at the GC) and the pulsar distance is $L=1$\,kpc; a sub-figure at the left bottom corner shows the detail of the pattern, and we use several grey dashed lines to represent 1$^\circ$ mesh grid. Right panel: $r=400$\,pc, $L=200$\,pc, $f=10^{-8}$\,Hz, and $\iota_{\rm E}=0$. The red solid line represents the envelope curve of $|\mathscr{P}^{+}_{\rm f}(f)|$ obtained by adopting the far-field approximation, i.e., $\rho = 1+ \cos \vartheta$. The red dashed line represents $\rho=\rho_0(1+\cos\vartheta)$ with $\rho_0={\rm max}(|\mathscr{P}^{+}(f)|)$ representing the maximum value of  $|\mathscr{P}^{+}(f)|$ over $\vartheta$ in the range from $0$ to $2\pi$. The near-field effect is clearly shown by the difference between the red solid curve and the outer envelope of the blue curve. The rapid oscillation of $|\mathscr{P}^{+}(f)|$ with $\vartheta$ may suggest a high angular resolution for the detection of individual GW sources. The right panel shown here is only for illustration purpose. See details in Section~\ref{subsubsec:nf}.
}
% 
%\label{fig:P_theta}
\label{fig:f4}
\end{figure*}

\begin{figure}
\centering
\includegraphics[width=0.5\textwidth]{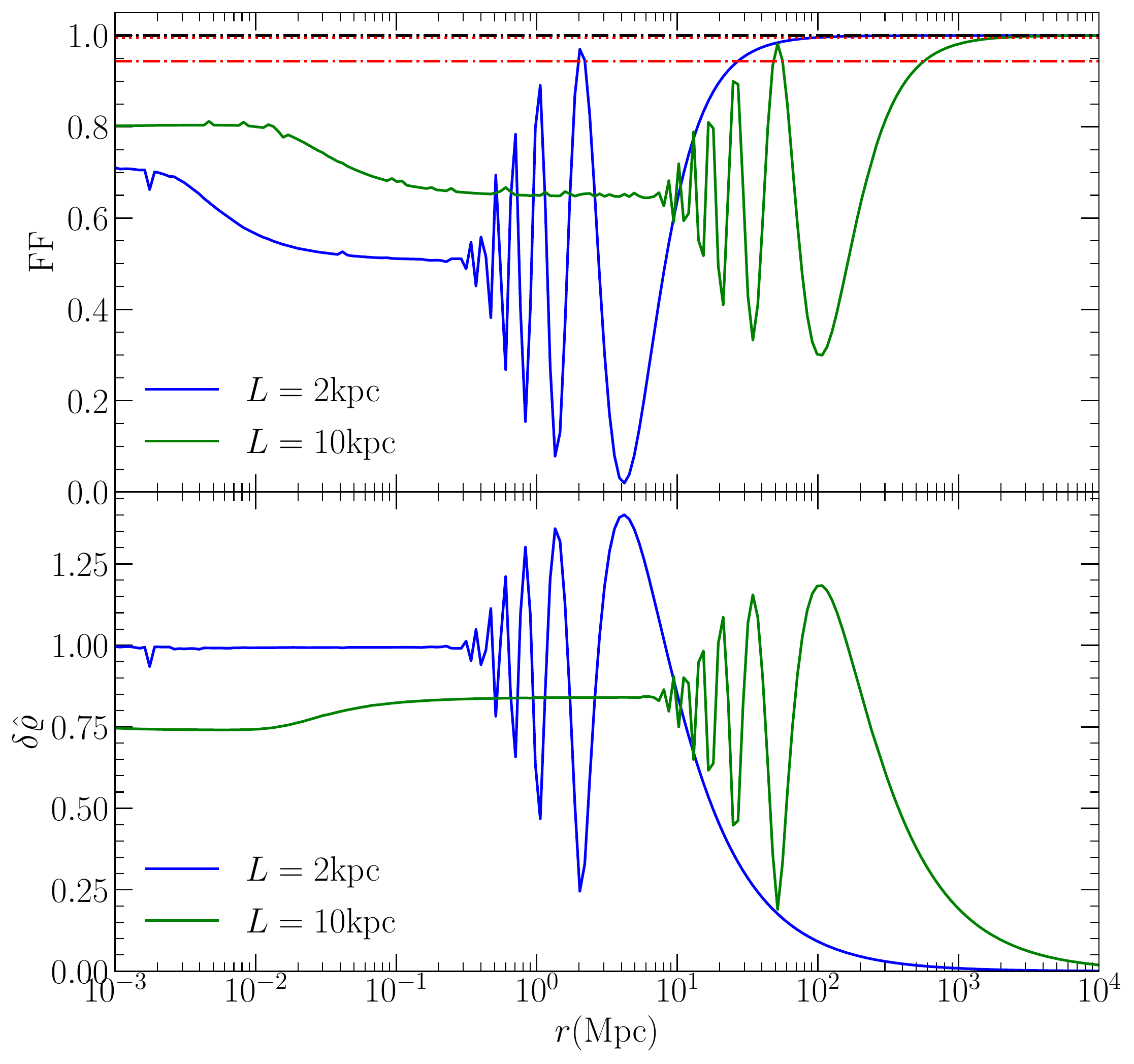}%PTAresidual FFT-8.ipynb
\caption{Top panel: Fitting factor (FF) as a function of the GW source distance (Eq.~\ref{eq:FF}) obtained by adopting the far-field approximation.  Bottom panel: $\delta\hat{\varrho}$ as a function of the GW source distance (Eq.~\ref{eq:delta_SNR}) obtained by adopting the far-field approximation. The MSP is assumed to be at $L=2$\,kpc (blue) or $L=10$\,kpc (green) with $\vartheta=\frac{\pi}{2}$. This figure shows that near-field effect can be significant for $L=2$\,kpc ($10$\,kpc) when the GW source distance $r\lesssim 27$\,Mpc ($572$\,Mpc) or $\lesssim 91$\,Mpc ($1.9$\,Gpc), with corresponding FF$\gtrsim 0.944$ (indicating by the red dotted-dashed line in top panel) or $\gtrsim 0.995$ (indicating by the red dotted line in top panel), if the SNR threshold is set to be $3$ or $10$. In the top panels, the black dotted dashed lines represent ${\rm FF}=1$.
}
% 
%\label{fig:FF}
\label{fig:f5}
\end{figure}

In this near-field regime, we define a coordinate system $(\hat{e}_1,\hat{e}_2,\hat{\Omega})$ (See Appendix \ref{sec:coordinate}), which rotates with GW propagation direction to make $\hat{\Omega}=(0, 0, 1)$ even for different $l$, and $\hat{p}=(0, \sin\gamma, \cos\gamma)$. The inclination angle $\zeta$ changes along the path of pulsar pulses. In this case, $\mathscr{P}^{a}(f)$ can be calculated numerically (for details, see Appendix~\ref{sec:concrete} and Fig.~\ref{fig:f1}a) and it also oscillates with $f$, similar to that shown for the case adopting the far-field approximation. 
To illustrate the behavior of $\mathscr{P}^{a}(f)$ as a function of frequency, we assume $\vartheta=\pi/2$, $L=1\rm kpc$, and $r=8\rm kpc$ (i.e., a source at the GC), $\iota_{\rm E}=\psi_{\rm E}=0$, thus $\mathscr{P}^{+}(f)\neq0$ and $\mathscr{P}^{\times}(f)=0$. For these settings, we adopt both the far-field approximation as described in~\ref{subsubsec:ff} and the general formulas that include the near-field effect to calculate $\mathscr{P}^{a}_{\rm f}(f)$ and $\mathscr{P}^{a}(f)$, respectively. Figure~\ref{fig:f3} shows the relative difference between $\mathscr{P}_{\rm f}^{+}(f)$ and $\mathscr{P}^{+}(f)$ which is significant at a large fraction of frequencies. 

The quantity $\mathscr{P}^{a}(f)$ also depends on the relative angle between the directions of MSP and GW source $\vartheta$. Figure~\ref{fig:f4} illustrates such an angular dependence for two assumed nearby GW sources in a plane of $\rho$ versus $\vartheta$. Here $\rho$ represents the absolute value of $\mathscr{P}^{+}(f)$. The red solid line represents the envelope curve of $|\mathscr{P}^{+}_{\rm f}(f)|$ by adopting the far-field approximation, i.e. $\rho = 1+ \cos \vartheta$, which is consistent with that studied in \citet{2011MNRAS.414.3251L} for the response pattern of distant sources. The red dashed line represents the curve $\rho= \rho_0 (1+ \cos \vartheta)$ with $\rho_0={\rm max}( |\mathscr{P}^{+}(f)| )$, which has similar shape with red solid line. As seen from this Figure, the near-field effect leads to the shape distortion of $|\mathscr{P}^{+}(f)|$ dependent on $\vartheta$. 
There are small white empty areas around $\rho=0$ in both panels of Figure~\ref{fig:f4}, different from that shown for the case adopting far-field approximation in \citet{2011MNRAS.414.3251L}. The reason is as follows. If the far-field approximation is valid, the amplitude and propagation direction of GWs are almost the same at different $Q$ on the path of pulses, respectively, and only the GW phases are different at different $Q$. At phases $2\pi n$ ($n$ is an integer), $\mathscr{P}^{a}_{\rm f}(f)$ vanishes even $\mathscr{F}^{a}\neq0$, as indicated by Equation~\eqref{eq:Pfar}. However, in the near-field regime, the GWs from the same source have different directions, amplitudes, and phases, at different $Q$ on the path of pulses. The superposition of GW effect at different parts of the path generally does not vanish, and thus $\mathscr{P}^{a}(f)$ is oscillating but cannot reach $0$ as shown in Figure~\ref{fig:f4}.

The small-scale spiky features (Fig.~\ref{fig:f4}) may enable precise localization of GW sources via PTA observations, if pulsar distances can be measured accurately by the timing parallax and the curvature of GW wavefront \citep{2011MNRAS.414...50D} or some other methods with an error not larger than the GW wavelength \citep[see][]{2011MNRAS.414.3251L}. The angular dependence is much different from the dependence on the frequency or pulsar distance, therefore, they can be distinguished. In principle, it is plausible that timing parallax can give accurate distance measurements for pulsars. In observations, however, pulsar distances may be difficult to be measured via timing parallax with an accuracy $\ll 0.1\%$ and the GW radiation may be not strictly monochromatic (due to the finite observation time span $T$), and thus $f$ and $L$ are degenerate with each other in the case adopting the far-field approximation. However, the degeneracy between $f$ and $L$ may be broken in the near-field regime since $\mathscr{P}^{a}(f)$ is not strictly periodic. The spiky features in the response of PTA can also help to locate the angular position of a GW source within a lobe as these spiky lobes form many concentric circles in the sky and the intersection of the concentric circles of many PTA pulsars gives the source position \citep{2011MNRAS.414.3251L}. From Figure~\ref{fig:f4} (left-bottom sub-figure), we can see that the width of each lobe is $\lesssim 1^{\circ}$, thus the location of the GW source at $r=8$\,kpc may be identified with an accuracy of $1^\circ$. However, the precision may be significantly decreased because of the non-zero noises in the actual observations.

It is therefore important to adopt the general formalism presented in Section~\ref{sec:General} when considering the detection of nearby GW sources by PTAs, as adopting the (inaccurate) far-field approximation in such cases would introduce significant errors in the waveform templates (as the production of $\mathscr{P}^{a}_{\rm f}(f)$ and $\tilde{h}(f)$; for an example, see Fig.~\ref{fig:f8} below) and thus in the parameter estimates.

If inaccurate GW templates are used to match, then it leads to a decline of SNR or wrong parameter estimations. Such an effect can be described by the \emph{fitting factor} (FF) defined as \citep{2008PhRvD..77j4017A} 
\be
{\rm FF} \equiv \frac{\{s|h\}}{\sqrt{\{s|s\}\{h|h\} }},
\label{eq:FF}
\ee
where $s$ is the actual signal, $h$ is the template, and  $\{\cdot|\cdot\}$ represents the inner product defined as
\be
\{s|h\}\equiv 4\Re\left( \int_0^{\infty}\frac{\tilde{s}^*(f)\tilde{h}(f)}{S_{\rm n}(f)}{\rm d}f\right),
\ee
and $\Re(Z)$ represents the real part of $Z$. When $s=h$, we obtain the optimal SNR, i.e., $\varrho_{\rm opt}=\sqrt{\{s|s\}}$, while if $h$ is close to $s$, but not equal to $s$, we obtain the actual SNR with template $h$ as $\varrho=\sqrt{\{s|h\}}\simeq\sqrt{\rm FF} \varrho_{\rm opt}$.

In our example, we regard the accurate waveform in the near field (at distance $r$) as $s$, and regard the waveform in the far-field approximation as $h$. We can then define a threshold for FF as 
\be
{\rm FFS}\equiv 1-\frac{1}{2\varrho_{\rm th} ^2}
\label{eq:FFS}
\ee
to show the significance of the near-field effect (see \citealt{2008PhRvD..78l4020L, 2019ApJ...887..210F})\footnote{Note that this threshold FF is valid only when $s\simeq h$. If the difference between $s$ and $h$ is too large, this criterion would be invalid.}. This criterion is ${\rm FFS}= 0.944$ when adopting a threshold of SNR as $\varrho_{\rm th}=3$, and $0.995$ when adopting $\varrho_{\rm th}=10$, respectively. As an example, we calculate FF of the templates resulting from the far-field approximation and those after considering the near-field effect, respectively. In reality, each pulsar in a PTA has a different direction and distance that leads to different waveform to match in the near-field regime. To clearly show the dependence of FF (or $\delta\varrho$, whose definition can be seen from Eq.~\ref{eq:delta_SNR} at the end of this subsection.) on each variable, for simplicity, we assume the same GW waveform is adopted to calculate the FF (or $\delta\varrho$) for all PTA pulsars. According to Equation~\eqref{eq:FF}, we obtain FF$=0.54$ for a monochromatic GW signal with $h_0=10^{-15}$ and $f=10^{-8}$\,Hz from the GC, monitored by the SKA-PTA with properties listed in Table~\ref{tab:parameter} (for simplicity, we set $L=1$\,kpc). This small FF value means that the adoption of inaccurate templates under the far-field approximation, without considering the near-field effect, leads to a significant SNR decline (e.g., by a factor of $1.36$ for the above case) and a less good match.

Figure~\ref{fig:f5} (top panel) shows FF for those GW sources with similar properties but located at different distances $r$, monitored by a PTA with MSPs at $L=2$\,kpc or $L=10$\,kpc. It is clear that the near-field effect is important at least for GW sources at $r \lesssim$ a few Mpc and it should be considered when considering the detectability and extracting the GW signal of nearby sources in the PTA data. For different parameters settings (e.g., $\vartheta$), the results may be quantitatively different, but we can still obtain similar near-field effect qualitatively (as seen in Appendix~\ref{sec:angles}). We defer a more comprehensive investigation of the errors in the parameter estimates induced by ignoring the near-field effect to future work. 

If the difference between waveform $s$ and $h$ is significant, i.e. $s\sim h$ is incorrect, we can describe the difference of $s$ and $h$ by a relative quantity as \citep{2022PhRvD.106b3018G} 
\be
\delta\hat{\varrho}\equiv\sqrt{\frac{\{s-h|s-h\}}{\{s|s\}}}.
\label{eq:delta_SNR}
\ee
When $s\equiv h$, $\delta\hat{\varrho}=0$. Figure~\ref{fig:f5} also shows $\delta\hat{\varrho}$ as the function of source distance $r$ (bottom panel). The larger $\delta\hat{\varrho}$ is, the larger difference between $s$ and $h$. Despite the differences in the definitions of ${\rm FFS}$ and $\delta \hat{\varrho}$, they give similar results on the significance of the near-field effect as a function the GW source distance (top and bottom panels).

\subsection{Overlap Reduction Function $\Gamma_{12}(f)$ for individual sources}

In this subsection, we show the difference of overlap reduction function between far-field approximation and near-field regime. 

\subsubsection{The Far-Field Approximation}

According to Equation~\eqref{eq:ORF}, the absolute value of overlap reduction function (ORF) is given by
\be
|\Gamma_{12}(f)|=\left|\beta_{12}\sum_a\mathscr{P}^{*a}_1(f)\mathscr{P}^a_2(f)\right|.
\ee
For simplicity, we calculate the ORF values in a case assuming that two MSPs are located in different directions ($\vartheta$) but the same plane (e.g., $yOz$, see Fig.~\ref{fig:f1}).\footnote{For cases that they are located in different planes, the conclusions are similar.} For such a setting,
\be
|\Gamma(f)|\propto(1+\cos\vartheta_1)(1+\cos\vartheta_2) \left| \sin \left( \frac{\Delta\Phi_1}{2} \right) \sin \left( \frac{\Delta\Phi_2}{2} \right) \right|,
\ee
where $\Delta\Phi_1=-2\pi fL_1(1-\cos\vartheta_1)/c$, $\Delta\Phi_2=-2\pi fL_2(1-\cos\vartheta_2)/c$. The frequency dependence of ORF is the product of two sinusoidal oscillations. If one of the angle $\vartheta_i$ is fixed, the angular dependence is the same as that of $|\mathscr{P}^{a}_{\rm f}(f)|$.

\subsubsection{The Near-Field Regime}
\label{sec:chi_NF}

\begin{figure}
\centering
\includegraphics[width=0.5\textwidth]{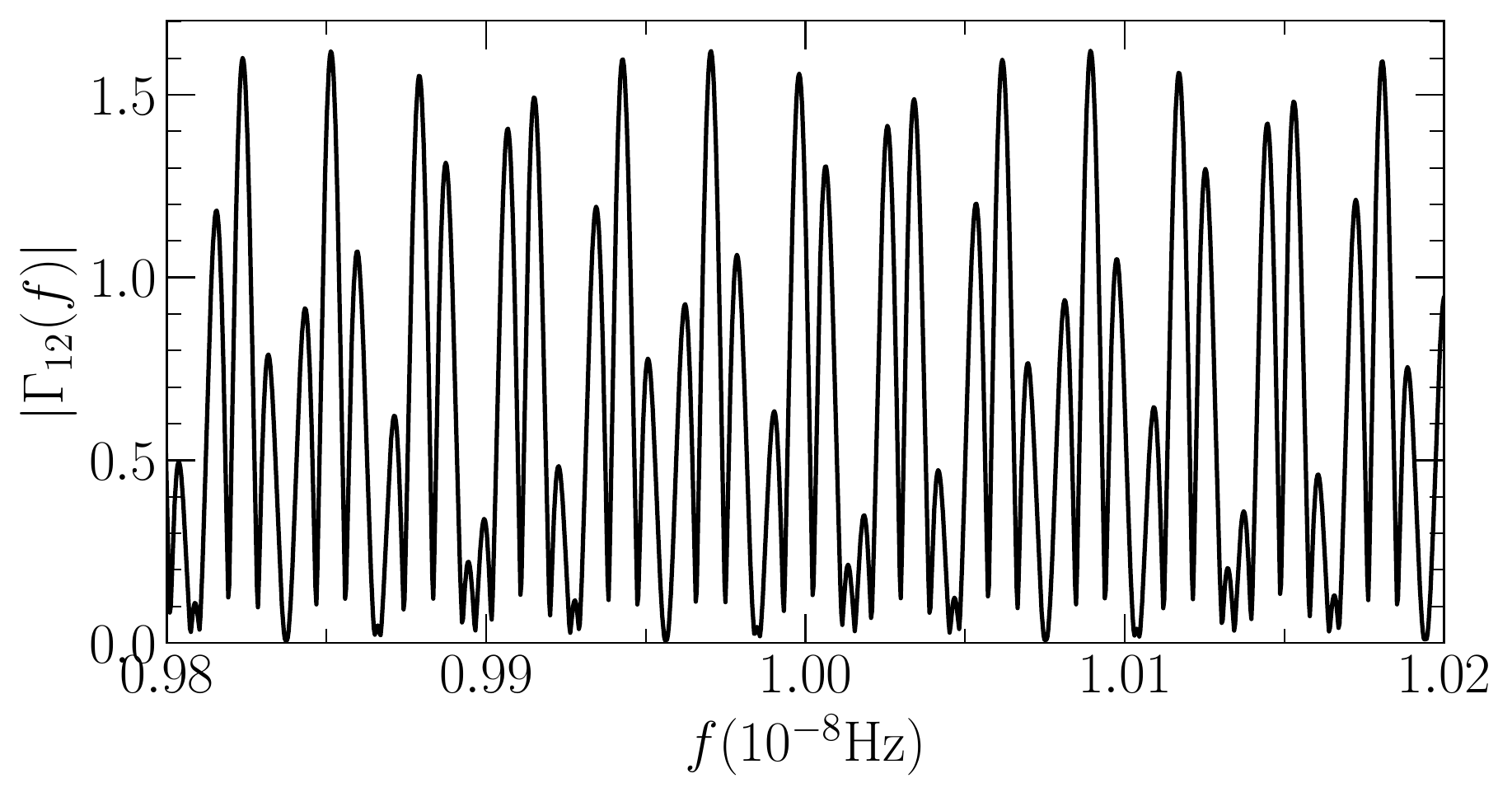}%PTAresidual FFT.ipynb
\caption{Overlap reduction function (ORF) $|\Gamma_{12}(f)|$ (see section~\ref{sec:chi_NF}) dependence on $f$. Here we set $\beta_{12}=1$, $L_1= L_2 = 1$\,kpc and $\vartheta_1=\frac{\pi}{2}$, $\vartheta_2=\frac{\pi}{4}$.
}
% 
%\label{fig:ORF_f}
\label{fig:f6}
\end{figure}

To illustrate the frequency dependence of ORF $\Gamma_{12}(f)$ in the near-field regime, we investigate a simple example by setting $\beta_{12}=1$ and $L_1= L_2 = 1$\,kpc, $\vartheta_1=\frac{\pi}{2}$, $\vartheta_2=\frac{\pi}{4}$. Figure~\ref{fig:f6} shows $|\Gamma_{12}(f)|$ as a function of $f$ in this case, which is rapidly oscillating with $f$ with some structures. And he angular dependence of ORF with the angle $\vartheta_1$ (or $\vartheta_2$) in the near-field regime is similar to Figure~\ref{fig:f4} when the other angle $\vartheta_2$ (or $\vartheta_1$) is fixed.

\subsection{The geometric factor $\chi$}
\label{subsec:chi}

According to the definition of $\chi$ given in Equation~\eqref{eq:chidefine}, $\chi$ for a PTA with $N_{\rm p}$ MSPs where $N_{\rm p}$ ($\gg 1$) and PTA MSPs are uniformly distributed on the sky is  
\be
\chi_{\rm f}=\sqrt{\int^{\pi}_0\frac{{\rm d}\psi}{\pi} \frac{1}{2N_{\rm p}}\sum^{N_{\rm p}}_{j=1}
\left|e^{i2\pi fL_j(1+\hat{\Omega}\cdot\hat{p}_j)/c}-1\right|^2
\left(\mathscr{F}^+_j+\mathscr{F}^\times_j\right)^2},
\label{eq:chi_far}
\ee
when the far-field approximation is valid. If not averaging over $\psi$, we may define $\chi_{\rm f}'$ as
\be
\chi_{\rm f}' \equiv\sqrt{ \frac{1}{2N_{\rm p}}\sum^{N_{\rm p}}_{j=1}
\left|e^{i2\pi fL_j(1+\hat{\Omega}\cdot\hat{p}_j)/c}-1\right|^2
\left(\mathscr{F}^+_j+\mathscr{F}^\times_j\right)^2}.
\ee
By averaging over all MSPs, the above equation may be further approximated as
\be
\chi_{\rm f} \simeq \sqrt{\frac{1}{N_{\rm p}}\sum^{N_{\rm  p}}_{j=1}(\mathscr{F}^+_j+\mathscr{F}^\times_j)^2},
\ee
since $fL/c\gg 1$ and the average of the function $|e^{i\theta}-1|^2$ over all $\theta$ ($\in (0, 2\pi]$) is $2$. \citet{2015CQGra..32e5004M} define $\chi$ by approximating the above equation to an average over all directions if the number of MSPs is large and its sky distribution is uniform and set $\iota=0\arcdeg$, $\psi=0\arcdeg$ \footnote{The misprint in Equation (11) in \citet{2015CQGra..32e5004M} is corrected and
$$ 
\chi \simeq \sqrt{\int_{\phi=0}^{2 \pi} \int_{\theta=0}^{\pi} \frac{\sin \theta \mathrm{d} \theta \mathrm{d} \phi}{4 \pi} \left(\frac{1}{2} \frac{\hat{p}_{i} \hat{p}_{j}\left(A^{+} H_{i j}^{+}+A^{ \times} H_{i j}^{ \times}\right)}{1+\hat{\Omega} \cdot \hat{p}}\right)^{2}}=\frac{1}{\sqrt{3}}.
 $$}
\be
\chi_{\rm f}'' \simeq \sqrt{\frac{1}{4\pi}\int d^2\hat{\Omega}_{\hat{p}}(\mathscr{F}^++\mathscr{F}^\times)^2}=\frac{1}{\sqrt{3}},
\ee
where $d^2\hat{\Omega}_{\hat{p}}$ is the solid angle corresponding to pulsar positions. 

For individual sources, the inclination angle $\iota_{\rm E}$ can be any value. To estimate the average SNR, we define the mean $\bar{\chi}_{\rm f}$ by averaging $\chi_{\rm f}$ on all possible $\iota_{\rm E}$, 
\be
\bar{\chi}_{\rm f} \equiv \sqrt{\int^{1}_{-1}\frac{{\rm d}\cos\iota}{2}\chi^2_{\rm f}(\iota)}\approx0.365, 
\label{eq:chibar1}
\ee
in which case, $\iota=\iota_{\rm E}$.

In the near-field regime, we can also define
\be
\chi_{\rm n}\equiv \sqrt{\int^{\pi}_0\frac{{\rm d}\psi_{\rm E}}{\pi}\frac{1}{2N_{\rm p}}\sum^{N_{\rm p}}_{j=1}|\mathscr{P}^{+}_j(f)+\mathscr{P}^{\times}_j(f)|^2},
\label{eq:chi_near}
\ee
which may be also approximated as 
\be
\chi_{\rm n}\simeq \sqrt{\int^{\pi}_0\frac{{\rm d}\psi_{\rm E}}{\pi}\int \frac{{\rm d}^2\hat{\Omega}_{\hat{p}}}{8\pi}|\mathscr{P}^{+}(f)+\mathscr{P}^{\times}(f)|^2},
\label{eq:chibar2}
\ee
by averaging over the whole sky. 

For individual sources, the inclination angle $\iota_{\rm E}$ can be any value. To estimate the average SNR, we can also define the mean $\bar{\chi}_{\rm n}$ by averaging $\chi_{\rm n}$ on all possible $\iota_{\rm E}$,  
\be
\bar{\chi}_{\rm n}= \sqrt{\int^{1}_{-1} \frac{{\rm d}\cos\iota_{\rm E}}{2}\int^{\pi}_0\frac{{\rm d}\psi_{\rm E} }{\pi} \int \frac{  {\rm d}^2 \hat{\Omega}_{\hat{p}}}{8\pi} |\mathscr{P}^{+}(f)+\mathscr{P}^{\times}(f)|^2}.
\label{eq:chi}
\ee
The calculation details and the orientation dependence of $\chi$ can be seen in Appendix~\ref{sec:concrete}.
We adopt $\bar{\chi}_{\rm n}$ in our SNR calculation and compare the difference of $\bar{\chi}_{\rm n}$ and $\bar{\chi}_{\rm f}$ as follows. 

\begin{figure*}
\centering
\includegraphics[width=\textwidth]{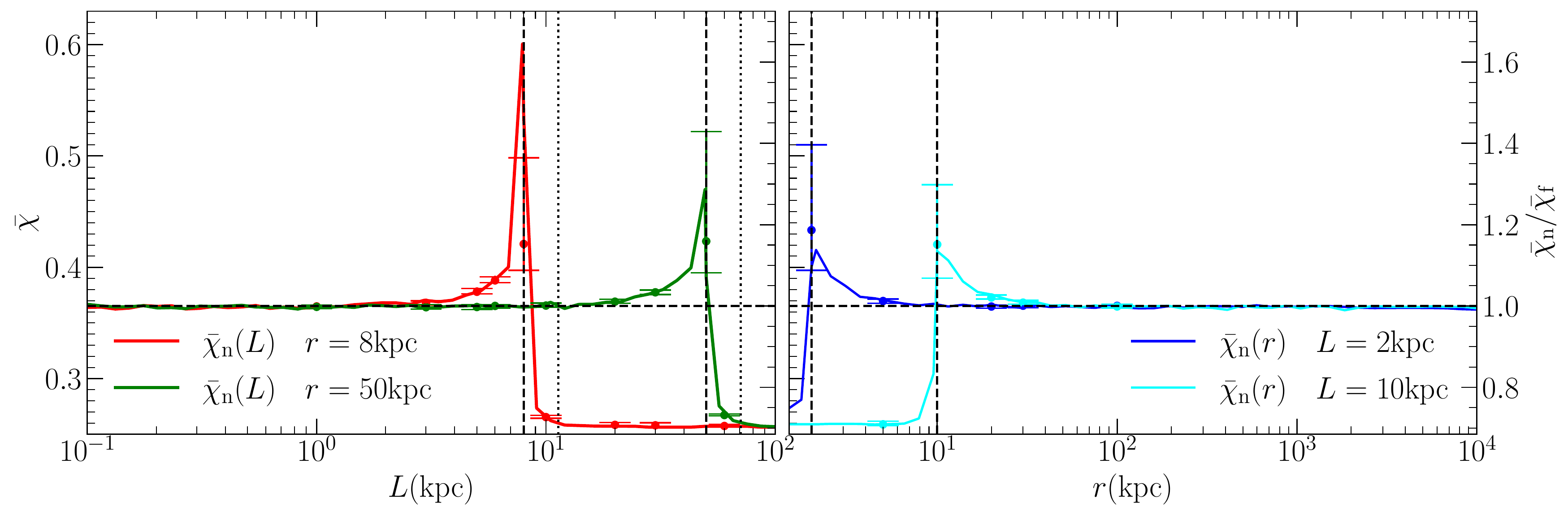}%PTAresidual.ipynb/PTAresidual_chi.ipynb
\caption{Left panel: $\bar{\chi}(L)$: $\bar{\chi}$ as a function of the PTA pulsar distance $L$ (Eq.~\ref{eq:chi}) in the near-field regime for a source at a distance of $r=8$\,kpc (red solid line) and $r=50$\,kpc (green solid line), respectively. Right panel: $\bar{\chi}(r)$: $\bar{\chi}$ as a function of the GW source distance $r$ in the near-field regimes for a PTA obtained by averaging results from 50000 different parameters sets, with pulsars randomly distributed on sky at a distance of $L=2$\,kpc (blue solid line) and $L=10$\,kpc (cyan solid line), respectively. The black dashed horizontal line shows $\bar{\chi}$ obtained by adopting the far-field approximation, i.e., $\bar{\chi}_{\rm f}=0.365$. The ratio $\bar{\chi}_{\rm n}/\bar{\chi}_{\rm f}$ as a function of $L$ is indicated by the right vertical axis, where averaged $\bar{\chi}_{\rm f}\simeq0.365$. The top and bottom horizontal axes indicate source distance $r$ and pulsar distance $L$, respectively. Different realizations for the sky locations of the PTA pulsars may result in different $\bar{\chi}_{\rm  n}$, especially when $L$ is comparable to $r$. To illustrate this, the uncertainties due to different realizations for some cases with fixed $r$ but different $L$ and with fixed $L$ but different $r$ are shown in both the left and right panels (solid circles with errorbars). The solid circles and its associated errorbars represent the median and the $16\%$ to $84\%$ range of $\bar{\chi}_{\rm n}$ obtained from $50$ realizations of the PTA pulsar sky locations (see Section~\ref{sec:chi_NF}).
}
% 
%\label{fig:chi}
\label{fig:f7}
\end{figure*}

If adopting the optimal templates for detection, the differences of SNR estimations between the near-field regime cases and those adopting the far-field approximation are at least partly represented by the differences in $\bar{\chi}$. It is worthy to note here that a low value of $\rm FF$ reflects the SNR decline due to the utilization of inaccurate GW templates, while the variation of $\bar{\chi}$ reflects the SNR change due to the geometric configuration of different PTA pulsars from the configuration in the far-field approximation (which is separated from the effects of inaccurate GW templates). The $\bar{\chi}$ represents the average response of stochastic uniform distributed PTA to GW source. We illustrate the dependence of such differences on the distance of GW sources below. We first consider a GW source located at the GC ($r=8$\,kpc) with a typical PTA frequency $f= 10^{-8}$\,Hz (or $10^{-7}$\,Hz; the $\chi$ value is independent of $f$), with which the distance between MSP and Earth $L$ is much larger than the GW wavelength $\lambda_{\rm GW}\sim 1-0.1$\,pc. We calculate $\bar{\chi}_{\rm n}$ and $\bar{\chi}_{\rm f}$ according to Equations~\eqref{eq:chi_near} and \eqref{eq:chi_far} for the near-field and far-field regimes, respectively, for PTAs with different typical pulsar distance $L$. (The averaged $\bar{\chi}_{\rm f}\simeq 0.365$.) Figure~\ref{fig:f7} shows the resulting $\bar{\chi}_{\rm n}$ (solid lines,left panel), $\bar{\chi}_{\rm f}$ (black dashed line), and their ratio versus pulsar distance $L$ with a fixed source distance $r$, either $8$\,kpc (GC distance) or $50$\,kpc (LMC distance), and also $\bar{\chi}_{\rm n}$ (solid lines, right panel) as a function of the GW source distance $r$, with a fixed MSP distance $L$, either $2$\,kpc or $10$\,kpc. According to this Figure and our calculation results, a number of conclusions are summarized as follows.
\begin{itemize}
\item $\bar{\chi}_{\rm n}$ is roughly $0.365$ when $L$ is a factor of about $4$ times smaller than $r$ (or $L\ll r$), which is exactly the cases that the far-field approximation works. $\bar{\chi}_{\rm n}$ begins to increase when $L$ is larger and reaches a maximum value when $L\simeq r$, then $\bar{\chi}_{\rm n}$ declines rapidly to a value of $\simeq0.26$ when $L$ becomes larger than $r$ and this value is even less than $\bar{\chi}_{\rm f}$ (see Fig.~\ref{fig:f7}). The change of $\bar{\chi}_{\rm n}$ with $L$ suggests that the near-field effect is significant when $L$ is comparable to or larger than the GW source distance. 

\item The large value of $\bar{\chi}_{\rm n}$ at $L\simeq r$ is mainly due to the contribution from PTA pulsars with small $\vartheta$. For small $\vartheta$, $\bar{\chi}_{\rm n}$ can be inversely proportional to $\vartheta$, as seen from Equation~\eqref{eq:chipulsarGW} in Appendix~\ref{sec:Kocsis}.

\item  When $L\gg r$ or $r\ll L$, $\bar{\chi}_{\rm n}$ becomes flat and can be smaller than $\bar{\chi}_{\rm f}$ as shown in Figure~\ref{fig:f7}. The main reason is as follows. The contribution from those PTA pulsars with small $\vartheta$ becomes small due to that the responses in the pulses of a pulsar to the GW signals at the near side of the pulse path is (partly) canceled by those at the far side of the pulse path. 

\item Both $\bar{\chi}_{\rm n}$ and $\bar{\chi}_{\rm f}$ do not depend on the GW frequency. The reason is that $f$ is only included in the phase factor of $\mathscr{P}^{a}(f)$, which is averaged over $\vartheta$ of many different MSPs in the calculations of $\bar{\chi}$. 
\end{itemize}

One should keep in mind that the $\bar{\chi}$ here is a mean geometric factor averaged over PTA MSPs. In real observations, the exact near-field effect depends on the properties of those PTA MSPs and the position of the source. 

\section{Applications}
\label{sec:Application}

In this Section, we apply the theoretical framework presented in Section~\ref{sec:PTA} to estimate SNRs of some hypothetical MBBHs in the GC and the center of LMC, 
monitored by current and future PTAs, thus check whether they can be detected by PTAs, if any.   

\subsection{Monochromatic GW Signals}
\label{sec:App_mono}

The GW from MBBHs in the PTA band is almost monochromatic since the frequency variation rate is negligible. In this case, the GW strain can be approximated as Equation~\eqref{eq:hplus} and \eqref{eq:hcross}, where the GW phase $\Phi(t)=2\pi f_0t$, GW amplitude
$h_0=4({G}\mathcal{M}_{\rm c})^{\frac{5}{3}}(\pi f_0)^{\frac{2}{3}}/{c}^4r$
with $\mathcal{M}_{\rm c}=\left(M_{\bullet,1} M_{\bullet,2}\right)^{3/5}/M_{\bullet\bullet}^{1/5}$, $M_{\bullet\bullet}=M_{\bullet,1}+M_{\bullet,2}$, $M_{\bullet,1}$ and $M_{\bullet,2}$ are the masses of two components. The Fourier transforms of the GW strains are
\begin{eqnarray}
\tilde{h}_{+,\rm o}(f) & = & h_0(f_0)(\delta(f-f_0)+ \delta(f+f_0))/2,  \\ 
\tilde{h}_{\times,\rm o}(f) & = & h_0(f_0) (\delta(f-f_0)- \delta(f+f_0))/2.
\end{eqnarray}
Only the half with $f>0$ of the frequency spectrum appears in the integral for estimating SNR (see Eqs.~\eqref{eq:SNR_mono_m} and \eqref{eq:SNR_mono_c} below), thus the RMS strain is 
\be
\left|\tilde{h}_{a,\textrm{o}}(f)\right|=\frac{h_0(f_0)\delta(f-f_0)}{\sqrt{2}}.
\label{eq:hfdelta}
\ee
In reality, the total observation time $T$ cannot be infinite, and thus $\delta(f)$ should be replaced by $\delta_T(f)=\sin(\pi fT)/(\pi f)$ \citep{2015CQGra..32e5004M}.

\begin{figure*}
\centering
\includegraphics[width=0.8\textwidth]{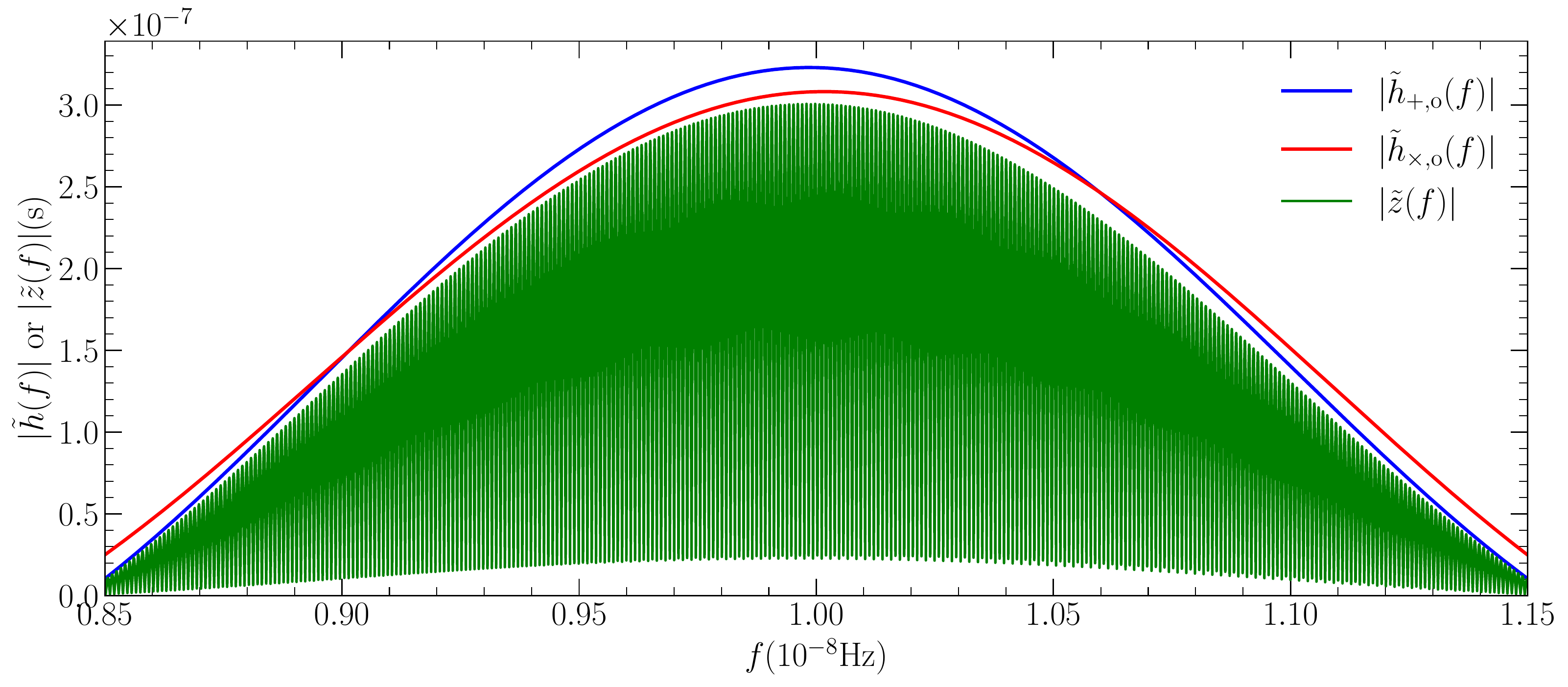}%PTAresidual FFT-8.ipynb
\caption{$|\tilde{h}_{a,\textrm{o}}(f)|$ (Eq.~\ref{eq:hfdelta}) and $|\tilde{z}(f)|$ (Eq.~\ref{eq:z_f_mono}) (both in unit of second) obtained from the Fast Fourier Transformation (FFT) of a monochromatic GW source located at $r=8$\,kpc with amplitude $h_0=10^{-15}$ and frequency $f=10^{-8}$\,Hz in the frequency domain ``observed'' by a PTA with $L=1$\,kpc over a time span of $T=20$\,yr. The $|\tilde{z}(f)|$ shown here is a modulated GW signal spectrum. 
This modulation occurs only when the pulsar term (in the term of $\mathcal{P}^{+}(f)$ in Eq.~\ref{eq:z_f_mono}) is considered and the results obtained in the near-field case are different from those obtained in the case adopting the far-field approximation.  
See section~\ref{sec:App_mono} for details.
}
% 
%\label{fig:z(f)_h(f)}
\label{fig:f8}
\end{figure*}

Figure~\ref{fig:f8} shows $|\tilde{h}_{\rm o}(f)|$ and $|\tilde{z}(f)|$ obtained for an example MBBH with $h_0=10^{-15}$ at $f_0=10^{-8}$\,Hz, according to Equation~\eqref{eq:z=ph} by assuming $\vartheta=\pi/2$, $L=1\rm kpc$, $r=8\rm kpc$, and $\iota_{\rm E}=\psi_{\rm E}=0$ (see Section~\ref{subsubsec:nf}). In this case, $\mathscr{P}^{\times}(f)=0$, therefore, 
\be
|\tilde{z}(f)|=|\mathscr{P}^{+}(f)||\tilde{h}_{+,\rm o}(f)|.
\label{eq:z_f_mono}
\ee

The frequency spectrum of redshift $|\tilde{z}(f)|$ is a modulated GW signal spectrum due to the oscillation of $\mathscr{P}^{a}(f)$. This modulation is important in the matched filtering. If the far-field approximation is adopted and only the Earth term is considered, the waveform template is not modulated as it should be, which may lead to a significant decline of SNR estimated for nearby GW sources and large errors in parameter estimations of the GW system. For the SKA-PTA, we obtain the fitting factor FF$=0.758$ if the un-modulated $|\tilde{h}_{a,\textrm{o}}(f)|$ is adopted to match $|\tilde{z}(f)|$. This demonstrates the importance of the near-field effect, otherwise it would lead to SNR underestimation with a factor $\sqrt{\rm FF}\approx0.87$. 

\subsection{PTA Noises}

The noises for PTA detection of individual GW sources can be divided into three main parts: the red noise, the shot noise, and the confusion from the GWB \citep{2015MNRAS.451.2417R, 2019MNRAS.485..248G}.  Here we do not consider the (intrinsic) red noises of pulsars though they are practically important. The main reason is that there are larger uncertainties in these red noises and their behaviour currently are not fully understood \cite[e.g.,][]{2020MNRAS.497.3264G,2016MNRAS.458.2161L}. Generally, the PSD of the GWB strain contributed by the shot noise is described as \citep{2011gwpa.book.....C}
\be
S_{\rm n,s}(f)=8\pi^2\sigma^2 f^2\Delta t,
\label{eq:S_ns}
\ee
where $\sigma$ is the root mean square (RMS) of pulsar timing noise and $\Delta t$ is the mean cadence of the PTA observations. The strain of GWB due to GW radiation from numerous distant MBBHs can be described as \citep[][c.f., \citealt{2015PhRvD..91h4055S}]{chen2020dynamical}
\be
h_{\rm b}=\mathcal{A}\frac{(f/1{\rm yr}^{-1})^{-2/3}}{[1+(f_{\rm bend}/f)^{\kappa_{\rm  gw}\gamma_{\rm gw}}]^{1/(2\gamma_{\rm gw})}}.
\ee

We adopt $\log\mathcal{A}\sim-15.70$, $f_{\rm bend} =2.45\times 10^{-10}$\,Hz, $\kappa_{\rm gw}=3.74$, $\gamma_{\rm gw}=0.19$, which are the median values for the GWB predictions in \citet{chen2020dynamical}.
The total noise for individual PTA sources is then 
\be
S_{\rm n}(f)=S_{\rm n,s}+\frac{h_{\rm b}^2}{f},
\label{eq:noise}
\ee
or 
\be
h_{\rm n}(f)=\sqrt{fS_{\rm n,s}+h_{\rm b}^2}.
%
%\eqno{\eqref{eq:noise}}
%
\ee
In the calculation of the SNR in this paper, we  consider the GWB as a source of noise. It is also plausible to only consider the shot noise to give optimistic SNR estimates after the GWB is well modelled and extracted from the PTA data.  For the cross-correlation method, both GWB and signal from an individual source are cross correlated, though they have different spectra. It is necessary to distinguish them from each other by using the matched-filtering method to extract individual signals and power-law modelling of the GWB. Combining the cross-correlation and matched filtering methods together, it is possible to obtain even higher SNR for single sources.
%}

\subsection{SNR}

We further derive the formulas to estimate SNR for monochromatic GW sources monitored by a coherent network of PTA within a limited time duration ($T$) as follows.

\begin{itemize}
\item \textbf{Matched-Filtering Method}:
the expected SNR can be roughly estimated as (combining Eqs.~\ref{eq:SNRmatched} and \ref{eq:hfdelta}), 
\bea
\varrho^2 &=& N_{\rm p}\int^{\infty}_{0}df \frac{2\chi^2h_0^2(f_0)\delta^2_T(f-f_0)}{S_{\rm n}(f)}, \nonumber \\
&\approx&N_{\rm p} \frac{2\chi^2h_0^2(f_0)T}{S_{\rm n}(f_0)},
%
%\label{eq:SNR_mono_m}
%
\eea
where $\delta_T(f)=\sin(\pi fT)/(\pi f)$ and $\delta_T(0)=T$ \citep{2015CQGra..32e5004M}
or equivalently,  
\be
\varrho^2 \approx N_{\rm p}\frac{\chi^2h_{\rm c}^2}{h_{\rm n}^2},
\label{eq:SNR_mono_m}
\ee
where the characteristic strain $h_{\rm c}=h_0 \sqrt{2fT}$, and $h_{\rm n}^2 =fS_{\rm n}$ \citep[see][]{2015CQGra..32a5014M}. 
 
\item \textbf{Cross-Correlation Method}:
the expected SNR can be roughly estimated as
(combining Eqs.~\ref{eq:SNRcross} and \ref{eq:hfdelta})
\bea
\varrho^2&=&\frac{N_{\rm p}(N_{\rm p}-1)}{T}\int^{\infty}_{0}df \frac{\chi^4h_0^4(f_0)\delta_T^4(f-f_0)}{S^2_{\rm n}(f)} \nonumber \\
&\approx&N_{\rm p}(N_{\rm p}-1) \frac{\chi^4h_0^4(f_0)T^2}{S^2_{\rm n}(f_0)},
%
%\label{eq:SNR_mono_c}
%
\eea
or equivalently,  
\be
\varrho^2 \approx \frac{N_{\rm p}(N_{\rm p}-1)}{4}\frac{\chi^4h_{\rm c}^4}{h_{\rm n}^4}.
\label{eq:SNR_mono_c}
\ee
\end{itemize}
 
Once the properties of a PTA ($\sigma$, $\Delta t$, $N_{\rm p}$) and a GW source ($h_0$ and $f_0$) are given, we can estimate its expected SNRs according to Equations~\eqref{eq:SNR_mono_m} and \eqref{eq:SNR_mono_c}. Note that the cross-correlation method usually results in a substantially higher SNR than that from the matched-filtering method if the SNR of one detection is higher than a threshold value of $3$ according to the above two Equations.

Table~\ref{tab:parameter} lists the (assumed) properties of a few current PTAs (EPTA/NANOGrav/PPTA, and IPTA), the  CPTA \citep{2016ASPC..502...19L}, and SKA-PTA \citep{2015CQGra..32a5014M, 2010CQGra..27h4016S}. We further note that a population of pulsars may exist within $1$\,pc from the GC and SKA may discover up to $100$ pulsars in the vicinity of Sgr A* according to recent model predictions \citep[e.g.,][]{2004ApJ...615..253P, 2014ApJ...784..106Z}. If some of those pulsars are stable MSPs, they may be monitored to form a special PTA (denoted as GC PTA, see also Tab.~\ref{tab:parameter}), which may be useful in detecting the nano-Hertz GW signals from the GC \citep[see][]{2012ApJ...752...67K}.   

It is worthy to note here that our SNR formulas are somewhat different from those in \citet{2015PhRvD..92f3010H} and \citet{2015CQGra..32e5004M}. 

First, the PSD $S_{\rm n}$ defined by Equation~\eqref{eq:noise} \citep[see also][]{2011gwpa.book.....C} is different from that defined in these two works. In \citet{2015CQGra..32e5004M}, the adopted shot noise $S_{\rm n}=2\sigma^2\Delta t$ is the PSD of the timing residuals, while $S_{\rm n}=24\pi^2\Delta t\sigma^2f^2$ in \citet{2015PhRvD..92f3010H} (from \citealt{2013PhRvD..88l4032T}) is the PSD of the GW strain. The shot noise (see Eq.~\ref{eq:S_ns}) we adopt is similar to \citet{2015PhRvD..92f3010H} and \citet{2013PhRvD..88l4032T}, but with a different coefficient of $8\pi^2$. The influence from the GWB is included in our PSD $S_{\rm n}$, but not in \citet{2015PhRvD..92f3010H} and \citet{2013PhRvD..88l4032T}. 
Second, we do not make approximations like those in \citet{2015CQGra..32e5004M}, in which the power of $4$ in their Equation~(12) was replaced by $2$, and then extended to the low- and high-frequency limits. We keep the accurate expression to calculate sensitivity curves numerically.
Third, the dependence of SNR $\varrho$ on $N_{\rm p}$ in this paper is a little different from that in \citet{2015PhRvD..92f3010H}. For simple superposition of coherent networks \citep[such as][]{wang2014a, wang2015coherent, 2015MNRAS.451.2417R}, we have $\varrho^2\propto N_{\rm p}h_{\rm c}^2/h_{\rm n}^2$; while if the cross-correlation method was adopted for each pulsar pairs, we have $\varrho^2\propto N_{\rm p}(N_{\rm p}-1)h_{\rm c}^4/h_{\rm n}^4$ \citep{2015CQGra..32e5004M}. However, \citet{2015PhRvD..92f3010H} obtained $\varrho^2\propto N_{\rm p}(N_{\rm p}-1) h_{\rm c}^2/h_{\rm n}^2$, which may overestimate the sensitivity of a PTA when $N_{\rm p}$ is high if setting $\varrho\equiv 1$ to define the PTA sensitivity. 
We also note here that the difference between the SNRs given by the matched-filtering method and the cross-correlation method results partly from the different definitions of the signals. The former is defined to be linear in the GW strain amplitude, while the latter as a quantity quadratic in the strain amplitude or linear in the GW power, which is the difference between these two methods \citep[see also][]{book:1417639}.

\subsection{Hypothetical GC MBBH}

We assume that there exists a circular intermediate mass BH(IMBH)-MBH binary in the GC ($r=8$\,kpc) with a total mass of $M_{\bullet\bullet}= 4.4\times 10^6M_\odot$ as given by observations \citep[e.g.,][]{2016ApJ...830...17B, 2017ApJ...837...30G, 2003ApJ...596.1015S, RevModPhys.82.3121, 2019Sci...365..664D} and the mass ratio is $q= M_{\bullet,2}/ M_{\bullet,1}$, though the probability for the existence of an MBBH with large $q$ (e.g., $>0.001$) in the GC may be little \citep[see][]{2020A&A...636L...5G}. We also assume that the GW frequency is either  $f_0=10^{-8}$, $3\times10^{-8}$, or $10^{-7}\rm Hz$, in the PTA band. Therefore, the GW strain and characteristic GW strain are 
\be
\begin{split}
h_0\approx & 8.2 \times 10^{-15} \frac{q}{(1+q)^{2}} \left(\frac{f_0}{10^{-8}\rm Hz}\right)^{\frac{2}{3}}\\
& \times \left(\frac{r}{8\rm kpc}\right)^{-1}\left(\frac{M_{\bullet\bullet}}{4.4\times10^6M_\odot}\right)^{\frac{5}{3}},
\end{split}
\label{eq:h0}
\ee
and
\be
\begin{split}
h_{\rm c} \approx & 2.0\times10^{-14} \frac{q}{(1+q)^{2}} \left(\frac{f_0}{10^{-8}\rm Hz}\right)^{\frac{7}{6}}\\
& \times \left(\frac{T}{10\rm yr}\right)^{\frac{1}{2}} \left(\frac{r}{8\rm kpc}\right)^{-1}\left(\frac{M_{\bullet\bullet}}{4.4\times10^6M_\odot}\right)^{\frac{5}{3}},
\end{split}
\label{eq:hc}
\ee
respectively. Adopting the mean of the distances of pulsars ($L$) for EPTA/NANOGrav/PPTA, IPTA, CPTA, and SKA-PTA as $\approx 2$\,kpc \citep{2005AJ....129.1993M}, $\bar{\chi}_{\rm n} \sim 0.368$ 
(see Fig.~\ref{fig:f7}), almost the same as that given by the far-field approximation $\bar{\chi}_{\rm f}$. We then estimate the SNRs for these different PTAs as listed in Table~\ref{tab:parameter}. It appears that the current PTAs (NANOGrav/EPTA/PPTA) are not likely to detect the GW signal with $f\sim 10^{-8}-10^{-7}$\,Hz from a hypothetical MBBH in the GC with $q\sim 0.01$ or less.

If the IPTA can include more pulsars (e.g., $N_{\rm p}=200$), observe more frequently (e.g., $\Delta t= 0.01$\,yr) with a higher timing precision ($\sigma_t=30$\,ns), like the row for IPTA$^{\rm opt}$ in Table \ref{tab:parameter}, such GW signals may be detectable in the frequency range of $f\sim 10^{-8}-10^{-7}$\,Hz  with SNR $\varrho>3$. The CPTA may be only able to detect the GW signal from a hypothetical MBBH with $q$ as small as $\sim 0.01$ at $f \sim 3\times10^{-8}$\,Hz with $\varrho> 3$. If the mass ratio $q\ll 0.01$, the  IPTA$^{\rm opt}$ and CPTA are not expected to detect such GW sources with substantially large SNR, but the SKA-PTA may be able to detect the GW signal from a hypothetical MBBH with $q$ as small as $\lesssim 0.001$ with a SNR $\varrho\gtrsim 3$.

We also estimate the expected SNRs for hypothetical MBBHs in the GC monitored by a possible PTA composed of MSPs close to it \citep[cf.,][]{2012ApJ...752...67K}. The properties of such a GC-PTA are assumed to be as the ``GC-PTA'' row listed in Table~\ref{tab:parameter}. From Equation~\eqref{eq:chipulsarGW}, we have
\be
\bar{\chi}\approx 2.92\times10^{3} \left(\frac{r_{\rm p}}{1\rm pc}\right)^{-1}.
\ee
The obtained SNRs via the GC PTA are high enough even if only $10$ MSPs with timing noises of $100$\,ns can be detected and applied. With such a GC-PTA, even the GW signals from a BH with mass down to several hundred times of solar masses rotating around the central MBH may be also detectable. This suggests that MSPs, if existing in the vicinity of the GC MBH, should be useful in detecting/constraining low frequency GWs emitted from  IMBHs or even stellar mass BHs rotating around the GC MBH. 

\begin{table*}
\caption{
Estimated SNR values for hypothetical GW sources at different frequencies monitored by current, future, and hypothetical PTAs with assumed properties by using different SNR estimation methods.
}
\begin{center}
\begin{tabular}{c|ccccc|ccc|ccc | ccc}
\hline \hline
\multirow{2}{*}{PTAs} & \multirow{2}{*}{$N_{\rm p}$} &  $\sigma$ & $T$ & $\Delta t$ &  $r_{\rm p}$ & $M_{\bullet\bullet}$ & \multirow{2}{*}{$q $} & \multirow{2}{*}{Location}  &\multicolumn{3}{c}{\multirow{2}{*}{SNR$_{\rm MF}(f)$}} \vline & \multicolumn{3}{c}{\multirow{2}{*}{SNR$_{\rm CC}(f)$}} \\ \cline{3-6} \cline{7-7}
& & (ns) & (yr) & (yr) & (pc) & ($M_\odot$) 
& & &  &   &  &  &   & \\ \hline
Single PTA & 20 & 100 & 20 & 0.04 & $\cdots$ & $4.26\cdot 10^6$ & $0.01$ & GC 
& 0.41 & 0.31  & 0.21 & 0.08 & 0.05  & 0.02\\
IPTA & 49 & 100 & 20 & 0.04 & $\cdots$ & $4.26\cdot 10^6$ & $0.01$ & GC 
& 0.64 & 0.49  & 0.33 & 0.20 & 0.12  & 0.05\\
IPTA$^{\rm opt}$ & 200 & 30 & 20 & 0.01 & $\cdots$ & $4.26\cdot 10^6$ & $0.01$ & GC 
& 3.09 & 6.32  & 4.38 & 4.77 & 19.9  & 9.55\\
CPTA & 100 & 20 & 20 & 0.04 & $\cdots$ & $4.26\cdot 10^6$ & $0.01$ & GC 
& 2.10 & 3.40 & 2.32 & 2.19 & 5.75 & 2.64\\
SKA & $10^3$ & 10 & 20 & 0.04 & $\cdots$ & $4.26\cdot 10^6$ & $0.01$ & GC 
& 7.13 & 20.4  & 14.7 & 25.4 & 207.3  & 107.6\\
SKA$^{\rm opt}$ & $10^3$ & 10 & 20 & 0.01 & $\cdots$ & $4.26\cdot 10^6$ & $0.001$  & GC 
& 0.74 & 3.49  & 2.98 & 0.27 & 6.08 & 4.45\\
GC-PTA & 10 & $100$ & 20 & 0.02 & 1 & $4.26\cdot 10^6$ & $0.0001$  & GC 
& 21.8 & 17.7 & 11.9 & 226 & 149  & 67.0\\
LMCC-PTA1  & 20 & 100 & 10 & 0.02 & 1 & $2.4 \cdot 10^4$ & $0.1$  & LMC 
& 3.70 & 4.55 & 2.48  & 6.65 & 10.1 & 2.99 \\
LMCC-PTA2  & 5 & 100 & 10 & 0.02& 0.1 & $2.4 \cdot 10^4$ & $0.1$ & LMC 
& 18.5 & 22.8 & 12.4 & 153 & 232  & 68.6 \\ \hline \hline
\end{tabular}
\end{center}
\tablecomments{
Columns from left to right lists the name of the PTAs, number of pulsars ($N_{\rm p}$), pulsar timing precision ($\sigma$) in unit of ns, total observational period $T_{\rm obs}$ in unit of year,  time interval for each observation $\Delta t$ in unit of year, distance of the pulsar to the GW source $r_{\rm p}$ in unit of pc, total MBBH mass $M_{\bullet\bullet}$ in unit of solar mass, mass ratio $q$, location of the GW source,  estimated SNR values for those GW sources at frequency $f=10^{-8}$, $3\times 10^{-8}$, and $10^{-7}$\,Hz, by the matched filtering method (MF) and the cross-correlation method (CC), respectively.  
}
\label{tab:parameter}
\end{table*}

\subsection{LMC}

An MBH with $M_{\bullet,1} \approx2 \times 10^4M_\odot$ is suggested to exist in the center of LMC \citep{2017ApJ...846...14B}. There was also tentative evidence for the existence of an MBBH in LMC center, e.g., hypervelocity star ejected from the LMC \citep[e.g.,][]{2019MNRAS.483.2007E}. Suppose there exists another IMBH with mass $M_{\bullet,2}\approx2\times10^3M_\odot$ rotating around the central MBH $M_1$, and the GW emission from such a binary system is at a frequency either of $f_0\approx10^{-8}$, $3\times10^{-8}$, or $10^{-7}\rm Hz$. The distance from the LMC to  Earth is about $r=49.97$\,kpc \citep{2013Natur.495...76P}. Thus the GW strain received at the Earth can be obtained from Equations~\eqref{eq:h0} and \eqref{eq:hc}. In this case, it is difficult to detect the GW signal by current PTAs and even future SKA-PTA. However, if the hypothetical MBBH is monitored via a PTA composed of MSPs at the LMC center as listed in Table~\ref{tab:parameter}, then
\be
\bar{\chi}\approx 1.82\times10^{4}\left(\frac{r_{\rm p}}{1\rm pc}\right)^{-1}.
\ee
As shown in Table~\ref{tab:parameter} (the last two rows), as long as $20$ MSPs in the center of LMC with $r_{\rm p}\sim1$\,pc can be detected and applied to form a PTA, the GW signal can be detected with SNR$\approx 4-15$; if only $5$ MSPs at $r_{\rm p}\sim0.1$\,pc, the GW signal can be detected with a SNR $\approx 20-340$. Note that the farthest the pulsar is $59.7$\,kpc away from Earth in pulsar catalog\footnote{http://www.atnf.csiro.au/people/pulsar/psrcat} \citep{2005AJ....129.1993M}, and 21 pulsars in the LMC have been discovered \citep{2019ARA&A..57..417C}. It is possible that many MSPs in the LMC may be detected in the SKA era. However, it would be a challenge to get the high-precision timing demanded by the PTA to detect GWs.

\section{Conclusions}
\label{sec:Conclusion}

We investigate the detection of GWs emitted from nearby MBBHs via PTAs and introduce a general theoretical framework to study the near-field effect on detecting these MBBHs by utilizing the standard matched-filtering method and the cross-correlation method. We find that the traditional plane wave approximation adopted for faraway GW sources is not valid in the cases for detecting MBBHs at distances comparable or not much larger than the distances of PTA pulsars. In this framework, we derive new and general expressions for some physical quantities, such as the geometric factor $\chi$, the overlap reduction function, and the SNR estimators for both the matched-filtering and the cross-correlation detection methods. Our main conclusions are summarized as follows.

\begin{itemize}

\item The near-field effect is significant in extracting GW signals from nearby MBBHs via PTA observations, as the matched-filtering is sensitive to the exact GW waveform in the frequency domain. For the detection of such nearby MBBHs, an appropriate modification should be made on the GW templates used in the far-field approximation, otherwise it will lead to a underestimate of the SNR (e.g., up to a factor of $1.36$ for a MBBH in the GC; see Fig.~\ref{fig:f5}) and further large uncertainties in the estimation of the system parameters. 

\item Combining the small-scale spiky features of the angular distribution of the response of PTA and the GW parallax effects due to the curvature of the wavefronts of GWs, the spatial locations of nearby GW sources may be determined with high precision (e.g., $\lesssim 1\arcdeg$ as seen from Fig.~\ref{fig:f4}) and thus the degeneracy between GW frequency and MSP distances may be also broken in the near-field regime.

\item MSPs in the GC, if any, will be powerful probes to nano-Hertz GWs emitted from the GC. If some stable MSPs located around GC are discovered in the future, the GW signal from an MBBH (if any) in the GC can be detected with a high SNR even if only several suitable stable MSPs are observed. Similarly, a PTA composed of some stable MSPs in the LMC can also be used to detect the GW signal from an MBBH (if any) in the LMC.

\item For most known MSPs ($L\sim 1-2$\,kpc) currently adopted in PTAs, the near-field effect is significant if the MBBH distance $r \lesssim30$\,Mpc in actual detection. Many galaxies are located within this distance and they may have MBBHs in their centers as possible GW sources for PTAs \citep{2016MNRAS.459.1737S}, therefore, the near-field effect needs to be carefully considered when using PTA to search for such GW sources. If more MSPs with higher distances were adopted in future PTAs, the near-field effect could be significant for MBBHs at even larger distances.

\item  The angular correlation between the responses of different pulsars to an isotropic GWB contributed by isotropically distributed nearby sources is similar to the Hellings-Downs curve obtained by the far-field approximation { except for the value at $\theta_{12}=0^\circ$}.

\end{itemize}

For simplicity, in our analysis we have neglected some observational effects, such as high-order effects in real observations like the red noises in pulsar timings \citep{2020MNRAS.497.3264G, 2015MNRAS.451.2417R}, the post-Newtonian effects \citep{2012ApJ...752...67K}, etc. These effects should be considered carefully when extracting GW signals of MBBHs from the TOA data series of PTAs.
 
\acknowledgments
{%
\noindent
We thank the referee for helpful comments and suggestions. This work is partly supported by the National SKA Program of China (Grant No. 2020SKA0120101), National Key Program for Science and Technology Research and Development (Grant No. 2020YFC2201400, 2016YFA0400703/4), the National Natural Science Foundation of China (Grant No. 11721303, 11873056, 11991052, 12173001, 11690024), and the Strategic Priority Program of the Chinese Academy of Sciences (Grant No. XDB 23040100).
}

\appendix
\section{Two Coordinate Systems}
\label{sec:coordinate}

In general cases, the GW propagation directions $\hat{\Omega}$ are different at different $Q$ (see Fig.~\ref{fig:f1}). We can define two kinds of coordinate systems. One is such a frame rotating with $\hat{\Omega}$ in the pulsar-Earth-GW source plane, and $(\hat{e}_1,\hat{e}_2,\hat{\Omega})$ are taken as the $x$-, $y$-, $z$- axis bases. Here $\hat{e}_1$ is perpendicular to the pulsar-Earth-GW source plane, $\hat{e}_2$ is a unit vector in the pulsar-Earth-GW source plane perpendicular to $\hat{\Omega}$. This coordinate system is denoted as the $(\hat{e}_1,\hat{e}_2,\hat{\Omega})$ system. Rotating this coordinate system by a polarization angle $\psi$, we can obtain the coordinate system in traverse traceless (TT) gauge. It is convenient to calculate the antenna pattern function in such a coordinate system because $\hat{e}_1,\hat{e}_2,\hat{\Omega}$ are invariant, though the pulsar direction $\hat{p}=(0, \sin\gamma, \cos\gamma)$ does change. In this coordinate system, antenna pattern functions are expressed as
\be
F^+(l)=-\frac{\sin^2\gamma}{2(1+\cos\gamma)}=-\frac{1-\cos\gamma}{2},
\ee
\be
F^{\times}(l)=0.
\ee

Another one is the fixed coordinate system relative to the observer's sky. We choose $(\hat{e}_{1},\hat{e}_{2,\rm E},\hat{\Omega}_{\rm E})$ at Earth as the $x,y,z$ axis bases. We denote it as the $(\hat{e}_1,\hat{e}_{2,\rm E},\hat{\Omega}_{\rm E})$ coordinate system. In this coordinate system, the pulsar direction $\hat{p}=(0, \sin\vartheta, -\cos\vartheta)$ and $\hat{e}_{1}$ does not change, however, $\hat{e}_2$ and $\hat{\Omega}$ change with $Q$ and $\hat{e}_2=(0,\cos\zeta,-\sin\zeta)$, $\hat{\Omega}=(0,\sin\zeta,\cos\zeta)$. It is convenient to transfer this frame to the celestial coordinates. We can also obtain the same antenna pattern function in this coordinate system, if choosing the same bases. These two coordinate systems are approximately the same in the far-field regime, i.e., the GW source is faraway from PTA MSPs.

\section{Some Formulas for calculating $\chi$}
\label{sec:concrete}

According to the geometry illustrated in Figure~\ref{fig:f1}(a) for a general configuration of the GW source, Earth, and PTA MSP, we have
\be
r'^2=l^2+r^2-2l r\cos\vartheta,
\label{eq:vartheta}
\ee
\be
\cos\gamma=\frac{l-r\cos\vartheta}{\sqrt{l^2+r^2-2l r\cos\vartheta}},
\ee
and
\be
\cos\zeta=\frac{r-l\cos\vartheta}{\sqrt{l^2+r^2-2l r\cos\vartheta}},
\ee
where $\vartheta$ is the angle between the line of sight to the GW source and that to the pulsar. For a given GW source, $r'$ is a function of $l$ as $r$ and $\vartheta$ are fixed.

Since $l$ can be comparable to $r$, $\gamma$ and $\zeta$ may vary significantly for different points $Q$ along the propagation paths of pulses from pulsars to Earth. If  $\iota_{\rm E}=0\arcdeg$, we have $\iota=\zeta$ at any point $Q$ between the Earth and pulsar. Because $\hat{\Omega}$ is always located in the pulsar-Earth-GW source plane, even if $\iota_{\rm E}\neq 0^\circ$, as long as $\hat{n}$ is in the pulsar-Earth-GW source plane, $\hat{n}\times\hat{\Omega}$ is always parallel to $\hat{e}_1$. Thus we have the polarization angle $\psi=0$ for any point $Q$ in this case.
\begin{figure}
\centering
\includegraphics[width=0.6\textwidth]{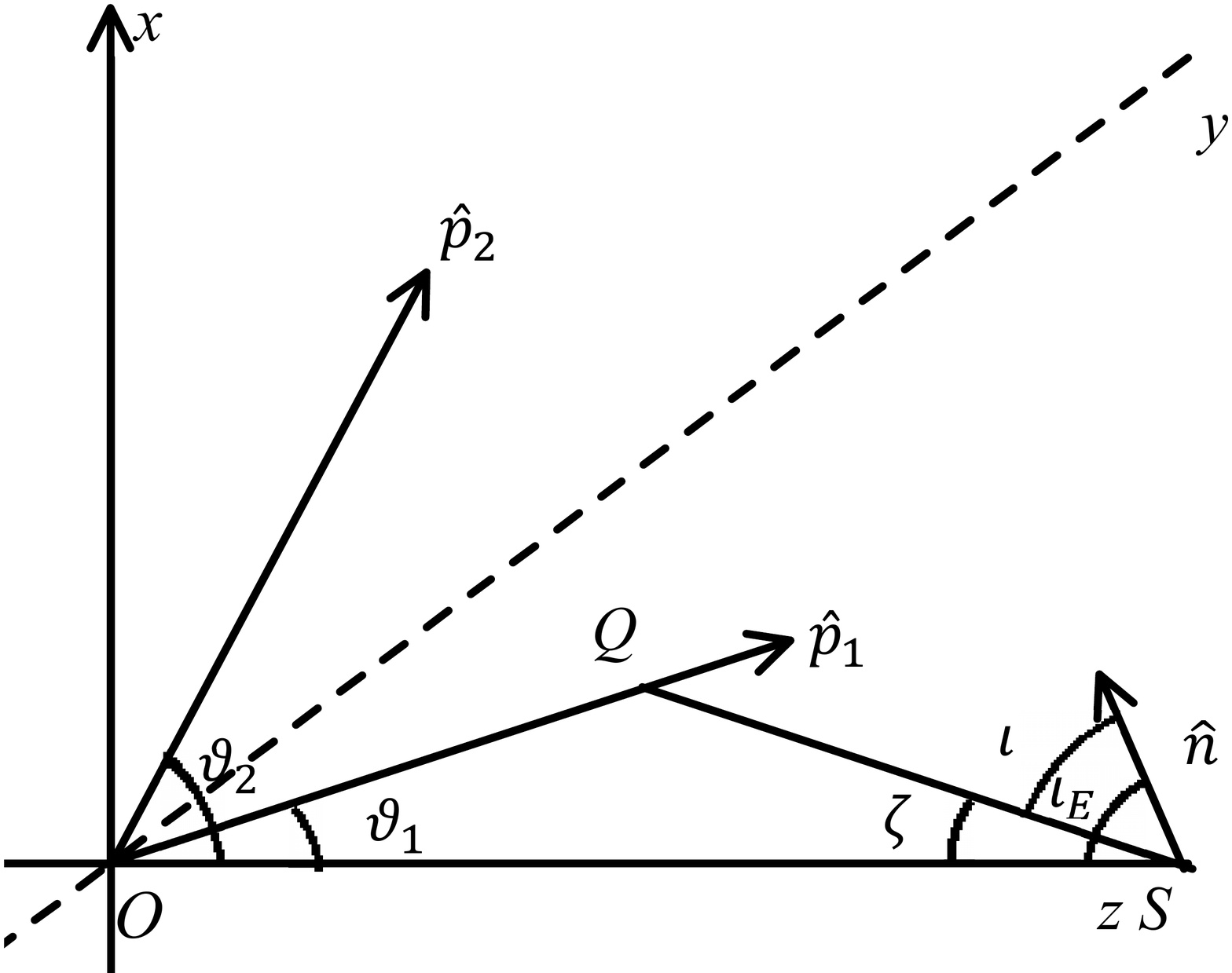}
\caption{
The 3-dimensional schematic for the configuration of a GW source and two pulsars, where two pulsars, GW sources ($S$) and Earth are not located in the same plane; Earth is locate at  the origin $O$ of coordinate system; and pulsar 1 is located in the plane $yOz$. It should be noted that the positive direction of $z$ axis points to left.
} 
% 
%\label{fig:sch3D}
\label{fig:f9}
\end{figure}

For the general case with $\psi_{\rm E}\neq 0^\circ$ and $\iota_{\rm E}\neq 0^\circ$ (see Fig.~\ref{fig:f9}), we denote the unit normal vector of the BBH orbital plane as $\hat{n}$. We use $\psi_{\rm E}$ and $\iota_{\rm E}$ to denote the angle between the pulsar-Earth-GW source plane and $\hat{n}$-$\hat{\Omega}_{\rm E}$ plane, and the angle between $\hat{\Omega}_{\rm E}$ and $\hat{n}$, respectively. From the spherical law of cosines and sines, we have
\be
\hat{n}\cdot\hat{\Omega}=\cos\iota=\cos\zeta\cos\iota_{\rm E}+\sin\zeta\sin\iota_{\rm E}\cos\psi_{\rm E},
\ee
\be
\sin\psi=\sin\psi_{\rm E}\frac{\sin\iota_{\rm E}}{\sin\iota},
\ee
and
\be
\cos\psi=\frac{\cos\iota\cos\zeta-\cos\iota_{\rm E}}{\sin\iota\sin\zeta}.
\ee

For a GW source with fixed $\iota_{\rm E}$, we can use an average over $\cos\vartheta$ and $\psi_{\rm E}$ to represent the PTA MSPs that uniformly distributed in the observer's sky. 

\begin{figure*}
\centering
\includegraphics[width=0.8\textwidth]{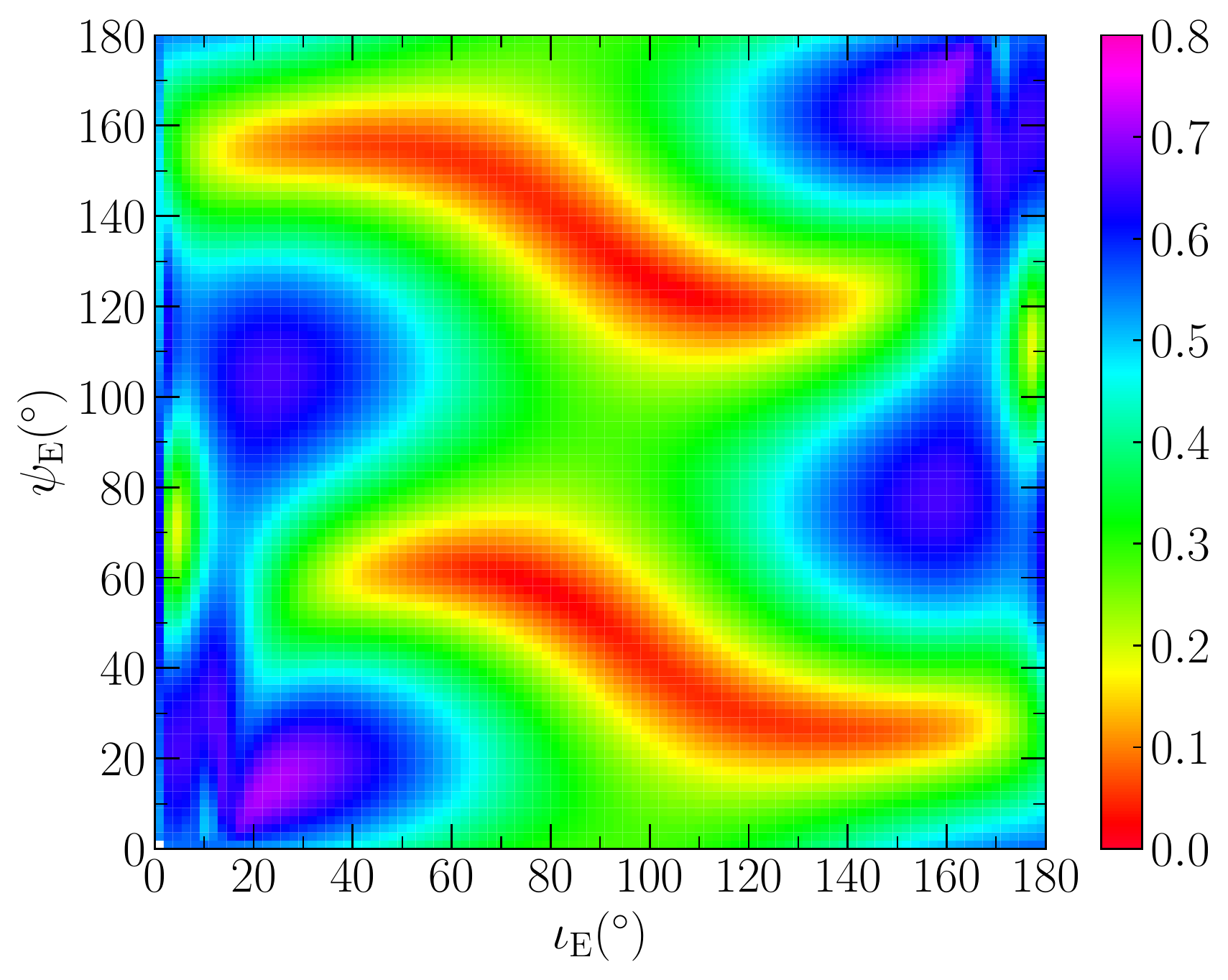}%chi_iota_psi.ipynb
\caption{Dependence of $\chi(\iota_{\rm E},\psi_{\rm E})$ on $\iota_{\rm E}$ and $\psi_{\rm E}$ for a GW source located at $r=8$\,kpc monitored by a PTA with MSP distances $L=2$\,kpc. Different colors represent different $\chi(\iota_{\rm E},\psi_{\rm E})$ value as indicated by the right color bar. 
}
% 
%\label{fig:chi_iota_psi}
\label{fig:f10}
\end{figure*}

To show the orientation dependence of $\chi$ with $\iota_{\rm E}$ and $\psi_{\rm E}$, we define 
\be
\chi(\iota_{\rm E},\psi_{\rm E})= \sqrt{\int \frac{  {\rm d} \hat{\Omega}_{\hat{p}}}{8\pi} |\mathscr{P}^{+}(f)+\mathscr{P}^{\times}(f)|^2}.
\ee
For illustration, Figure~\ref{fig:f10} shows $\chi(\iota_{\rm E},\psi_{\rm E})$ as the function of $\iota$ and $\psi$ for an example GW source with distance $r=8$\,kpc monitored by a PTA with $L=2$\,kpc.  Here we show the results for the region $\psi\in[0,\pi]$, and the results for the region $\psi\in[-\pi,0]$ has a similar pattern due to the symmetry. The mean value of $\chi$ can be given by
\be
\bar{\chi}= \sqrt{\int^{1}_{-1} \frac{{\rm d}\cos\iota_{\rm E}}{2}\int^{\pi}_0\frac{{\rm d}\psi_{\rm E} }{\pi}\int \frac{  {\rm d} \hat{\Omega}_{\hat{p}}}{8\pi} |\mathscr{P}^{+}(f)+\mathscr{P}^{\times}(f)|^2}.
\ee

\section{Celestial Coordinate System}
\label{sec:celestial}

In practice, only the celestial coordinates of pulsars and GW sources are known. These angles need to be expressed in a celestial coordinates system. From the spherical law of cosines, it's easy to obtain $$\cos\vartheta= \cos\delta \cos\delta_{\rm p} \cos(\alpha-\alpha_{\rm p}) + \sin\delta\sin\delta_{\rm p}$$ \citep[e.g.,][]{1987GReGr..19.1101W, 2011MNRAS.414.3251L, 2015MNRAS.449.1650Z, 2016MNRAS.461.1317Z}, where $\alpha$ and $\alpha_{\rm p}$ are the right ascensions of the GW source and the pulsar, respectively, $\delta$ and $\delta_{\rm p}$ are the declinations of the GW source and the pulsar, respectively. For convenience, we define that the normal vector $\hat{n}$ of GW source orbital plane points at a direction $(\alpha_{\rm n},\delta_{\rm n})$ in the celestial sphere. The pulsar ($\rm p$), GW source ($\rm S$), $\hat{n}$ ($\rm n$) three points in the celestial sphere can form a spherical triangle $\rm pSn$. Its three sides are given by
$$\overset{\frown}{\rm pS}=\vartheta,$$ 
$$\overset{\frown}{\mathrm{Sn}}= \arccos(\cos\delta \cos\delta_{\rm n} \cos(\alpha-\alpha_{\rm n}) + \sin\delta\sin\delta_{\rm n}),$$
and
$$\overset{\frown}{\rm pn}= \arccos(\cos\delta_{\rm n} \cos\delta_{\rm p} \cos(\alpha_{\rm n}-\alpha_{\rm p}) + \sin\delta_{\rm n}\sin\delta_{\rm p}).$$
We also define the angle at point S between sides $\overset{\frown}{\rm pS}$ and $\overset{\frown}{\rm Sn}$ of the triangle on the celestial sphere as $\angle \mathrm{pSn}$. It is easy to obtain that
$$\cos\angle \mathrm{pSn}=\frac{\cos \overset{\frown}{\rm pn}-\cos \overset{\frown}{\rm Sn}\cos\vartheta}{\sin \overset{\frown}{\rm Sn}\sin\vartheta},$$
thus polarization angle $\psi_{\rm E}$ can be expressed as $|\psi_{\rm E}|=\angle\rm pSn$ or $\pi-\angle\rm pSn$.

\section{Waveform Differences for Different Angle $\vartheta$}
\label{sec:angles}

\begin{figure}
\centering
\includegraphics[width=\textwidth]{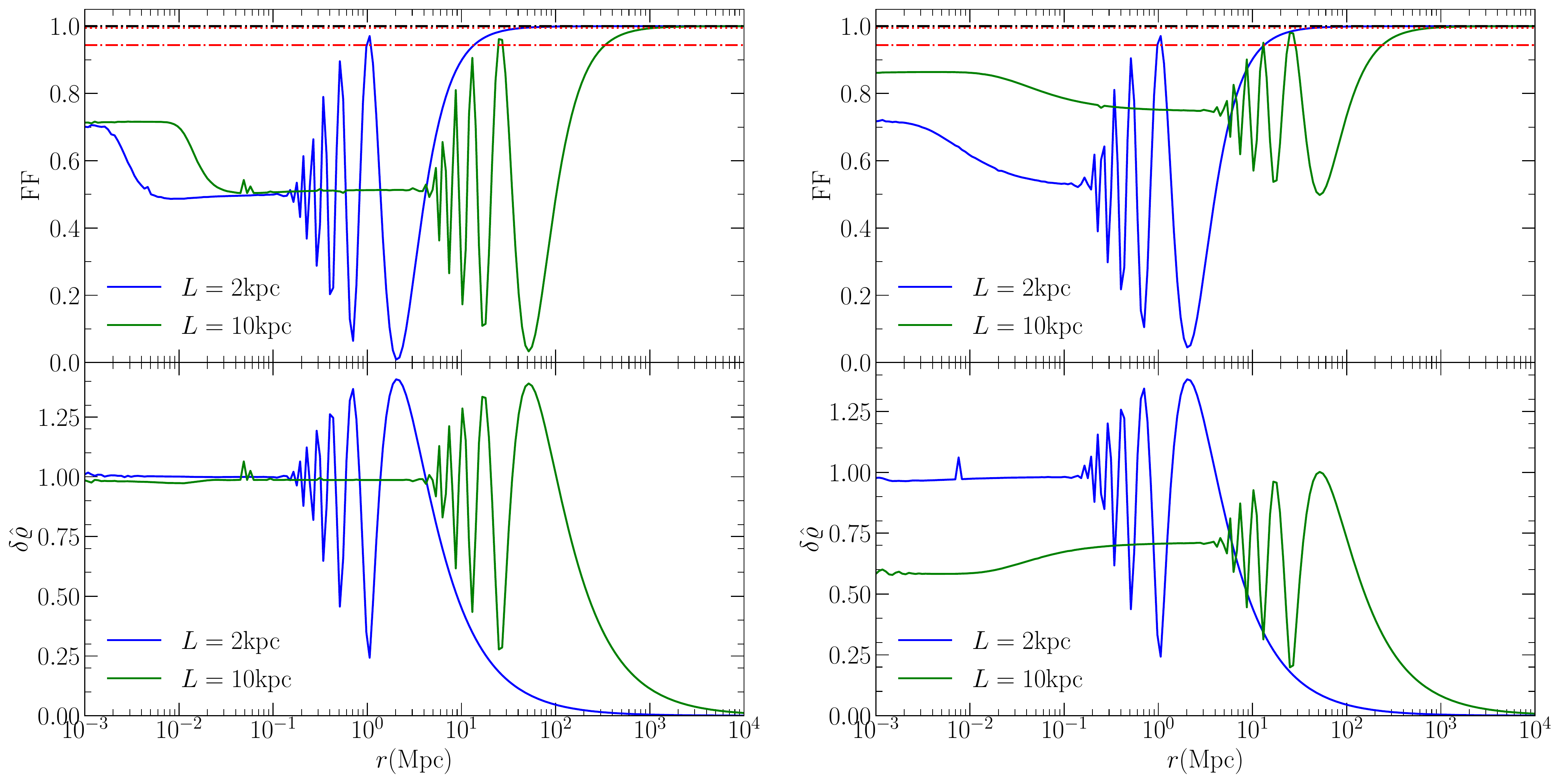}
\caption{Similar to Fig.~\ref{fig:f5}, except for $\vartheta=\frac{\pi}{4}$ in the left panel, and $\vartheta=\frac{3\pi}{4}$ in the right panel.}
%
%\label{fig:FF_13}
\label{fig:f11}
\end{figure}

We have shown the FF and $\delta\hat{\varrho}$ for the case with $\vartheta=\frac{\pi}{2}$ in the main text (see Fig.~\ref{fig:f5}). We also calculate the FF and $\delta\hat{\varrho}$ for cases with $\vartheta=\frac{\pi}{4}$ and $\frac{3\pi}{4}$, respectively (see Fig.~\ref{fig:f11}). For those cases with $L=1$\,kpc or $2$\,kpc, the FF and $\delta\hat{\varrho}$ of waveforms of a nearby source at $r=8$\,kpc by adopting the far-field approximation ($r=\infty$) are also summarized in Table~\ref{tab:diff_angles}. Although for different $\vartheta$, the resulting FF and $\delta\hat{\varrho}$ are different, qualitatively they all suggest that the near-field effect is important for GW sources with distances  (i.e., $r$) not much larger than the PTA pulsar distance (i.e., $L$), irrespective to the directions of PTA pulsars. Further, we also calculate the maximum distance $r$ (Mpc) that the near-field effect may be important (corresponding to a FF threshold of $0.944$ or $0.995$) for cases with given PTA pulsar distance $L$ and angle $\vartheta$ as listed Table~\ref{tab:FF_diff_angles}. The differences of these maximum distances for cases with different $\vartheta$ are about a factor of $3$ or less (see also Fig.~\ref{fig:f11}).

\begin{table}
\centering
\caption{FF (see Eq.~\ref{eq:FF}) and $\delta\hat{\varrho}$ (see Eq.~\ref{eq:delta_SNR}) obtained for several cases with different choices of $\vartheta$ and $(L, r)$. }
\begin{tabular}{c|c c|c c}    \hline\hline
$(L, r)$ & \multicolumn{2}{c}{($1$\,kpc, $8$\,kpc)} \vline & 
\multicolumn{2}{c}{($2$\,kpc, $8$\,kpc)} 
\\ \hline 
$\vartheta$ & FF & $\delta\hat{\varrho}$ & FF & $\delta\hat{\varrho}$\\ \hline
$\pi/4$ & 0.490 & 1.00 & 0.487 & 1.00 \\ %\hline
$\pi/2$ & 0.537 & 0.999 & 0.581 & 0.993 \\ %\hline
$3\pi/4$ & 0.572 & 0.995 & 0.635 & 1.01 \\ \hline\hline
\end{tabular}
\label{tab:diff_angles}
\end{table}

\begin{table}
\centering
\caption{The maximum distance $r$ (Mpc) obtained for those cases with different angles $\vartheta$ for which FF$<$FFS (see definitions of FF and FFS in Eqs.~\eqref{eq:FF} and \eqref{eq:FFS}). 
}
\begin{tabular}{c|c c|c c}  \hline\hline
$L$ & \multicolumn{2}{c}{2\,kpc} \vline & \multicolumn{2}{c}{10\,kpc}  \\ \hline 
$\vartheta$  & FF$<$0.944 & FF$<$0.995 & FF$<$0.944 & FF$<$0.995 \\ \hline
$\pi/4$ & 13.7 & 46.0 & 338 & 1140 \\ %\hline
$\pi/2$ & 27.1 & 91.3 & 572 & 1933 \\ %\hline
$3\pi/4$ & 13.5 & 45.4 & 241 & 821 \\ \hline\hline
\end{tabular}
\label{tab:FF_diff_angles}
\end{table}

\section{Pulsars Located Around GW Sources}
\label{sec:Kocsis}

If all PTA MSPs are located around the GW sources, similar to the configuration studied in \citet{2012ApJ...752...67K} with all MSPs located in the neighborhood of the GC, it can be also regarded as a special case of the frame work considered in section~\ref{sec:General}. In this case, we have $\vartheta \simeq 0$ and $L\approx r$, $r'\approx r-l$, and we define the position vector from the MSP to the GW source as  $\vec{r}_{\rm p}=\vec{r}-L\hat{p}$ and the distance between them is $r_{\rm p}=|\vec{r}_{\rm p}|$. Thus Equation~\eqref{eq:z=ph} can be rewritten as
\be
\tilde{z}(f)=\sum_a\tilde{h}_a(f)\int^L_0 dl \mathscr{F}^a(l){\frac{d}{dl}} \left(\frac{rA_a(\iota)}{(r-l)A_a(\iota_{\rm E})}\right)
\approx\sum_a\tilde{h}_a(f)\mathscr{F}^a(0) \cdot \left(\frac{rA_a(\iota_{\rm p})}{r_{\rm p}A_a(\iota_{\rm E})}-1\right),
\ee
where the following approximations are adopted,
\begin{equation}
\int^L_0 dl \mathscr{F}^a(l)\frac{d}{dl}\left(\frac{rA_a(\iota)}{(r-l)A_a(\iota_{\rm E})}\right)
=\mathscr{F}^a(L)\frac{rA_a(\iota_{\rm p})}{r_{\rm p}A_a(\iota_{\rm E})} -\mathscr{F}^a(0) -\int^L_0\frac{rA_a(\iota)}{(r-l)A_a(\iota_{\rm E})}\frac{d\mathscr{F}^a(l)}{dl}dl,
\label{eq:integ1}
\end{equation}
and
\begin{eqnarray}
\int^L_0\frac{rA_a(\iota)}{(r-l)A_a(\iota_{\rm E})}\frac{d\mathscr{F}^a(l)}{dl}dl
&=&\int^{L-\epsilon}_0\frac{rA_a(\iota)}{(r-l)A_a(\iota_{\rm E})} \frac{d \mathscr{F}^a(l)}{dl}dl+\int^L_{L-\epsilon}\frac{rA_a(\iota)}{(r-l)A_a(\iota_{\rm E})} \frac{d \mathscr{F}^a(l) }{dl}dl  \\
\label{eq:integ2u}
&\approx & 0+\frac{r}{r_{\rm p}}\frac{A_a(\iota_{\rm p})}{A_a(\iota_{\rm E})}\left[\mathscr{F}^a(L)-\mathscr{F}^a(0)\right], 
\label{eq:integ2l}
\end{eqnarray}
where $\epsilon$ is a small quantity relative to $L$ but greater than $r_{\rm p}$. As $\frac{r}{r-l}\ll\frac{r}{\epsilon}<\frac{r}{r_{\rm p}}$, the first integral at the right hand side of Equation~\eqref{eq:integ2u} is $< \frac{r}{\epsilon}(\mathscr{F}^a(L-\epsilon)-\mathscr{F}^a(0))$ and $\mathscr{F}^a(l)$ does not change much in the integration range of $l$ from $0$ to $L-\epsilon$. Therefore, the first integral is small compared with the second one and thus can be ignored. For the second integral, we make an approximation $\frac{r}{r-l}\approx\frac{r}{r_{\rm p}}$ between $L-\epsilon$ and $L$. Although the variation of $\iota$ may be significant between $L-\epsilon$ and $L$ that depends on $\psi_{\rm E}$, $\iota_{\rm E}$, $r_{\rm p}$, etc. As an approximation, we may use inclination $\iota_{\rm p}$ at MSPs to replace all $\iota$ because the integral may be dominated by the contribution from $l\approx L$, and we also adopt the approximation $\mathscr{F}^a(L-\epsilon)\approx\mathscr{F}^a(0)$ to get Equation~\eqref{eq:integ2l}. Thus 
\be
\mathscr{P}^{a}(f)\approx \mathscr{F}^a(0) A_a(\iota_{\rm E}) \cdot \left(\frac{r}{r_{\rm p}}\frac{A_a(\iota_{\rm p})}{A_a(\iota_{\rm E})}-1\right)\approx \mathscr{F}^aA_a(\iota_{\rm p})\frac{r}{r_{\rm p}}.
\label{eq:Pfapp}
\ee
For a GC-PTA, if $r_{\rm p}\sim 1$\,pc, $|\frac{r}{r_{\rm p}}|\sim10^3\gg1$, and  $\frac{A_a(\iota_{\rm p})}{A_a(\iota_{\rm E})} \sim O(1)$ on average, therefore, the pulsar term is much greater than the Earth term, and the Earth term can be ignored. This geometrical configuration is the same as the case considered in \citet{2012ApJ...752...67K}. The above equations are derived for a single pulsar, and the pulsar term is dominant in Equation~\eqref{eq:Pfapp}. Note that the approximation of taking only one single pulsar term may be inaccurate (see \citealt{2012ApJ...752...67K}), because the combined effect from a PTA should be averaged over all the different pulsars around the source applied in the PTA. Adopting the general framework presented in Section~\ref{sec:General} of the present paper, we obtain
\be
\bar{\chi}\approx\frac{r}{r_{\rm p}}\bar{\chi}_{\rm f}\approx0.365\frac{r}{r_{\rm p}}
\label{eq:chipulsarGW}
\ee
by averaging over $(\psi_{\rm E}, \iota_{\rm E})$ for different pulsars.

\section{Relevant geometry for calculations of the angular correlation function in the near-field regime}
\label{sec:ORF_GWB}

For a pair of PTA pulsars (denoted by $i=1$ and $2$, respectively), we set a coordinate system for them so that they are located at $\hat{p}_1=(0,0,1)$ and $\hat{p}_2=(0,\sin\theta_{12},\cos\theta_{12})$ with $\theta_{12}$ denoting the angle between their directions. We set $-\hat{\Omega}_{\rm E}=(\sin\theta_{\rm gw}\cos\phi_{\rm gw},\sin\theta_{\rm gw}\sin\phi_{\rm gw},\cos\theta_{\rm gw})$,
$\hat{n}=(\sin\theta_{\rm n}\cos\phi_{\rm n},\sin\theta_{\rm n}\sin\phi_{\rm n},\cos\theta_{\rm n})$ with $\theta_{\rm gw}$ and $\phi_{\rm gw}$ represent the polar angle and azimuthal angle of the direction of GW source, and $\theta_{\rm n}$ and $\phi_{\rm n}$ represent the polar angle and azimuthal angle of the direction of normal vector $\hat{n}$ of the orbital plane in this coordinate system. Then we have
$$
\cos\iota_{\rm E}=\hat{\Omega}_{\rm E}\cdot\hat{n},
$$
$$
\cos\vartheta_{i}=-\hat{p}_{i}\cdot\hat{\Omega}_{\rm E},
$$
$$
\sin\psi_{\mathrm{E},i}={\rm sign}[(\hat{p}_i\times\hat{\Omega}_{\rm E})\cdot\hat{n}]\frac{|(\hat{p}_{i}\times\hat{\Omega}_{\rm E})\times(\hat{n}\times\hat{\Omega}_{\rm E})|}{|\hat{p}_{i}\times\hat{\Omega}_{\rm E}||\hat{n}\times\hat{\Omega}_{\rm E}|},
$$
$$
\cos\psi_{\mathrm{E},i}=\frac{(\hat{p}_{i}\times\hat{\Omega}_{\rm E})\cdot(\hat{n}\times\hat{\Omega}_{\rm E})}{|\hat{p}_{i}\times\hat{\Omega}_{\rm E}||\hat{n}\times\hat{\Omega}_{\rm E}|},
$$
for $i=1,2$. Thus we can transform these angles ($\theta_{12}$, $\theta_{\rm gw}$, $\phi_{\rm gw}$, $\theta_{\rm n}$, $\phi_{\rm n}$) into $(\iota_{\rm E},\psi_{\mathrm{E},i},\vartheta_i)$ for $i=1,2$. According to Appendix~\ref{sec:concrete}, we can calculate $\iota$, $\psi$ and so on.
For each pulsar $i$, we can calculate $\mathscr{P}^a_i(f)$ for $a=+,\times$, once $(\iota_{\rm E},\psi_{\mathrm{E},i},\vartheta_i)$, $L$, and the distances of GW sources $r$ are given. We can then obtain $\bar{\Gamma}^{\rm b}_{12}(f)$ according to Equation~\eqref{eq:aveORF}.  
When $\iota_{\rm E}=\frac{\pi}{2}$, $A_{\times}=\cos\iota_{\rm E}=0$ is a singularity in the numerical integration of equation~\eqref{eq:aveORF}. To avoid this singularity, we excise  $|\cos\iota_{\rm E}|<10^{-2}$ part in the calculation. 
Another way to avoid the singularity is to re-define the ORF as 
$$
\frac{\beta^{\rm b}_{12}}{4\pi}\int d^2\hat{n}\int d^2\hat{\Omega}_{\rm E}\sum_a\mathscr{P}_1^{*a}(f,\hat{\Omega}_{\rm E})\mathscr{P}_2^{a}(f,\hat{\Omega}_{\rm E}),
$$
instead of Equation~\eqref{eq:aveORF}. The resulting ORF shape is also similar to that obtained from Equation~\eqref{eq:aveORF} except for a normalization difference of $\sim 2.5$. 

Besides using spherical triangle relation to obtain $\psi$ as shown in Appendix~\ref{sec:concrete}, we also have another way to calculate $\psi$ according to vector expressions.
If $(\hat{p}\times\hat{\Omega}_{\rm E})\cdot\hat{n}>0$, $\psi\in(0,\pi)$, $\sin\psi>0$, $\psi={\rm angle}(\hat{n}\times\hat{\Omega},\hat{p}\times\hat{\Omega})$, where ${\rm angle}(\boldsymbol{a},\boldsymbol{b})$ is a function defined to represent the angle between vectors $\boldsymbol{a}$ and $\boldsymbol{b}$.
If $(\hat{p}\times\hat{\Omega}_{\rm E})\cdot\hat{n}<0$, $\psi\in(-\pi,0)$, $\sin\psi<0$, $\psi={\rm angle}(\hat{n}\times\hat{\Omega},-\hat{p}\times\hat{\Omega})-\pi$. The coordinates of $\hat{\Omega}$ can be obtained from rotating $\hat{\Omega}_{\rm E}$ by angle $\zeta$ around axis $\hat{p}\times\hat{\Omega}_{\rm E}$. Then we have
$$\sin\psi_{i}={\rm sign}[(\hat{p}_i\times\hat{\Omega})\cdot\hat{n}]\frac{|(\hat{p}_{i}\times\hat{\Omega})\times(\hat{n}\times\hat{\Omega})|}{|\hat{p}_{i}\times\hat{\Omega}||\hat{n}\times\hat{\Omega}|},$$
$$\cos\psi_{i}=\frac{(\hat{p}_{i}\times\hat{\Omega})\cdot(\hat{n}\times\hat{\Omega})}{|\hat{p}_{i}\times\hat{\Omega}||\hat{n}\times\hat{\Omega}|}.$$

\bibliographystyle{yahapj}
\bibliography{refer}

\begin{thebibliography}{}
\expandafter\ifx\csname natexlab\endcsname\relax\def\natexlab#1{#1}\fi
\providecommand{\url}[1]{\href{#1}{#1}}
\providecommand{\dodoi}[1]{doi:~\href{http://doi.org/#1}{\nolinkurl{#1}}}
\providecommand{\doeprint}[1]{\href{http://ascl.net/#1}{\nolinkurl{http://ascl.net/#1}}}
\providecommand{\doarXiv}[1]{\href{https://arxiv.org/abs/#1}{\nolinkurl{https://arxiv.org/abs/#1}}}

\bibitem[{{Abuter} {et~al.}(2020){Abuter}, {Amorim}, {Baub{\"o}ck}, {Berger},
  {Bonnet}, {Brandner}, {Cardoso}, {Cl{\'e}net}, {de Zeeuw}, {Dexter},
  {Eckart}, {Eisenhauer}, {F{\"o}rster Schreiber}, {Garcia}, {Gao}, {Gendron},
  {Genzel}, {Gillessen}, {Habibi}, {Haubois}, {Henning}, {Hippler}, {Horrobin},
  {Jim{\'e}nez-Rosales}, {Jochum}, {Jocou}, {Kaufer}, {Kervella}, {Lacour},
  {Lapeyr{\`e}re}, {Le Bouquin}, {L{\'e}na}, {Nowak}, {Ott}, {Paumard},
  {Perraut}, {Perrin}, {Pfuhl}, {Rodr{\'\i}guez-Coira}, {Shangguan},
  {Scheithauer}, {Stadler}, {Straub}, {Straubmeier}, {Sturm}, {Tacconi},
  {Vincent}, {von Fellenberg}, {Waisberg}, {Widmann}, {Wieprecht}, {Wiezorrek},
  {Woillez}, {Yazici}, \& {Zins}}]{2020A&A...636L...5G}
{Abuter}, R., {Amorim}, A., {Baub{\"o}ck}, M., {et~al.} 2020, \aap, 636, L5,
  \dodoi{10.1051/0004-6361/202037813}

\bibitem[{{Ajith} {et~al.}(2008){Ajith}, {Babak}, {Chen}, {Hewitson},
  {Krishnan}, {Sintes}, {Whelan}, {Br{\"u}gmann}, {Diener}, {Dorband},
  {Gonzalez}, {Hannam}, {Husa}, {Pollney}, {Rezzolla}, {Santamar{\'{\i}}a},
  {Sperhake}, \& {Thornburg}}]{2008PhRvD..77j4017A}
{Ajith}, P., {Babak}, S., {Chen}, Y., {et~al.} 2008, \prd, 77, 104017,
  \dodoi{10.1103/PhysRevD.77.104017}

\bibitem[{{Anholm} {et~al.}(2009){Anholm}, {Ballmer}, {Creighton}, {Price}, \&
  {Siemens}}]{2009PhRvD..79h4030A}
{Anholm}, M., {Ballmer}, S., {Creighton}, J.~D.~E., {Price}, L.~R., \&
  {Siemens}, X. 2009, \prd, 79, 084030, \dodoi{10.1103/PhysRevD.79.084030}

\bibitem[{{Antoniadis} {et~al.}(2022){Antoniadis}, {Arzoumanian}, {Babak},
  {Bailes}, {Bak Nielsen}, {Baker}, {Bassa}, {B{\'e}csy}, {Berthereau},
  {Bonetti}, {Brazier}, {Brook}, {Burgay}, {Burke-Spolaor}, {Caballero},
  {Casey-Clyde}, {Chalumeau}, {Champion}, {Charisi}, {Chatterjee}, {Chen},
  {Cognard}, {Cordes}, {Cornish}, {Crawford}, {Cromartie}, {Crowter}, {Dai},
  {DeCesar}, {Demorest}, {Desvignes}, {Dolch}, {Drachler}, {Falxa}, {Ferrara},
  {Fiore}, {Fonseca}, {Gair}, {Garver-Daniels}, {Goncharov}, {Good}, {Graikou},
  {Guillemot}, {Guo}, {Hazboun}, {Hobbs}, {Hu}, {Islo}, {Janssen}, {Jennings},
  {Johnson}, {Jones}, {Kaiser}, {Kaplan}, {Karuppusamy}, {Keith}, {Kelley},
  {Kerr}, {Key}, {Kramer}, {Lam}, {Lamb}, {Lazio}, {Lee}, {Lentati}, {Liu},
  {Luo}, {Lynch}, {Lyne}, {Madison}, {Main}, {Manchester}, {McEwen}, {McKee},
  {McLaughlin}, {Mickaliger}, {Mingarelli}, {Ng}, {Nice}, {Os{\l}owski},
  {Parthasarathy}, {Pennucci}, {Perera}, {Perrodin}, {Petiteau}, {Pol},
  {Porayko}, {Possenti}, {Ransom}, {Ray}, {Reardon}, {Russell}, {Samajdar},
  {Sampson}, {Sanidas}, {Sarkissian}, {Schmitz}, {Schult}, {Sesana},
  {Shaifullah}, {Shannon}, {Shapiro-Albert}, {Siemens}, {Simon}, {Smith},
  {Speri}, {Spiewak}, {Stairs}, {Stappers}, {Stinebring}, {Swiggum}, {Taylor},
  {Theureau}, {Tiburzi}, {Vallisneri}, {van der Wateren}, {Vecchio},
  {Verbiest}, {Vigeland}, {Wahl}, {Wang}, {Wang}, {Wang}, {Witt}, {Zhang}, \&
  {Zhu}}]{2022MNRAS.510.4873A}
{Antoniadis}, J., {Arzoumanian}, Z., {Babak}, S., {et~al.} 2022, \mnras, 510,
  4873, \dodoi{10.1093/mnras/stab3418}

\bibitem[{{Apostolatos} {et~al.}(1994){Apostolatos}, {Cutler}, {Sussman}, \&
  {Thorne}}]{1994PhRvD..49.6274A}
{Apostolatos}, T.~A., {Cutler}, C., {Sussman}, G.~J., \& {Thorne}, K.~S. 1994,
  \prd, 49, 6274, \dodoi{10.1103/PhysRevD.49.6274}

\bibitem[{{Arzoumanian} {et~al.}(2014){Arzoumanian}, {Brazier},
  {Burke-Spolaor}, {Chamberlin}, {Chatterjee}, {Cordes}, {Demorest}, {Deng},
  {Dolch}, {Ellis}, {Ferdman}, {Garver-Daniels}, {Jenet}, {Jones}, {Kaspi},
  {Koop}, {Lam}, {Lazio}, {Lommen}, {Lorimer}, {Luo}, {Lynch}, {Madison},
  {McLaughlin}, {McWilliams}, {Nice}, {Palliyaguru}, {Pennucci}, {Ransom},
  {Sesana}, {Siemens}, {Stairs}, {Stinebring}, {Stovall}, {Swiggum},
  {Vallisneri}, {van Haasteren}, {Wang}, {Zhu}, \& {NANOGrav
  Collaboration}}]{2014ApJ...794..141A}
{Arzoumanian}, Z., {Brazier}, A., {Burke-Spolaor}, S., {et~al.} 2014, \apj,
  794, 141, \dodoi{10.1088/0004-637X/794/2/141}

\bibitem[{{Arzoumanian} {et~al.}(2020){Arzoumanian}, {Baker}, {Blumer},
  {B{\'e}csy}, {Brazier}, {Brook}, {Burke-Spolaor}, {Chatterjee}, {Chen},
  {Cordes}, {Cornish}, {Crawford}, {Cromartie}, {Decesar}, {Demorest}, {Dolch},
  {Ellis}, {Ferrara}, {Fiore}, {Fonseca}, {Garver-Daniels}, {Gentile}, {Good},
  {Hazboun}, {Holgado}, {Islo}, {Jennings}, {Jones}, {Kaiser}, {Kaplan},
  {Kelley}, {Key}, {Laal}, {Lam}, {Lazio}, {Lorimer}, {Luo}, {Lynch},
  {Madison}, {McLaughlin}, {Mingarelli}, {Ng}, {Nice}, {Pennucci}, {Pol},
  {Ransom}, {Ray}, {Shapiro-Albert}, {Siemens}, {Simon}, {Spiewak}, {Stairs},
  {Stinebring}, {Stovall}, {Sun}, {Swiggum}, {Taylor}, {Turner}, {Vallisneri},
  {Vigeland}, {Witt}, \& {Nanograv Collaboration}}]{2020ApJ...905L..34A}
{Arzoumanian}, Z., {Baker}, P.~T., {Blumer}, H., {et~al.} 2020, \apjl, 905,
  L34, \dodoi{10.3847/2041-8213/abd401}

\bibitem[{{Arzoumanian} {et~al.}(2021{\natexlab{a}}){Arzoumanian}, {Baker},
  {Blumer}, {B{\'e}csy}, {Brazier}, {Brook}, {Burke-Spolaor}, {Charisi},
  {Chatterjee}, {Chen}, {Cordes}, {Cornish}, {Crawford}, {Cromartie},
  {Decesar}, {Degan}, {Demorest}, {Dolch}, {Drachler}, {Ellis}, {Ferrara},
  {Fiore}, {Fonseca}, {Garver-Daniels}, {Gentile}, {Good}, {Hazboun},
  {Holgado}, {Islo}, {Jennings}, {Jones}, {Kaiser}, {Kaplan}, {Kelley}, {Key},
  {Laal}, {Lam}, {W. Lazio}, {Lorimer}, {Liu}, {Luo}, {Lynch}, {Madison},
  {McEwen}, {McLaughlin}, {Mingarelli}, {Ng}, {Nice}, {Olum}, {Pennucci},
  {Pol}, {Ransom}, {Ray}, {Romano}, {Sardesai}, {Shapiro-Albert}, {Siemens},
  {Simon}, {Siwek}, {Spiewak}, {Stairs}, {Stinebring}, {Stovall}, {Sun},
  {Swiggum}, {Taylor}, {Turner}, {Vallisneri}, {Vigeland}, {Wahl}, {Witt}, \&
  {NANOGRAV Collaboration}}]{2021ApJ...923L..22A}
---. 2021{\natexlab{a}}, \apjl, 923, L22, \dodoi{10.3847/2041-8213/ac401c}

\bibitem[{{Arzoumanian} {et~al.}(2021{\natexlab{b}}){Arzoumanian}, {Baker},
  {Brazier}, {Brook}, {Burke-Spolaor}, {Becsy}, {Charisi}, {Chatterjee},
  {Cordes}, {Cornish}, {Crawford}, {Cromartie}, {Decesar}, {Demorest}, {Dolch},
  {Elliott}, {Ellis}, {Ferrara}, {Fonseca}, {Garver-Daniels}, {Gentile},
  {Good}, {Hazboun}, {Islo}, {Jennings}, {Jones}, {Kaiser}, {Kaplan}, {Kelley},
  {Key}, {Lam}, {Lazio}, {Luo}, {Lynch}, {Ma}, {Madison}, {McLaughlin},
  {Mingarelli}, {Ng}, {Nice}, {Pennucci}, {Pol}, {Ransom}, {Ray},
  {Shapiro-Albert}, {Siemens}, {Simon}, {Spiewak}, {Stairs}, {Stinebring},
  {Stovall}, {Swiggum}, {Taylor}, {Vallisneri}, {Vigeland}, {Witt}, \&
  {Nanograv Collaboration}}]{2021ApJ...914..121A}
{Arzoumanian}, Z., {Baker}, P.~T., {Brazier}, A., {et~al.} 2021{\natexlab{b}},
  \apj, 914, 121, \dodoi{10.3847/1538-4357/abfcd3}

\bibitem[{{Babak} \& {Sesana}(2012)}]{2012PhRvD..85d4034B}
{Babak}, S., \& {Sesana}, A. 2012, \prd, 85, 044034,
  \dodoi{10.1103/PhysRevD.85.044034}

\bibitem[{{Begelman} {et~al.}(1980){Begelman}, {Blandford}, \&
  {Rees}}]{1980Natur.287..307B}
{Begelman}, M.~C., {Blandford}, R.~D., \& {Rees}, M.~J. 1980, \nat, 287, 307,
  \dodoi{10.1038/287307a0}

\bibitem[{{Blair} {et~al.}(2015){Blair}, {Ju}, {Zhao}, {Wen}, {Chu}, {Fang},
  {Cai}, {Gao}, {Lin}, {Liu}, {Wu}, {Zhu}, {Reitze}, {Arai}, {Zhang},
  {Flaminio}, {Zhu}, {Hobbs}, {Manchester}, {Shannon}, {Baccigalupi}, {Gao},
  {Xu}, {Bian}, {Cao}, {Chang}, {Dong}, {Gong}, {Huang}, {Ju}, {Luo}, {Qiang},
  {Tang}, {Wan}, {Wang}, {Xu}, {Zang}, {Zhang}, {Lau}, \&
  {Ni}}]{2015SCPMA..58.5748B}
{Blair}, D., {Ju}, L., {Zhao}, C., {et~al.} 2015, Science China Physics,
  Mechanics, and Astronomy, 58, 5748, \dodoi{10.1007/s11433-015-5748-6}

\bibitem[{{Boehle} {et~al.}(2016){Boehle}, {Ghez}, {Sch{\"o}del}, {Meyer},
  {Yelda}, {Albers}, {Martinez}, {Becklin}, {Do}, {Lu}, {Matthews}, {Morris},
  {Sitarski}, \& {Witzel}}]{2016ApJ...830...17B}
{Boehle}, A., {Ghez}, A.~M., {Sch{\"o}del}, R., {et~al.} 2016, \apj, 830, 17,
  \dodoi{10.3847/0004-637X/830/1/17}

\bibitem[{{Boyce} {et~al.}(2017){Boyce}, {L{\"u}tzgendorf}, {van der Marel},
  {Baumgardt}, {Kissler-Patig}, {Neumayer}, \& {de
  Zeeuw}}]{2017ApJ...846...14B}
{Boyce}, H., {L{\"u}tzgendorf}, N., {van der Marel}, R.~P., {et~al.} 2017,
  \apj, 846, 14, \dodoi{10.3847/1538-4357/aa830c}

\bibitem[{Brazier {et~al.}(2016)Brazier, Lassus, Petiteau, Possenti, Sesana,
  Vecchio, Lyne, Christy, Perera, Stappers, Tiburzi, Bassa, Mingarelli,
  Perrodin, Reardon, Champion, Nice, Madison, Stinebring, Fonseca, Graikou,
  Desvignes, Hobbs, Jones, Shaifullah, Theureau, Janssen, Cognard, Stairs,
  Mckee, Simon, Ellis, Swiggum, Cordes, Gair, Hessels, Wang, Liu, Stovall, Lee,
  Guillemot, Lentati, Levin, Toomey, Wen, Burgay, Kerr, Kramer, Vallisneri,
  McLaughlin, Gonzalez, Keith, Lam, Bhat, Garver-Daniels, Palliyaguru, Brem,
  Gentile, Lazarus, Rosado, Demorest, Lasky, Karuppusamy, Smits, Spiewak, van
  Haasteren, Ferdman, Shannon, Caballero, Manchester, Lynch, Babak,
  Burke-Spolaor, Chatterjee, Dai, Osłowski, Sanidas, Chamberlin, Ransom,
  Taylor, McWilliams, Dolch, Lazio, Pennucci, van Straten, Zhu, Siemens, You,
  Zhu, Wang, Arzoumanian, \& Verbiest}]{10.1093/mnras/stw347}
Brazier, A., Lassus, A., Petiteau, A., {et~al.} 2016, Monthly Notices of the
  Royal Astronomical Society, 458, 1267, \dodoi{10.1093/mnras/stw347}

\bibitem[{{Charisi} {et~al.}(2022){Charisi}, {Taylor}, {Runnoe}, {Bogdanovic},
  \& {Trump}}]{2022MNRAS.510.5929C}
{Charisi}, M., {Taylor}, S.~R., {Runnoe}, J., {Bogdanovic}, T., \& {Trump},
  J.~R. 2022, \mnras, 510, 5929, \dodoi{10.1093/mnras/stab3713}

\bibitem[{{Chen} {et~al.}(2021{\natexlab{a}}){Chen}, {Caballero}, {Guo},
  {Chalumeau}, {Liu}, {Shaifullah}, {Lee}, {Babak}, {Desvignes},
  {Parthasarathy}, {Hu}, {van der Wateren}, {Antoniadis}, {Bak Nielsen},
  {Bassa}, {Berthereau}, {Burgay}, {Champion}, {Cognard}, {Falxa}, {Ferdman},
  {Freire}, {Gair}, {Graikou}, {Guillemot}, {Jang}, {Janssen}, {Karuppusamy},
  {Keith}, {Kramer}, {Liu}, {Lyne}, {Main}, {McKee}, {Mickaliger}, {Perera},
  {Perrodin}, {Petiteau}, {Porayko}, {Possenti}, {Samajdar}, {Sanidas},
  {Sesana}, {Speri}, {Stappers}, {Theureau}, {Tiburzi}, {Vecchio}, {Verbiest},
  {Wang}, {Wang}, \& {Xu}}]{2021MNRAS.508.4970C}
{Chen}, S., {Caballero}, R.~N., {Guo}, Y.~J., {et~al.} 2021{\natexlab{a}},
  \mnras, 508, 4970, \dodoi{10.1093/mnras/stab2833}

\bibitem[{{Chen} {et~al.}(2020){Chen}, {Yu}, \& {Lu}}]{chen2020dynamical}
{Chen}, Y., {Yu}, Q., \& {Lu}, Y. 2020, \apj, 897, 86,
  \dodoi{10.3847/1538-4357/ab9594}

\bibitem[{{Chen} {et~al.}(2021{\natexlab{b}}){Chen}, {Yuan}, \&
  {Huang}}]{2021SCPMA..6420412C}
{Chen}, Z.-C., {Yuan}, C., \& {Huang}, Q.-G. 2021{\natexlab{b}}, Science China
  Physics, Mechanics, and Astronomy, 64, 120412,
  \dodoi{10.1007/s11433-021-1797-y}

\bibitem[{{Cordes} \& {Chatterjee}(2019)}]{2019ARA&A..57..417C}
{Cordes}, J.~M., \& {Chatterjee}, S. 2019, \araa, 57, 417,
  \dodoi{10.1146/annurev-astro-091918-104501}

\bibitem[{{Creighton} \& {Anderson}(2011)}]{2011gwpa.book.....C}
{Creighton}, J., \& {Anderson}, W. 2011, {Gravitational-Wave Physics and
  Astronomy: An Introduction to Theory, Experiment and Data Analysis.} (Wiley
  -VCH Verlag GmbH \& Co. KGaA)

\bibitem[{{Deng} \& {Finn}(2011)}]{2011MNRAS.414...50D}
{Deng}, X., \& {Finn}, L.~S. 2011, \mnras, 414, 50,
  \dodoi{10.1111/j.1365-2966.2010.17913.x}

\bibitem[{{Detweiler}(1979)}]{1979ApJ...234.1100D}
{Detweiler}, S. 1979, \apj, 234, 1100, \dodoi{10.1086/157593}

\bibitem[{{Do} {et~al.}(2019){Do}, {Hees}, {Ghez}, {Martinez}, {Chu}, {Jia},
  {Sakai}, {Lu}, {Gautam}, {O{\textquoteright}Neil}, {Becklin}, {Morris},
  {Matthews}, {Nishiyama}, {Campbell}, {Chappell}, {Chen}, {Ciurlo},
  {Dehghanfar}, {Gallego-Cano}, {Kerzendorf}, {Lyke}, {Naoz}, {Saida},
  {Sch{\"o}del}, {Takahashi}, {Takamori}, {Witzel}, \&
  {Wizinowich}}]{2019Sci...365..664D}
{Do}, T., {Hees}, A., {Ghez}, A., {et~al.} 2019, Science, 365, 664,
  \dodoi{10.1126/science.aav8137}

\bibitem[{{D'Orazio} \& {Loeb}(2021)}]{2021PhRvD.104f3015D}
{D'Orazio}, D.~J., \& {Loeb}, A. 2021, \prd, 104, 063015,
  \dodoi{10.1103/PhysRevD.104.063015}

\bibitem[{{Ellis} {et~al.}(2012){Ellis}, {Siemens}, \&
  {Creighton}}]{2012ApJ...756..175E}
{Ellis}, J.~A., {Siemens}, X., \& {Creighton}, J.~D.~E. 2012, \apj, 756, 175,
  \dodoi{10.1088/0004-637X/756/2/175}

\bibitem[{{Erkal} {et~al.}(2019){Erkal}, {Boubert}, {Gualandris}, {Evans}, \&
  {Antonini}}]{2019MNRAS.483.2007E}
{Erkal}, D., {Boubert}, D., {Gualandris}, A., {Evans}, N.~W., \& {Antonini}, F.
  2019, \mnras, 483, 2007, \dodoi{10.1093/mnras/sty2674}

\bibitem[{{Fang} {et~al.}(2019){Fang}, {Chen}, \&
  {Huang}}]{2019ApJ...887..210F}
{Fang}, Y., {Chen}, X., \& {Huang}, Q.-G. 2019, \apj, 887, 210,
  \dodoi{10.3847/1538-4357/ab510e}

\bibitem[{Genzel {et~al.}(2010)Genzel, Eisenhauer, \&
  Gillessen}]{RevModPhys.82.3121}
Genzel, R., Eisenhauer, F., \& Gillessen, S. 2010, Rev. Mod. Phys., 82, 3121,
  \dodoi{10.1103/RevModPhys.82.3121}

\bibitem[{{Gillessen} {et~al.}(2017){Gillessen}, {Plewa}, {Eisenhauer}, {Sari},
  {Waisberg}, {Habibi}, {Pfuhl}, {George}, {Dexter}, {von Fellenberg}, {Ott},
  \& {Genzel}}]{2017ApJ...837...30G}
{Gillessen}, S., {Plewa}, P.~M., {Eisenhauer}, F., {et~al.} 2017, \apj, 837,
  30, \dodoi{10.3847/1538-4357/aa5c41}

\bibitem[{{Girma} \& {Loeb}(2018)}]{2018MNRAS.tmp.2529G}
{Girma}, E., \& {Loeb}, A. 2018, \mnras, \dodoi{10.1093/mnras/sty2643}

\bibitem[{{Goldstein} {et~al.}(2019){Goldstein}, {Sesana}, {Holgado}, \&
  {Veitch}}]{2019MNRAS.485..248G}
{Goldstein}, J.~M., {Sesana}, A., {Holgado}, A.~M., \& {Veitch}, J. 2019,
  \mnras, 485, 248, \dodoi{10.1093/mnras/stz420}

\bibitem[{{Goncharov} {et~al.}(2020){Goncharov}, {Zhu}, \&
  {Thrane}}]{2020MNRAS.497.3264G}
{Goncharov}, B., {Zhu}, X.-J., \& {Thrane}, E. 2020, \mnras, 497, 3264,
  \dodoi{10.1093/mnras/staa2081}

\bibitem[{{Goncharov} {et~al.}(2021){Goncharov}, {Shannon}, {Reardon}, {Hobbs},
  {Zic}, {Bailes}, {Cury{\l}o}, {Dai}, {Kerr}, {Lower}, {Manchester}, {Mandow},
  {Middleton}, {Miles}, {Parthasarathy}, {Thrane}, {Thyagarajan}, {Xue}, {Zhu},
  {Cameron}, {Feng}, {Luo}, {Russell}, {Sarkissian}, {Spiewak}, {Wang}, {Wang},
  {Zhang}, \& {Zhang}}]{2021ApJ...917L..19G}
{Goncharov}, B., {Shannon}, R.~M., {Reardon}, D.~J., {et~al.} 2021, \apjl, 917,
  L19, \dodoi{10.3847/2041-8213/ac17f4}

\bibitem[{{Gourgoulhon} {et~al.}(2019){Gourgoulhon}, {Le Tiec}, {Vincent}, \&
  {Warburton}}]{2019arXiv190302049G}
{Gourgoulhon}, E., {Le Tiec}, A., {Vincent}, F.~H., \& {Warburton}, N. 2019,
  arXiv e-prints.
\newblock \doarXiv{1903.02049}

\bibitem[{{Guo} \& {Lu}(2022)}]{2022PhRvD.106b3018G}
{Guo}, X., \& {Lu}, Y. 2022, \prd, 106, 023018,
  \dodoi{10.1103/PhysRevD.106.023018}

\bibitem[{{Hawking} \& {Israel}(1989)}]{1989thyg.book.....H}
{Hawking}, S.~W., \& {Israel}, W. 1989, {Three Hundred Years of Gravitation},
  704

\bibitem[{{Hellings} \& {Downs}(1983)}]{1983ApJ...265L..39H}
{Hellings}, R.~W., \& {Downs}, G.~S. 1983, \apjl, 265, L39,
  \dodoi{10.1086/183954}

\bibitem[{{Huerta} {et~al.}(2015){Huerta}, {McWilliams}, {Gair}, \&
  {Taylor}}]{2015PhRvD..92f3010H}
{Huerta}, E.~A., {McWilliams}, S.~T., {Gair}, J.~R., \& {Taylor}, S.~R. 2015,
  \prd, 92, 063010, \dodoi{10.1103/PhysRevD.92.063010}

\bibitem[{{Jaranowski} {et~al.}(1996){Jaranowski}, {Kokkotas}, {Kr{\'o}lak}, \&
  {Tsegas}}]{1996CQGra..13.1279J}
{Jaranowski}, P., {Kokkotas}, K.~D., {Kr{\'o}lak}, A., \& {Tsegas}, G. 1996,
  Classical and Quantum Gravity, 13, 1279, \dodoi{10.1088/0264-9381/13/6/004}

\bibitem[{{Joshi} {et~al.}(2018){Joshi}, {Arumugasamy}, {Bagchi},
  {Bandyopadhyay}, {Basu}, {Dhanda Batra}, {Bethapudi}, {Choudhary}, {De},
  {Dey}, {Gopakumar}, {Gupta}, {Krishnakumar}, {Maan}, {Manoharan}, {Naidu},
  {Nandi}, {Pathak}, {Surnis}, \& {Susobhanan}}]{2018JApA...39...51J}
{Joshi}, B.~C., {Arumugasamy}, P., {Bagchi}, M., {et~al.} 2018, Journal of
  Astrophysics and Astronomy, 39, 51, \dodoi{10.1007/s12036-018-9549-y}

\bibitem[{{Kocsis} {et~al.}(2012){Kocsis}, {Ray}, \& {Portegies
  Zwart}}]{2012ApJ...752...67K}
{Kocsis}, B., {Ray}, A., \& {Portegies Zwart}, S. 2012, \apj, 752, 67,
  \dodoi{10.1088/0004-637X/752/1/67}

\bibitem[{{Kramer} \& {Champion}(2013)}]{2013CQGra..30v4009K}
{Kramer}, M., \& {Champion}, D.~J. 2013, Classical and Quantum Gravity, 30,
  224009, \dodoi{10.1088/0264-9381/30/22/224009}

\bibitem[{Lazio(2013)}]{Lazio2013SKA}
Lazio, T. J.~W. 2013, Classical and Quantum Gravity, 30, 224011.
\newblock \url{http://stacks.iop.org/0264-9381/30/i=22/a=224011}

\bibitem[{{Lee}(2016)}]{2016ASPC..502...19L}
{Lee}, K.~J. 2016, in Astronomical Society of the Pacific Conference Series,
  Vol. 502, Frontiers in Radio Astronomy and FAST Early Sciences Symposium
  2015, ed. L.~{Qain} \& D.~{Li}, 19

\bibitem[{{Lee} {et~al.}(2011){Lee}, {Wex}, {Kramer}, {Stappers}, {Bassa},
  {Janssen}, {Karuppusamy}, \& {Smits}}]{2011MNRAS.414.3251L}
{Lee}, K.~J., {Wex}, N., {Kramer}, M., {et~al.} 2011, \mnras, 414, 3251,
  \dodoi{10.1111/j.1365-2966.2011.18622.x}

\bibitem[{{Lentati} {et~al.}(2016){Lentati}, {Shannon}, {Coles}, {Verbiest},
  {van Haasteren}, {Ellis}, {Caballero}, {Manchester}, {Arzoumanian}, {Babak},
  {Bassa}, {Bhat}, {Brem}, {Burgay}, {Burke-Spolaor}, {Champion}, {Chatterjee},
  {Cognard}, {Cordes}, {Dai}, {Demorest}, {Desvignes}, {Dolch}, {Ferdman},
  {Fonseca}, {Gair}, {Gonzalez}, {Graikou}, {Guillemot}, {Hessels}, {Hobbs},
  {Janssen}, {Jones}, {Karuppusamy}, {Keith}, {Kerr}, {Kramer}, {Lam}, {Lasky},
  {Lassus}, {Lazarus}, {Lazio}, {Lee}, {Levin}, {Liu}, {Lynch}, {Madison},
  {McKee}, {McLaughlin}, {McWilliams}, {Mingarelli}, {Nice}, {Os{\l}owski},
  {Pennucci}, {Perera}, {Perrodin}, {Petiteau}, {Possenti}, {Ransom},
  {Reardon}, {Rosado}, {Sanidas}, {Sesana}, {Shaifullah}, {Siemens}, {Smits},
  {Stairs}, {Stappers}, {Stinebring}, {Stovall}, {Swiggum}, {Taylor},
  {Theureau}, {Tiburzi}, {Toomey}, {Vallisneri}, {van Straten}, {Vecchio},
  {Wang}, {Wang}, {You}, {Zhu}, \& {Zhu}}]{2016MNRAS.458.2161L}
{Lentati}, L., {Shannon}, R.~M., {Coles}, W.~A., {et~al.} 2016, \mnras, 458,
  2161, \dodoi{10.1093/mnras/stw395}

\bibitem[{{Lindblom} {et~al.}(2008){Lindblom}, {Owen}, \&
  {Brown}}]{2008PhRvD..78l4020L}
{Lindblom}, L., {Owen}, B.~J., \& {Brown}, D.~A. 2008, \prd, 78, 124020,
  \dodoi{10.1103/PhysRevD.78.124020}

\bibitem[{Maggiore(2008)}]{book:1417639}
Maggiore, M. 2008, Gravitational waves vol.1 Theory and Experiments (Oxford
  University Press).
\newblock
  \url{http://gen.lib.rus.ec/book/index.php?md5=ee1879513fb76a776528f459e6fbbc31}

\bibitem[{{Manchester} {et~al.}(2005){Manchester}, {Hobbs}, {Teoh}, \&
  {Hobbs}}]{2005AJ....129.1993M}
{Manchester}, R.~N., {Hobbs}, G.~B., {Teoh}, A., \& {Hobbs}, M. 2005, \aj, 129,
  1993, \dodoi{10.1086/428488}

\bibitem[{{Manchester} \& {IPTA}(2013)}]{2013CQGra..30v4010M}
{Manchester}, R.~N., \& {IPTA}. 2013, Classical and Quantum Gravity, 30,
  224010, \dodoi{10.1088/0264-9381/30/22/224010}

\bibitem[{Manchester {et~al.}(2013)Manchester, Hobbs, Bailes, Coles,
  Van~Straten, Keith, Shannon, Bhat, Brown, Burkespolaor,
  {et~al.}}]{manchester2013the}
Manchester, R.~N., Hobbs, G., Bailes, M., {et~al.} 2013, Publications of the
  Astronomical Society of Australia, 30, 17

\bibitem[{{McGrath} \& {Creighton}(2021)}]{2021MNRAS.505.4531M}
{McGrath}, C., \& {Creighton}, J. 2021, \mnras, 505, 4531,
  \dodoi{10.1093/mnras/stab1417}

\bibitem[{{McLaughlin}(2013)}]{2013CQGra..30v4008M}
{McLaughlin}, M.~A. 2013, Classical and Quantum Gravity, 30, 224008,
  \dodoi{10.1088/0264-9381/30/22/224008}

\bibitem[{Mingarelli(2015)}]{mingarelli2015gravitational}
Mingarelli, C.~M. 2015, Gravitational wave astrophysics with pulsar timing
  arrays (Springer)

\bibitem[{{Mingarelli} {et~al.}(2017){Mingarelli}, {Lazio}, {Sesana}, {Greene},
  {Ellis}, {Ma}, {Croft}, {Burke-Spolaor}, \& {Taylor}}]{2017NatAs...1..886M}
{Mingarelli}, C. M.~F., {Lazio}, T. J.~W., {Sesana}, A., {et~al.} 2017, Nature
  Astronomy, 1, 886, \dodoi{10.1038/s41550-017-0299-6}

\bibitem[{{Moore} {et~al.}(2015{\natexlab{a}}){Moore}, {Cole}, \&
  {Berry}}]{2015CQGra..32a5014M}
{Moore}, C.~J., {Cole}, R.~H., \& {Berry}, C.~P.~L. 2015{\natexlab{a}},
  Classical and Quantum Gravity, 32, 015014,
  \dodoi{10.1088/0264-9381/32/1/015014}

\bibitem[{{Moore} {et~al.}(2015{\natexlab{b}}){Moore}, {Taylor}, \&
  {Gair}}]{2015CQGra..32e5004M}
{Moore}, C.~J., {Taylor}, S.~R., \& {Gair}, J.~R. 2015{\natexlab{b}}, Classical
  and Quantum Gravity, 32, 055004, \dodoi{10.1088/0264-9381/32/5/055004}

\bibitem[{{Nan} {et~al.}(2011){Nan}, {Li}, {Jin}, {Wang}, {Zhu}, {Zhu},
  {Zhang}, {Yue}, \& {Qian}}]{2011IJMPD..20..989N}
{Nan}, R., {Li}, D., {Jin}, C., {et~al.} 2011, International Journal of Modern
  Physics D, 20, 989, \dodoi{10.1142/S0218271811019335}

\bibitem[{{Perera} {et~al.}(2019){Perera}, {DeCesar}, {Demorest}, {Kerr},
  {Lentati}, {Nice}, {Os{\l}owski}, {Ransom}, {Keith}, {Arzoumanian}, {Bailes},
  {Baker}, {Bassa}, {Bhat}, {Brazier}, {Burgay}, {Burke-Spolaor}, {Caballero},
  {Champion}, {Chatterjee}, {Chen}, {Cognard}, {Cordes}, {Crowter}, {Dai},
  {Desvignes}, {Dolch}, {Ferdman}, {Ferrara}, {Fonseca}, {Goldstein},
  {Graikou}, {Guillemot}, {Hazboun}, {Hobbs}, {Hu}, {Islo}, {Janssen},
  {Karuppusamy}, {Kramer}, {Lam}, {Lee}, {Liu}, {Luo}, {Lyne}, {Manchester},
  {McKee}, {McLaughlin}, {Mingarelli}, {Parthasarathy}, {Pennucci}, {Perrodin},
  {Possenti}, {Reardon}, {Russell}, {Sanidas}, {Sesana}, {Shaifullah},
  {Shannon}, {Siemens}, {Simon}, {Spiewak}, {Stairs}, {Stappers}, {Swiggum},
  {Taylor}, {Theureau}, {Tiburzi}, {Vallisneri}, {Vecchio}, {Wang}, {Zhang},
  {Zhang}, {Zhu}, \& {Zhu}}]{2019MNRAS.490.4666P}
{Perera}, B.~B.~P., {DeCesar}, M.~E., {Demorest}, P.~B., {et~al.} 2019, \mnras,
  490, 4666, \dodoi{10.1093/mnras/stz2857}

\bibitem[{{Pfahl} \& {Loeb}(2004)}]{2004ApJ...615..253P}
{Pfahl}, E., \& {Loeb}, A. 2004, \apj, 615, 253, \dodoi{10.1086/423975}

\bibitem[{{Pietrzy{\'n}ski} {et~al.}(2013){Pietrzy{\'n}ski}, {Graczyk},
  {Gieren}, {Thompson}, {Pilecki}, {Udalski}, {Soszy{\'n}ski}, {Koz{\l}owski},
  {Konorski}, {Suchomska}, {Bono}, {Moroni}, {Villanova}, {Nardetto},
  {Bresolin}, {Kudritzki}, {Storm}, {Gallenne}, {Smolec}, {Minniti}, {Kubiak},
  {Szyma{\'n}ski}, {Poleski}, {Wyrzykowski}, {Ulaczyk}, {Pietrukowicz},
  {G{\'o}rski}, \& {Karczmarek}}]{2013Natur.495...76P}
{Pietrzy{\'n}ski}, G., {Graczyk}, D., {Gieren}, W., {et~al.} 2013, \nat, 495,
  76, \dodoi{10.1038/nature11878}

\bibitem[{{Portegies Zwart} {et~al.}(2006){Portegies Zwart}, {Baumgardt},
  {McMillan}, {Makino}, {Hut}, \& {Ebisuzaki}}]{2006ApJ...641..319P}
{Portegies Zwart}, S.~F., {Baumgardt}, H., {McMillan}, S.~L.~W., {et~al.} 2006,
  \apj, 641, 319, \dodoi{10.1086/500361}

\bibitem[{{Ransom} {et~al.}(2019){Ransom}, {Brazier}, {Chatterjee}, {Cohen},
  {Cordes}, {DeCesar}, {Demorest}, {Hazboun}, {Lam}, {Lynch}, {McLaughlin},
  {Ransom}, {Siemens}, {Taylor}, \& {Vigeland}}]{2019BAAS...51g.195R}
{Ransom}, S., {Brazier}, A., {Chatterjee}, S., {et~al.} 2019, in Bulletin of
  the American Astronomical Society, Vol.~51, 195.
\newblock \doarXiv{1908.05356}

\bibitem[{{Robson} {et~al.}(2018){Robson}, {Cornish}, \&
  {Liu}}]{2018arXiv180301944R}
{Robson}, T., {Cornish}, N., \& {Liu}, C. 2018, arXiv e-prints.
\newblock \doarXiv{1803.01944}

\bibitem[{{Romano} \& {Cornish}(2017)}]{2017LRR....20....2R}
{Romano}, J.~D., \& {Cornish}, N.~J. 2017, Living Reviews in Relativity, 20, 2,
  \dodoi{10.1007/s41114-017-0004-1}

\bibitem[{{Rosado} {et~al.}(2015){Rosado}, {Sesana}, \&
  {Gair}}]{2015MNRAS.451.2417R}
{Rosado}, P.~A., {Sesana}, A., \& {Gair}, J. 2015, \mnras, 451, 2417,
  \dodoi{10.1093/mnras/stv1098}

\bibitem[{{Sampson} {et~al.}(2015){Sampson}, {Cornish}, \&
  {McWilliams}}]{2015PhRvD..91h4055S}
{Sampson}, L., {Cornish}, N.~J., \& {McWilliams}, S.~T. 2015, \prd, 91, 084055,
  \dodoi{10.1103/PhysRevD.91.084055}

\bibitem[{{Sazhin}(1978)}]{1978SvA....22...36S}
{Sazhin}, M.~V. 1978, \sovast, 22, 36

\bibitem[{{Sch{\"o}del} {et~al.}(2003){Sch{\"o}del}, {Ott}, {Genzel}, {Eckart},
  {Mouawad}, \& {Alexander}}]{2003ApJ...596.1015S}
{Sch{\"o}del}, R., {Ott}, T., {Genzel}, R., {et~al.} 2003, \apj, 596, 1015,
  \dodoi{10.1086/378122}

\bibitem[{{Schutz} \& {Ma}(2016)}]{2016MNRAS.459.1737S}
{Schutz}, K., \& {Ma}, C.-P. 2016, \mnras, 459, 1737,
  \dodoi{10.1093/mnras/stw768}

\bibitem[{{Sesana}(2013)}]{2013CQGra..30x4009S}
{Sesana}, A. 2013, Classical and Quantum Gravity, 30, 244009,
  \dodoi{10.1088/0264-9381/30/24/244009}

\bibitem[{{Sesana} \& {Vecchio}(2010)}]{2010CQGra..27h4016S}
{Sesana}, A., \& {Vecchio}, A. 2010, Classical and Quantum Gravity, 27, 084016,
  \dodoi{10.1088/0264-9381/27/8/084016}

\bibitem[{{Sesana} {et~al.}(2009){Sesana}, {Vecchio}, \&
  {Volonteri}}]{2009MNRAS.394.2255S}
{Sesana}, A., {Vecchio}, A., \& {Volonteri}, M. 2009, \mnras, 394, 2255,
  \dodoi{10.1111/j.1365-2966.2009.14499.x}

\bibitem[{{Smits} {et~al.}(2009){Smits}, {Lorimer}, {Kramer}, {Manchester},
  {Stappers}, {Jin}, {Nan}, \& {Li}}]{2009A&A...505..919S}
{Smits}, R., {Lorimer}, D.~R., {Kramer}, M., {et~al.} 2009, \aap, 505, 919,
  \dodoi{10.1051/0004-6361/200911939}

\bibitem[{{Takekawa} {et~al.}(2019){Takekawa}, {Oka}, {Iwata}, {Tsujimoto}, \&
  {Nomura}}]{2019ApJ...871L...1T}
{Takekawa}, S., {Oka}, T., {Iwata}, Y., {Tsujimoto}, S., \& {Nomura}, M. 2019,
  \apjl, 871, L1, \dodoi{10.3847/2041-8213/aafb07}

\bibitem[{{Taylor}(2021)}]{2021arXiv210513270T}
{Taylor}, S.~R. 2021, arXiv e-prints, arXiv:2105.13270.
\newblock \doarXiv{2105.13270}

\bibitem[{{Taylor} {et~al.}(2016){Taylor}, {Huerta}, {Gair}, \&
  {McWilliams}}]{2016ApJ...817...70T}
{Taylor}, S.~R., {Huerta}, E.~A., {Gair}, J.~R., \& {McWilliams}, S.~T. 2016,
  \apj, 817, 70, \dodoi{10.3847/0004-637X/817/1/70}

\bibitem[{{Taylor} {et~al.}(2019){Taylor}, {Burke-Spolaor}, {Baker}, {Charisi},
  {Islo}, {Kelley}, {Madison}, {Simon}, \& {Vigeland}}]{2019arXiv190308183T}
{Taylor}, S.~R., {Burke-Spolaor}, S., {Baker}, P.~T., {et~al.} 2019, arXiv
  e-prints.
\newblock \doarXiv{1903.08183}

\bibitem[{{Thrane} \& {Romano}(2013)}]{2013PhRvD..88l4032T}
{Thrane}, E., \& {Romano}, J.~D. 2013, \prd, 88, 124032,
  \dodoi{10.1103/PhysRevD.88.124032}

\bibitem[{{Tsuboi} {et~al.}(2017){Tsuboi}, {Kitamura}, {Tsutsumi}, {Uehara},
  {Miyoshi}, {Miyawaki}, \& {Miyazaki}}]{2017ApJ...850L...5T}
{Tsuboi}, M., {Kitamura}, Y., {Tsutsumi}, T., {et~al.} 2017, \apjl, 850, L5,
  \dodoi{10.3847/2041-8213/aa97d3}

\bibitem[{{van Haasteren}(2014)}]{2014gwdd.book.....V}
{van Haasteren}, R. 2014, {Gravitational Wave Detection and Data Analysis for
  Pulsar Timing Arrays} (Springer)

\bibitem[{{Wahlquist}(1987)}]{1987GReGr..19.1101W}
{Wahlquist}, H. 1987, General Relativity and Gravitation, 19, 1101,
  \dodoi{10.1007/BF00759146}

\bibitem[{{Wang} \& {Mohanty}(2017)}]{2017PhRvL.118o1104W}
{Wang}, Y., \& {Mohanty}, S.~D. 2017, Physical Review Letters, 118, 151104,
  \dodoi{10.1103/PhysRevLett.118.151104}

\bibitem[{Wang {et~al.}(2014)Wang, Mohanty, \& Jenet}]{wang2014a}
Wang, Y., Mohanty, S.~D., \& Jenet, F.~A. 2014, The Astrophysical Journal, 795,
  96

\bibitem[{Wang {et~al.}(2015)Wang, Mohanty, \& Jenet}]{wang2015coherent}
---. 2015, The Astrophysical Journal, 815, 125

\bibitem[{{Yu}(2002)}]{2002MNRAS.331..935Y}
{Yu}, Q. 2002, \mnras, 331, 935, \dodoi{10.1046/j.1365-8711.2002.05242.x}

\bibitem[{{Yu} {et~al.}(2007){Yu}, {Lu}, \& {Lin}}]{2007ApJ...666..919Y}
{Yu}, Q., {Lu}, Y., \& {Lin}, D.~N.~C. 2007, \apj, 666, 919,
  \dodoi{10.1086/520622}

\bibitem[{{Yu} \& {Tremaine}(2003)}]{2003ApJ...599.1129Y}
{Yu}, Q., \& {Tremaine}, S. 2003, \apj, 599, 1129, \dodoi{10.1086/379546}

\bibitem[{{Zhang} {et~al.}(2014){Zhang}, {Lu}, \& {Yu}}]{2014ApJ...784..106Z}
{Zhang}, F., {Lu}, Y., \& {Yu}, Q. 2014, \apj, 784, 106,
  \dodoi{10.1088/0004-637X/784/2/106}

\bibitem[{{Zhu} {et~al.}(2016){Zhu}, {Wen}, {Xiong}, {Xu}, {Wang}, {Mohanty},
  {Hobbs}, \& {Manchester}}]{2016MNRAS.461.1317Z}
{Zhu}, X.-J., {Wen}, L., {Xiong}, J., {et~al.} 2016, \mnras, 461, 1317,
  \dodoi{10.1093/mnras/stw1446}

\bibitem[{{Zhu} {et~al.}(2014){Zhu}, {Hobbs}, {Wen}, {Coles}, {Wang},
  {Shannon}, {Manchester}, {Bailes}, {Bhat}, {Burke-Spolaor}, {Dai}, {Keith},
  {Kerr}, {Levin}, {Madison}, {Os{\l}owski}, {Ravi}, {Toomey}, \& {van
  Straten}}]{2014MNRAS.444.3709Z}
{Zhu}, X.~J., {Hobbs}, G., {Wen}, L., {et~al.} 2014, \mnras, 444, 3709,
  \dodoi{10.1093/mnras/stu1717}

\bibitem[{{Zhu} {et~al.}(2015){Zhu}, {Wen}, {Hobbs}, {Zhang}, {Wang},
  {Madison}, {Manchester}, {Kerr}, {Rosado}, \& {Wang}}]{2015MNRAS.449.1650Z}
{Zhu}, X.-J., {Wen}, L., {Hobbs}, G., {et~al.} 2015, \mnras, 449, 1650,
  \dodoi{10.1093/mnras/stv381}

\end{thebibliography}
\end{CJK*}
\end{document}